\documentclass[12pt,letterpaper,onecolumn,oneside]{IEEEtran/IEEEtran}
\ifCLASSINFOpdf
\else
\fi
\usepackage[T5]{fontenc}
\newcommand{\bt}[1]{\textbf{#1}}
\newcommand{\tr}{\text{tr}}
\newcommand{\ve}{\text{vec}}
\newcommand{\re}{\text{Re}}
\newcommand{\im}{\text{Im}}
\usepackage{amsthm}
\usepackage{fancyhdr}
\usepackage{setspace}
\usepackage{algorithm}
\usepackage{stfloats}
\usepackage{times}
\usepackage{textcomp}
\usepackage{eso-pic}
\usepackage{textpos}
\usepackage{eso-pic}
\usepackage{algpseudocode}
\usepackage{amsmath}
\usepackage{amssymb}
\usepackage{amsmath}
\usepackage{graphicx}
\usepackage{subfigure}
\usepackage{caption}
\usepackage{cases}
\usepackage{datenumber}
\setdatetoday
\addtocounter{datenumber}{0}
\setdatebynumber{\thedatenumber}
\makeatletter
\def\ps@headings{%
\def\@oddhead{\mbox{}\scriptsize\rightmark \hfil \thepage}%
\def\@evenhead{\scriptsize\thepage \hfil \leftmark\mbox{}}%
\def\@oddfoot{\scriptsize \@date\hfil DRAFT}%
\def\@evenfoot{\scriptsize DRAFT\hfil \@date}}
\makeatother
\newtheorem{theorem}{{\bf Theorem}}

\newtheorem{proposition}{{\bf Proposition}}

\newtheorem{definition}{{\bf Definition}}

\begin{document}
\title{Secure Transmissions Using Artificial Noise in MIMO Wiretap Interference Channel: A Game Theoretic Approach}
\author{Peyman~Siyari,
        Marwan~Krunz,
        and~Diep~N.~Nguyen,
\thanks{P. Siyari and M. Krunz are with the Department
of Electrical and Computer Engineering, University of Arizona (e-mail:\{psiyari, krunz\}@ email.arizona.edu).

Diep N. Nguyen is with the Faculty of Engineering and Information Technology, University of Technology Sydney, Australia (email: Diep.Nguyen@uts.edu.au).}}
\markboth{}
{\tiny Siyari \MakeLowercase{\textit{et al.}:} Secure Transmissions Using Artificial Noise in MIMO Wiretap Interference Channel: A Game Theoretic Approach}
\maketitle
\begin{abstract}
We consider joint optimization of artificial noise (AN) and information signals in a MIMO wiretap interference network, wherein the transmission of each link may be overheard by several MIMO-capable eavesdroppers. Each information signal is accompanied with AN, generated by the same user to confuse nearby eavesdroppers. Using a noncooperative game, a distributed optimization mechanism is proposed to maximize the secrecy rate of each link. The decision variables here are the covariance matrices for the information signals and ANs. However, the nonconvexity of each link's optimization problem (i.e., best response) makes conventional convex games inapplicable, even to find whether a Nash Equilibrium (NE) exists. To tackle this issue, we analyze the proposed game using a relaxed equilibrium concept, called \textit{quasi-Nash equilibrium} (QNE). Under a constraint qualification condition for each player's problem, the set of QNEs includes the NE of the proposed game. We also derive the conditions for the existence and uniqueness of the resulting QNE. It turns out that the uniqueness conditions are too restrictive, and do not always hold in typical network scenarios. Thus, the proposed game often has multiple QNEs, and convergence to a QNE is not always guaranteed. To overcome these issues, we modify the utility functions of the players by adding several specific terms to each utility function. The modified game converges to a QNE even when multiple QNEs exist. Furthermore, players have the ability to select a desired QNE that optimizes a given social objective (e.g., sum-rate or secrecy sum-rate). Depending on the chosen objective, the amount of signaling overhead as well as the performance of resulting QNE can be controlled. Simulations show that not only we can guarantee the convergence to a QNE, but also due to the QNE selection mechanism, we can achieve a significant improvement in terms of secrecy sum-rate and power efficiency, especially in dense networks.
\end{abstract}
\IEEEpeerreviewmaketitle
\section{Introduction}
\IEEEPARstart{P}{hysical}-layer (PHY-layer) security provides a cost-efficient alternative to cryptographic methods in scenarios where the use of the latter is either impractical or expensive. One of the common settings for PHY-layer security is the wiretap channel. In this channel, a node (Alice) wishes to transmit messages securely to a legitimate receiver (Bob) in the presence of one or more eavesdroppers (Eve). Most PHY-layer security techniques for the wiretap channel are based on an information-theoretic definition of security, namely, the \textit{secrecy capacity}, defined as the largest amount of information that can be confidentially communicated between Alice and Bob \cite{wyner}.

Over the last decade, several PHY-layer security techniques have been proposed. Some of these techniques rely on the use of artificial noise (AN) as a friendly jamming (FJ) signal \cite{goel}. In this method, Alice uses multiple antennas to generate an FJ signal along with the information signal, increasing the interference at Eve but without affecting Bob. The authors in \cite{goel} proposed a simple version of this technique, which relies on MIMO zero-forcing to ensure that the FJ signal falls in the null-space of the channel between Alice and Bob. The interest in using AN for a single link is driven by pragmatic considerations, and not necessarily due to its optimality. In fact, it was shown in \cite{wornell} that in one-Eve scenario, the optimal approach for securing a single link with the knowledge of Eve's location is not to use AN. Complementing the classic AN approach in \cite{goel}, which relies on transmitting the AN in the null-space of the legitimate channel, it was shown in \cite{general} that adding AN to both the legitimate channel and its null-space can further improve the secrecy rate of a link. In the case of multiple eavesdroppers, it was shown in \cite{winkin} that the use of AN can significantly improve the secrecy rate compared to the case when AN is not used.

In a multi-link scenario, where several transmitters wish to convey their messages simultaneously to several legitimate receivers (see Fig. \ref{sysmodel}), the FJ signal of each transmitter must be designed to not interfere with other unintended (but legitimate) receivers in the network. This can be quite challenging when only limited or no coordination is possible between links. Therefore, providing PHY-layer secrecy has to be done in a distributed yet noninterfering manner. Interference management for PHY-layer security involves two conflicting factors. On the one hand, the AN from one transmitter degrades the respective information signals at unintended (but legitimate) receivers. On the other hand, AN also increases the interference at eavesdroppers, and is hence useful in terms of improving the security of the communications. The idea of using interference in networks to provide secrecy was first discussed in \cite{ozan}. Several subsequent works exploited this idea in other applications, such as the coexistence of different protocols on the same channel. For instance, the authors in \cite{kalan} considered a two-link SISO interference network, which resembles a coexistence scenario. They showed that with a careful power control design for both links, one link can assist the other in providing a rate demand guarantee as well as secure the transmission by increasing interference on a single-antenna eavesdropper. For the case of two transmitter-receiver-eavesdropper triples, the authors in \cite{letter} proposed a cooperative beamforming approach to achieve maximum secure degree of freedom for both users. In fact, given the knowledge of co-channel interference at the receivers, a cooperative transmission alignment scheme between transmitters is established such that their respective receivers will get interference-free signals and the eavesdropper corresponding to each link will receive interference. 

In this paper, we consider a peer-to-peer multi-link interference network in which the transmission on each link can be overheard by several external eavesdroppers. Perhaps the closest works to our scenario are \cite{kalan} and \cite{letter}. Apart from the fact that both of these works consider only a two-user scenario, which limits their applicability, in \cite{kalan} one of the users generates only interference to provide PHY-layer security for the other user, so providing the PHY-layer secrecy of the former user is overlooked. In contrast, our work provides PHY-layer security for all users. Moreover, although the work in \cite{letter} considers providing secrecy for both users (in a slightly different network than the one we consider), it requires a significant amount of signaling (i.e., coordination) between the two users. In this paper, we limit the amount of coordination as much as possible. 

In our system model, we assume that the transmission of each information signal is accompanied with AN. Each node in the network is equipped with multiple antennas. Our goal is to design a framework through which the co-channel interference at each legitimate receiver is minimized while the aggregate interference at external eavesdroppers remains high. Because nodes cannot cooperate with each other in our settings, each link independently tries to maximize its secrecy rate by designing the covariance matrices (essentially, the precoders) of its information signal and AN. This independent secrecy optimization can be modeled via noncooperative game theory. Specifically, we design a game-theoretic framework in which the utility of each player (i.e., link) is his secrecy rate, and the player's strategy is to optimize the covariance matrices of information signal and AN. It turns out that finding the best response of each link requires solving a nonconvex optimization problem. Thus, the existence of a Nash Equilibrium (NE) cannot be proved using traditional concepts of convex (concave) games \cite{rosen}. Instead, we study the proposed game based on a relaxed equilibrium concept called \textit{quasi-Nash equilibrium} (QNE)  \cite{pangscut}. A QNE is a solution of a variational inequality (VI) \cite{pang} obtained under the K.K.T optimality conditions of the players' problems. We show that under a constraint qualification (CQ) condition for each player's problem, the set of QNEs also includes the NE. Sufficient conditions for the existence and uniqueness of the resulting QNE are provided. Then, an iterative algorithm is proposed to achieve the unique QNE.

Despite their attractiveness in terms of not requiring link coordination, the (Q)NEs of a purely noncooperative game are often inefficient in terms of the achievable sum-utility (i.e., secrecy sum-rate). Furthermore, the conditions that guarantee the uniqueness of the QNE are dependent on the channel gains between links. The random nature of channel gains greatly reduces the possibility of having a unique QNE, which further limits the effectiveness of the proposed noncooperative game. More specifically, the convergence to a QNE cannot be always achieved. This forces the links to terminate their iterative optimizations at some point, resulting in a low secrecy sum-rate. To overcome this issue, we introduce several modifications to the proposed game. Every modification appears as the addition of a term in the utility function of each player. These modifications allow us to not only guarantee the convergence of the game, but also give links the ability to selectively converge to a specific QNE among multiple QNEs. Selecting a particular QNE is done based on how much it satisfies a particular design criterion. We propose three possibilities for QNE selection, each providing different benefits and requiring a different amounts of communication overhead. The proposed QNE selection algorithm can improve the performance of the formerly proposed noncooperative game while keeping the communication overhead reasonably low.

The concept of QNE has been recently used in \cite{qnefull} in sum-rate maximization in cognitive radio users. However, no effort has been made to improve the performance of achieved QNEs. The work in \cite{jointsense} also considers the use of QNEs to jointly optimize the sensing and power allocation of cognitive radio users in the presence of primary users. Although in this work some improvements have been made on the performance of the resulting QNEs, they are specific to cognitive radios and thus not extendable to other networks. The framework we propose can be generalized to any similarly structured game. Overall, the contributions of this paper are as follows:
\begin{itemize}
\item
We propose a noncooperative game to model PHY-layer secrecy optimization in a multi-link MIMO wiretap interference network. Due to the nonconvexity of each player's optimization problem, the analysis of equilibria is done through the concept of QNE. We show that the set of QNEs includes NE as well.
\item
Because many network scenarios may involve multiple QNEs, the purely noncooperative games do not always guarantee the convergence to a unique QNE. Hence, we introduce several modifications to the proposed game to guarantee the convergence to a QNE. The modifications appear as the additional terms in the utility function of the players and keep the distributed nature of the noncooperative game.
\item
We show that the modified game allows users to select a QNE among multiple QNEs according to a design criterion. QNE selection makes it possible to improve the resulting secrecy sum-rate of the modified game compared to a purely noncooperative game.
\item
We show that the freedom in choosing the design criterion gives a degree of flexibility to the modified game. We propose three different choices for the design criterion, each of which requiring a different level of coordination between links and offering a different amount secrecy sum-rate improvement.
\end{itemize}

The rest of this paper is organized as follows. In Section \ref{sec2}, we introduce the system model. In Section \ref{sec3}, we formulate the optimization of information signal and AN as a noncooperative game. The conditions for the existence and uniqueness of the QNE are established in Section \ref{sec4}. In Section \ref{gra}, we modify the proposed noncooperative game, and introduce the theoretical aspects of our QNE selection method. In Section \ref{sec6} an algorithm that implements the QNE selection is given and practical considerations are discussed. We present a centralized algorithm as a measure of efficiency of our proposed game in Section \ref{seccent}. In Section \ref{sec7} simulation results assess the performance of our algorithms. Finally, Section \ref{sec8} concludes the paper.
\section{System Model}
\label{sec2}
\begin{figure}
\begin{center}
\includegraphics[scale = 0.4]{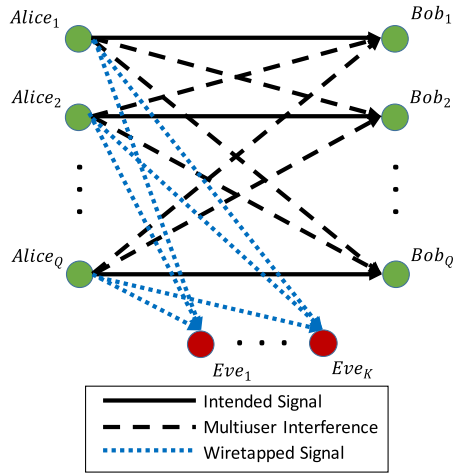}
\end{center}
\centering
\caption{System Model}
\label{sysmodel}
\end{figure}

Consider the network shown in Fig. \ref{sysmodel}, where $Q$ transmitters, $Q> 1$, communicate with $Q$ corresponding receivers. The $q$th transmitter is equipped with $N_{T_q}$ antennas, $q = 1,\dots,Q$. The $q$th receiver has $N_{R_q}$ antennas, $q = 1,\dots,Q$. The link between each transmit-receive (Alice-Bob) pair may experience interference from the other $Q-1$ links. There are $K$ noncolluding Eves overhearing the communications. The $k$th Eve, $k = 1,\dots,K$, has $N_{e,k}$ receive antennas\footnote{The treatment can be easily extended to colluding eavesdroppers by combining the $K$ Eves into one with $\sum_{k=1}^K N_{e,k}$ antennas.}. The received signal at the $q$th receiver, $\bt y_q$, is
\begin{equation}
\bt y_q = \tilde{\bt H}_{qq} \bt u_q+\sum_{\underset{r \neq q}{r = 1}}^Q\tilde{\bt H}_{rq} \bt u_r+\bt n_q,~q \in \mathbb Q
\end{equation}
where $\tilde{\bt H}_{rq}$ ($\tilde{\bt H}_{qq}$) denotes the $N_{R_q}\times N_{T_r}$ ($N_{R_q}\times N_{T_q}$) channel matrix between the $r$th ($q$th) transmitter and $q$th receiver, $\bt u_q$ is the $N_{T_q}\times 1$ vector of transmitted signal from the $q$th transmitter, $\bt n_q$ is the $ N_{R_q} \times 1$ vector of additive noise whose elements are i.i.d zero-mean circularly symmetric complex Gaussian distributed with unit variance, and $\mathbb Q \triangleq \{1,\dots,Q\}$. The term $\sum_{\underset{r \neq q}{r = 1}}^Q\tilde{\bt H}_{rq} \bt u_r$ is the multi-user interference (MUI). The received signal at the $k$th eavesdropper, $\bt z_k$, is expressed as 
\begin{equation}
\bt z_k = \sum_{q = 1}^Q\bt G_{qk}\bt u_q +\bt n_{e,k},~k \in \mathbb K
\end{equation}
where $\bt G_{qk}$ is the $N_{e,k}\times N_{T_q}$ channel matrix between the $q$th transmitter and the $k$th eavesdropper, $\bt n_{e,k}$ is the $N_{e,k}\times 1$ vector of additive noise at the $k$th eavesdropper, and $\mathbb K\triangleq \{1,\dots,K\}$.
The transmitted signal $\bt u_q$ has the following form:
\begin{equation}
\bt u_q \triangleq \bt s_q + \bt w_q
\end{equation}
where $\bt s_q$ is the information signal and $\bt w_q$ is the AN. We use the Gaussian codebook for the information signal and the Gaussian noise for the AN\footnote{Other practical codebooks for the information signal (e.g., QAM) can be approximated to a Gaussian codebook with a capacity gap (see \cite{goldsmith}).}. The matrices ${\bf \Sigma}_q$ and $\bt W_q$ indicate the covariance matrices of $\bt s_q $ and $\bt w_q$, respectively.

The $q$th link, $q \in \mathbb Q$, together with $K$ eavesdroppers form a compound wiretap channel for which the achievable secrecy rate of the $q$th link is written as \cite{compound1}:
\begin{equation}
\label{secrecy}
R^{sec}_{q}({\bf\Sigma}_q, \bt W_q) \triangleq C_q({\bf \Sigma}_q,\bt W_q)-\max_{k \in \mathbb K}C_{e,q,k}({\bf \Sigma}_q,\bt W_q),~q \in \mathbb Q
\end{equation}
where $C_q({\bf \Sigma}_q,\bt W_q)$ is the information rate and $C_{e,q,k}({\bf \Sigma}_q,\bt W_q)$ is the received rate at the $k$th eavesdropper, $k \in \mathbb K$, while eavesdropping on the $q$th link, $q \in \mathbb Q$. Specifically,
\begin{align}
\label{capleg}
&C_q({\bf \Sigma}_q,\bt W_q) \triangleq \ln\left|\bt I+ \bt M_{q}^{-1}\bt H_{qq}{\bf \Sigma_{q}}\bt H_{qq}^H\right| = \nonumber
\\
&\ln\left|\bt M_{q}+ \bt H_{qq}{\bf \Sigma_{q}}\bt H_{qq}^H\right|+\ln\left|\bt M_{q}^{-1}\right|
\end{align}
where $\bt M_q \triangleq \bt I + \bt H_{qq}\bt W_q \bt H^H_{qq}+\sum_{\underset{r\neq q}{r = 1}}^Q\bt H_{rq}\left({\bf \Sigma}_r+\bt W_r\right) \bt H^H_{rq}$ and
\begin{align}
\label{capeve}
&C_{e,q,k}({\bf \Sigma}_q,\bt W_q) \triangleq \ln\left|\bt I+ \bt M_{e,q,k}^{-1}\bt G_{qk}{\bf \Sigma_{q}}\bt G_{qk}^H\right| = \nonumber
\\
&\ln\left|\bt M_{e,q,k}+\bt G_{qk}{\bf \Sigma_{q}}\bt G_{qk}^H\right|+\ln\left|\bt M_{e,q,k}^{-1}\right|
\end{align}
where $\bt M_{e,q,k} \triangleq \bt I + \bt G_{qk}\bt W_q \bt G^H_{qk}+\sum_{\underset{r\neq q}{r = 1}}^Q\bt G_{rk}\left({\bf \Sigma}_r+\bt W_r\right) \bt G^H_{rk}$.
The term $\bt M_q$ is the covariance matrix of received interference at the $q$th receiver and $\bt M_{e,q,k}$ is the covariance matrix of interference received at the $k$th eavesdropper while eavesdropping on the $q$th link\footnote{Specifically, while eavesdropping on a user, an eavesdropper is treating interference as additive (colored) noise.}. Notice that both $\bt M_q$ and $\bt M_{e,q,k}$ include the information signal and AN of other $Q-1$ links. Furthermore, we require $\tr({\bf \Sigma}_q+\bt W_q) \le P_q$ for all $q \in \mathbb Q$, where $\tr(.)$ is the trace operator and $P_q$ is a positive value that represents the amount of power available (for both information and AN signals) at the $q$th transmitter.
\section{Problem Formulation}
\label{sec3}
We assume that the $q$th link, $q \in \mathbb Q$, optimizes its information and AN signals (through their covariance matrices ${\bf \Sigma}_q$ and $\bt W_q$) to maximize its own secrecy rate. The dynamics of such interaction between $Q$ links can be modeled as a noncooperative game where each player (i.e., link) uses his best strategy to maximize his own utility (i.e., secrecy rate) given the strategies of other players. The best response of each player can be found by solving the following optimization problem
\begin{align}
\label{srm}
\underset{{\bf \Sigma}_q, \bt W_q}{\text{maximize}~}& R^{sec}_{q}({\bf\Sigma}_q, \bt W_q) \nonumber
\\
\text{s.t.~~~~~~}&({\bf \Sigma}_q,\bt W_q)\in \mathcal{F}_q,~q \in \mathbb Q
\end{align} 
where $\mathcal{F}_q \triangleq \left\{({\bf \Sigma}_q,\bt W_q)|\tr({\bf \Sigma}_q+\bt W_q) \le P_q,~{\bf \Sigma}_q\succeq 0,~\bt W_q \succeq 0\right\}$ is the set of all Hermitian matrices $({\bf \Sigma}_q,\bt W_q)$ that are positive semi-definite (i.e., ${\bf \Sigma}_q\succeq 0,~\bt W_q \succeq 0$) and meet the link's power constraint. 

Unfortunately, problem \eqref{srm} is a nonconvex optimization problem. In the remainder of this section, we aim to find a tractable solution for this problem. To that end, we first mention the following identity for a positive definite matrix $\bt M_q$ of size $N_{R_q}$ \cite[Example 3.23]{boyd}:
\begin{equation}
\label{lem1}
\ln|\bt M_q^{-1}| = f(\bt S^*) = \max_{\bt S \in \mathbb{C}^{N_{R_q}\times N_{R_q}}, \bt S \succeq 0}f(\bt S)
\end{equation}
where $f(\bt S) \triangleq -\tr(\bt S \bt M_q)+\ln|\bt S|+N_{R_q}$ and $\bt S^* \triangleq \bt M_q^{-1}$ is the solution to the most RHS of \eqref{lem1}.
Applying the reformulation in \eqref{lem1} to the term $\ln|\bt M_q^{-1}|$ in \eqref{capleg} and $\ln\left|\bt M_{e,q,k}+\bt G_{qk}{\bf \Sigma_{q}}\bt G_{qk}^H\right|$ in \eqref{capeve}, \eqref{srm} can be rewritten as
\begin{align}
\label{reform}
\underset{{\bf\Sigma}_q,\bt W_q, \bt S_q}{\text{maximize}}&~f_q({\bf \Sigma}_q,\bt W_q,\left\{\bt S_{q,k}\right\}_{k=0}^K), \nonumber
\\
\text{s.t.~~~~~~}& ({\bf \Sigma}_q,\bt W_q)\in \mathcal{F}_q,~\bt S_{q,k} \succeq 0,~q \in \mathbb Q,~ k \in \mathbb \{0\} \cup K
\end{align}
where $\left\{\bt S_{q,k}\right\}_{k=0}^K = [\bt S_{q,0}^T,\dots,\bt S_{q,K}^T]^T$, and
\begin{subequations}
\begin{align}
\label{aa}
f_q({\bf \Sigma}_q,\bt W_q,\left\{\bt S_{q,k}\right\}_{k=0}^K)& \triangleq \varphi_{q}({\bf \Sigma}_q,\bt W_q,\bt S_{q,0}) \nonumber
\\
&-\max_{k \in \mathbb K}\varphi_{e,q,k}({\bf \Sigma}_q,\bt W_q,\bt S_{q,k})
\\
\label{bb}
\varphi_{q}({\bf \Sigma}_q,\bt W_q,\bt S_{q,0}) \triangleq -\mbox{t}&\mbox{r}(\bt S_{q,0}\bt M_q)+\ln \left|\bt S_{q,0}\right|+N_{R_q}\nonumber
\\
+\ln&\left|\bt M_q+\bt H_{qq}{\bf \Sigma_{q}}\bt H_{qq}^H\right|
\\
\label{cc}
\varphi_{e,q,k}({\bf \Sigma}_q,\bt W_q,\bt S_{q,k}) \triangleq&~\mbox{t}\mbox{r}(\bt S_{q,k}(\bt M_{e,q,k}+\bt G_{qk}{\bf \Sigma_{q}}\bt G_{qk}^H))\nonumber
\\
-&\ln \left|\bt S_{q,k}\right|-N_{e,k}-\ln\left|\bt M_{e,q,k}\right|.
\end{align}
\end{subequations}

Problem \eqref{reform} is nonconvex with respect to (w.r.t) $({\bf \Sigma}_q, \bt W_q,\{\bt S_{q,k}\}_{k = 0}^K)$. However, it is easy to verify that problem \eqref{reform} is convex w.r.t either $({\bf \Sigma}_q, \bt W_q)$ or $\{\bt S_{q,k}\}_{k = 0}^K$ (by checking its Hessian).  A stationary point to problem \eqref{srm} that satisfies its K.K.T optimality conditions then can be found by solving \eqref{reform} sequentially w.r.t $({\bf \Sigma}_q,\bt W_q)$ and $\{\bt S_{q,k}\}_{k=0}^K$ \cite[Section IV-B]{winkin}. Specifically, in one iteration, problem \eqref{reform} is solved w.r.t only $\{\bt S_{q,k}\}_{k = 0}^K$ to find an optimal solution $ \{\bt S_{q,k}^*\}_{k = 0}^K$. Next, with $\{\bt S_{q,k}^*\}_{k = 0}^K$ plugged in \eqref{aa}, the problem in \eqref{reform} is optimized w.r.t $({\bf \Sigma}_q, \bt W_q)$ to find an optimal solution $({\bf \Sigma}_q^*,\bt W_q^*)$. This Alternating Optimization (AO) cycle continues until reaching a convergence point. The $n$th iteration of AO, i.e., $({\bf \Sigma}^n_q,\bt W^n_q,\left\{\bt S^n_{q,k}\right\}_{k=0}^K)$, is as follows:
\begin{subequations}
\label{allss}
\begin{align}
\label{a}&({\bf \Sigma_q^n, \bt W_q^n}) = \arg \max_{({\bf \Sigma}_q,\bt W_q)\in \mathcal{F}_q}f_q({\bf \Sigma}_q,\bt W_q,\left\{\bt S^{n-1}_{q,k}\right\}_{k=0}^K)
\end{align}
\begin{align}
\label{b}&\bt S^n_{q,0} \triangleq \arg \max_{\bt S_{q,0}\succeq 0} \varphi_{q}({\bf \Sigma}^n_q,\bt W^n_q,\bt S_{q,0}) = ({\bt M_q^n})^{-1}= \nonumber
\\
&\Big(\bt I + \bt H_{qq}\bt W^n_q \bt H^H_{qq}+\sum_{\underset{r\neq q}{r = 1}}^Q\bt H_{rq}\left({\bf \Sigma}^0_r+\bt W^0_r\right) \bt H^H_{rq}\Big)^{-1}
\\
\label{c}&\bt S^n_{q,k} \triangleq \arg \max_{\bt S_{q,k}\succeq 0} \varphi_{e,q,k}({\bf \Sigma}^n_q,\bt W^n_q,\bt S_{q,k})  = \Big(\bt M^{n}_{e,q,k}+\bt G_{qk}{\bf \Sigma_{q}}^n\bt G_{qk}^H\Big)^{-1}\nonumber
\\
&=\Big(\bt I + \bt G_{qk}\bt ({\bf \Sigma}^n_q+W^n_q) \bt G^H_{qk}+\sum_{\underset{r\neq q}{r = 1}}^Q\bt G_{rk}\left({\bf \Sigma}^0_r+\bt W^0_r\right) \bt G^H_{rk}\Big)^{-1},~k\neq 0
\end{align}
\end{subequations}
where ${\bf \Sigma}_r^0$ and $\bt W_r^0$ (for $r \neq q$) denote the received interference components at the $q$th receiver prior to solving \eqref{reform}. Incorporating \eqref{b} and \eqref{c} in \eqref{a}, the solution to the convex problem \eqref{a} can be found using a convex optimization solver. Notice that in \eqref{b} and \eqref{c}, the users do not coordinate with each other in the middle of finding a stationary point for \eqref{reform}, for all $q \in \mathbb Q$. Hence, the terms ${\bf \Sigma}_r^0$ and $\bt W_r^0$, $r \neq q$ remain constant during the AO iterations. To solve problem \eqref{reform} faster, the authors in \cite{winkin} solved the smooth approximation of \eqref{srm} based on the log-sum-exp inequality \cite[chapter 3.1.5]{boyd}, which states that
\begin{equation}
\label{logsumexp}
\max\{a_1,\dots,a_K\} \le \frac{1}{\beta}\ln(\sum_{k = 1}^K e^{\beta a_k}) \le \max\{a_1,\dots, a_K\}+ \frac{1}{\beta}\ln K.
\end{equation}
where $a_k \in \mathbb{R}$ and $\beta > 0$. Applying \eqref{logsumexp} to \eqref{secrecy}, we can write problem \eqref{srm} as
\begin{align}
\label{secrecy2}
\underset{{\bf \Sigma}_q, \bt W_q}{\text{maximize}~}& \bar{R}_{s,q}({\bf\Sigma}_q, \bt W_q) \nonumber
\\
\text{s.t.~~~~~~}&({\bf \Sigma}_q,\bt W_q)\in \mathcal{F}_q,~q \in \mathbb Q
\end{align}
where
\begin{align}
\label{secrecy22}
\bar{R}_{s,q}({\bf\Sigma}_q, \bt W_q) \triangleq &C_q({\bf \Sigma}_q,\bt W_q)\nonumber
\\
-\frac{1}{\beta}\ln&(\sum_{k = 1}^K\exp\left\{\beta C_{e,q,k}({\bf \Sigma}_q,\bt W_q)\right\}),~q \in \mathbb Q.
\end{align}
Hence, we can do the same reformulation procedure taken in \eqref{reform} to end up with the following smooth reformulation \cite{winkin}:
{\begin{align}
\label{reform2}
\underset{{\bf\Sigma}_q,\bt W_q, \bt S_q}{\text{maximize}}&~\bar{f}_q({\bf \Sigma}_q,\bt W_q,\left\{\bt S_{q,k}\right\}_{k=0}^K), \nonumber
\\
\text{s.t.~~~~~~}& ({\bf \Sigma}_q,\bt W_q)\in \mathcal{F}_q,~\bt S_k \succeq 0,q \in \mathbb Q, k \in \mathbb K
\end{align}}
where
{\begin{align}
\label{reform3}
\bar{f}_q({\bf \Sigma}_q,\bt W_q,\left\{\bt S_{q,k}\right\}_{k=0}^K) \triangleq &\varphi_{q}({\bf \Sigma}_q,\bt W_q,\bt S_{q,0})\nonumber
\\
-\frac{1}{\beta}\ln&\Big(\sum_{k = 1}^Ke^{\beta \varphi_{e,q,k}({\bf \Sigma}_q,\bt W_q,\bt S_{q,k})}\Big).
\end{align}}
with $\varphi_q$ and $\varphi_{e,q,k}$ defined in \eqref{bb} and \eqref{cc}, respectively. Hence, the AO iteration in \eqref{a} changes to
{\begin{equation}
\label{reforma}
({\bf \Sigma_q^n, \bt W_q^n}) = \arg \max_{({\bf \Sigma}_q,\bt W_q)\in \mathcal{F}_q}\bar{f}_q({\bf \Sigma}_q,\bt W_q,\left\{\bt S^{n-1}_{q,k}\right\}_{k=0}^K),
\end{equation}}
while $\left\{\bt S^{n-1}_{q,k}\right\}_{k=0}^K$ remain the same as \eqref{b} and \eqref{c}\footnote{As far as optimality is concerned, it is shown in \cite{winkin} that in the single-user scenario, the limit point of AO iterations done using \eqref{reforma}, \eqref{b}, and \eqref{c} are very close to the solutions found from AO iterations done using \eqref{a}, \eqref{b}, and \eqref{c}.}. After plugging \eqref{b} and \eqref{c} into \eqref{reforma}, the solution to \eqref{reforma} at the $n$th iteration is computed using the Projected Gradient (PG) algorithm. The $l$th iteration of PG algorithm while solving \eqref{reforma} is as follows.
\begin{align}
\label{PG}
\left(\begin{array}{c} 
\hat{{\bf \Sigma}}_q^{n,l+1}
\\
\hat{\bt W}_q^{n,l+1}
\end{array}\right) = \texttt{Proj}_{\mathcal{F}_q}\left(\begin{array}{c}{\bf \Sigma}_q^{n,l}+\alpha_l\nabla_{{\bf \Sigma}_q}\bar{f}_q^{n,l}
\\
\bt W_q^{n,l}+\alpha_l\nabla_{\bt W_q}\bar{f}_q^{n,l}
\end{array}\right),
\\
\label{PG2}
\left(\begin{array}{c}
{\bf \Sigma}_q^{n,l+1}
\\
\bt W_q^{n,l+1}\end{array}\right) = \left(\begin{array}{c}
{\bf \Sigma}_q^{n,l}
\\
\bt W_q^{n,l}\end{array}\right) + \varepsilon_l\left(\begin{array}{c}
\hat{{\bf \Sigma}}_q^{n,l+1}-{\bf \Sigma}_q^{n,l}
\\
\hat{\bt W}_q^{n,l+1}-\bt W_q^{n,l}\end{array}\right),
\end{align}
where $\alpha_l$ and $\varepsilon_l$ are step sizes that can be determined using Wolfe conditions for PG method \cite{nonlinear}; $\texttt{Proj}_{\mathcal{F}_q}$ is the projection operator to the set $\mathcal{F}_q$, which can be written as
\begin{equation}
\label{projector}
\texttt{Proj}_{\mathcal{F}_q}\left(\begin{array}{c}\tilde{\bf \Sigma}
\\
\tilde{\bt W}\end{array}\right) = \min_{\bt W, {\bf \Sigma}\in \mathcal{F}_q}||\bt W- \tilde{\bt W}||^2_F+||{\bf \Sigma}- \tilde{{\bf \Sigma}}||^2_F;
\end{equation}
and {$(\nabla_{{\bf \Sigma}_q}\bar{f}_q^{n,l},\nabla_{\bt W_q}\bar{f}_q^{n,l}) = \left(\nabla_{{\bf \Sigma}_q}\bar{f}_q({\bf \Sigma}^{n,l}_q,\bt W^{n,l}_q,\left\{\bt S^{n-1}_{q,k}\right\}_{k=0}^K),\right.$
\\
$\left.\nabla_{\bt W_q}\bar{f}_q({\bf \Sigma}^{n,l}_q,\bt W_q^{n,l},\left\{\bt S^{n-1}_{q,k}\right\}_{k=0}^K)\right)$} where
\begin{subequations}
\label{every}
\begin{align}
&\nabla_{{\bf \Sigma}_q}\bar{f}_q({\bf \Sigma}^{n,l}_q,\bt W^{n,l}_q,\left\{\bt S^{n-1}_{q,k}\right\}_{k=0}^K) =\nonumber
\\
&\bt H^H_{qq}(\bt M^{n,l}_q+\bt H_{qq}{\bf \Sigma}_q^{n,l}\bt H_{qq}^H)^{-1}\bt H_{qq}-\sum_{k = 1}^K\rho_{q,k}^{n,l}\bt G_{q,k}^H\bt S_{q,k}^{n-1}\bt G_{q,k},
\\
&\bt M_{q}^{n,l} = \bt I + \bt H_{qq}\bt W^{n,l}_q \bt H^H_{qq}+\sum_{\underset{r\neq q}{r = 1}}^Q\bt H_{rq}\left({\bf \Sigma}^0_r+\bt W^0_r\right) \bt H^H_{rq},\
\\
\label{every1}
&\rho_{q,k}^{n,l} = \frac{e^{\beta \varphi_{e,q,k}({\bf \Sigma}_q^{n,l},\bt W_q^{n,l},\bt S_{q,k}^{n-1})}}{\sum_{j = 1}^Ke^{\beta \varphi_{e,q,j}({\bf \Sigma}_q^{n,l},\bt W_q^{n,l},\bt S_{q,j}^{n-1})}},
\end{align}
\begin{align}
\nabla_{\bt W_q}\bar{f}_q({\bf \Sigma}^{n,l}_q,\bt W_q^{n,l},\left\{\bt S^{n-1}_{q,k}\right\}_{k=0}^K) = \nonumber&
\\
\bt H_{qq}^H \Big((\bt M_q^{n,l}+\bt H_{qq}{\bf \Sigma}_{q}^{n,l}\bt H_{qq})^{-1}-&\bt S^{n-1}_{q,0} \Big) \bt H_{qq} +\nonumber
\\
\sum_{k = 1}^K \rho_{q,k}^{n,l}\bt G_{qk}^H\left(({\bt M_{e,q,k}^{n,l}})^{-1}\right.&\left.-\bt S^{n-1}_{q,k}\right)\bt G_{qk},
\\
\bt M_{e,q,k}^{n,l} = \bt I + \bt G_{qk}\bt W^{n,l}_q \bt G^H_{qk}+\sum_{\underset{r\neq q}{r = 1}}^Q&\bt G_{rk}\left({\bf \Sigma}^0_r+\bt W^0_r\right) \bt G^H_{rk}.
\end{align}
\end{subequations}
The projection in \eqref{projector} can be efficiently computed according to \cite[Fact 1]{winkin}.
We refer to the game where the actions of the players are defined by \eqref{reform2} as the proposed smooth game. Now that we have the response of each user, we can analyze the dynamics of the proposed smooth game.

A pseudo-code of the proposed smooth game mentioned so far is shown in Algorithm 1. As mentioned earlier, finding a stationary point for \eqref{reform2} for each user consists of two nested loops. The inner loop involves the gradient projection which is shown in \eqref{PG} and \eqref{PG2} (i.e., the loop in Line 6 of Algorithm 1). Once the optimal solution to inner loop is found, one AO iteration is done by recalculating $\{\bt S_{q,k}\}_{k = 0}^K$ according to \eqref{b} and \eqref{c} in the outer loop (i.e., Line 4). After the AO iterations converge to a stationary point, the users begin their transmissions using the computed precoders of information signal and AN\footnote{Although the optimization of covariance matrices of information signal and AN has been taken into account so far, the precoders can be found using eigenvalue decomposition.}. Therefore, one round of this competitive secrecy rate maximization is done. Notice that according to Line 2, the players will be notified of actions of each other (i.e., recalculate the received interference) only after the AO iterations has converged\footnote{Such procedure in Line 2 of Algorithm 1 also explains the reason why $\bt W_r^0$ and ${\bf \Sigma}_r^0$ in \eqref{allss} and \eqref{every} remain constant during AO iterations.}. The last round of the game will be the one where the convergence is reached.

\begin{algorithm}[H]
	\caption{Proposed Smooth Game}
	\label{alg1}
\footnotesize
\setstretch{1.5}
	{\bf Initialize:} ${\bf \Sigma}^{1,1}_q$, $\bt W_q^{1,1}$, $\tr({\bf \Sigma}^{1,1}_q+\bt W_q^{1,1}) < P_q$, $\forall q \in \mathbb Q$
	\begin{algorithmic}[1]
		\Repeat
		\\
		Each link $q$ computes $\bt M_q$, $\bt M_{e,q,k}$, $\forall k \in \mathbb K$ locally
		\For {q =1,\dots,Q}
		
		\For {n = 1,\dots}
		\\
		\hspace{10mm}Compute $\bt S^{n-1}_{q,k}$, $k = 0,\dots,K$
		\For {$l$ = 1,\dots}
		\\
		\hspace{15mm}Compute $\varphi_{e,q,k}({\bf \Sigma}^{n,l}_q,\bt W_q^{n,l},\bt S^{n-1}_{q,k})$, $\bt M^{n,l}_q$, $\bt M^{n,l}_{e,q,k}$, $\forall (q,k)$
		\\
		\hspace{15mm}Compute $({\bf \Sigma}^{n,l+1}_q, \bt W_q^{n,l+1})$ using \eqref{PG}-\eqref{every}\hspace{5mm}\% Use Wolfe conditions
		\EndFor
		\EndFor
		\EndFor
		\Until{Convergence to QNE}
	\end{algorithmic}
\end{algorithm}
\section{Existence and Uniqueness of the QNE}
\label{sec4}
Before we begin to analyze the existence and uniqueness of the QNE, we review fundamentals of variational inequality theory as the basis of our analyses.
\subsubsection*{Variational Inequality Theory}
Let $F: \mathcal{Q} \rightarrow \mathbb{R}^N$ be a vector-valued continuous real function, where $N>1$ and $\mathcal{Q} \subseteq \mathbb{R}^N$ is a nonempty, closed, and convex set. The variational inequality $\text{VI}(F,\mathcal{Q})$ is the problem of finding a vector $x^*$ such that
\begin{equation}
\label{videf}
(x-x^*)^TF(x^*)\ge0,~~\forall x\in\mathcal{Q}.
\end{equation}
The relation between variational inequality and game theory is summarized in the following theorem:
\begin{theorem}
\cite[Chapter 2]{pang} Consider $Q$ players in a noncooperative game with utility function $f_q(x)$ for the $q$th player (not to be confused with the $f_q$ defined in \eqref{reform}), where $x \in \mathcal {Q}$ and $x = [x_1,x_2,...,x_Q]^T$, $x_q$ is the $q$th player's strategy, and $f_q(x)$ is concave w.r.t $x_q$ for all $q$. The set $\mathcal Q$ is comprised of all strategy sets (i.e., $\mathcal{Q} = \prod_{q=1}^Q\mathcal{Q}_q$, where $\mathcal{Q}_q$ is the $q$th player's strategy set). Assuming the differentiability of $f_q(x)$ w.r.t $x_q$ and that $\mathcal{Q}_q$ is a closed and convex set for all $q$, the vector $x^*$ is the NE of the game if for $F(x) = [-\nabla_{x_1}f_1(x),-\nabla_{x_2}f_2(x),...,-\nabla_{x_Q}f_Q(x)]^T$ we have:
\begin{equation*}
(x-x^*)^TF(x^*)\ge0,~~\forall x\in\mathcal{Q}.
\end{equation*}\qed
\end{theorem}
\subsection{Variational Inequality in Complex Domain}
The theory of VI mentioned in \eqref{videf} assumes that $\mathcal{Q} \subseteq \mathbb{R}^n$. However, this assumption might not be of our interest because the strategies of the players in our proposed game are two complex matrices (i.e., ${\bf \Sigma}_q$ and $\bt W_q$). Therefore, an alternative definition for VI in complex domain is needed. We use the definitions derived by the authors in \cite{monotone} to define VI in complex domain.
\subsubsection*{Minimum Principle in Complex Domain}
Consider the following optimization
\begin{align}
\label{complexopt}
&\underset{\bt Z}{\text{minimize}}~~~ f(\bt Z) \nonumber
\\
&\text{s.t.}~~~ \bt Z \in \mathcal K
\end{align}
where $f: \mathcal K \rightarrow \mathbb R$ is convex and continuously differentiable on $\mathcal K$ where $\mathcal K \subseteq \mathbb C^{N^\prime \times N}$, $N^\prime >1$, and $N >1$. $X \in \mathcal K$ is an optimal solution to \eqref{complexopt} if and only if we have \cite[lemma23]{monotone}
\begin{equation}
\label{minprinc}
\left<\bt Z-\bt X,\nabla_{\bt Z}f(\bt X)\right> \ge 0,~ \forall \bt Z \in \mathcal K.
\end{equation}
where $\left<\bt A, \bt B\right> = \re\left(Tr\left(\bt A^H \bt B\right)\right)$.
\subsubsection{VI in Complex Domain}
Using the definition of minimum principle in complex domain, we can now define the VI problem in the domain of complex matrices. For a complex-valued matrix $F^\mathbb{C}(\bt Z): \mathcal K \rightarrow \mathbb C^{N^\prime \times N}$ where $\mathcal K \subseteq \mathbb C^{N^\prime \times N}$, the VI in the complex domain is the problem of finding a complex matrix $\bt Y$ such that the following is satisfied \cite[Definition 25]{monotone}
\begin{equation}
\label{compvi}
\left<\bt Z -\bt Y,F^{\mathbb C}(\bt Y)\right> \ge 0,~\forall \bt Z \in \mathcal{K}.
\end{equation}
\subsection{Quasi-Nash Equilibrium}
It should be emphasized that the optimization problem of each player mentioned in \eqref{secrecy2} is nonconvex. Hence, the solution found for each link by solving \eqref{reform2} at Line 10 of Algorithm 1 is only a stationary point of problem \eqref{secrecy2}. As a consequence, the traditional concepts of concave games used in proving the existence of a NE are not applicable here. Specifically, according to \cite{rosen}, the quasi-concavity of each player's utility w.r.t his strategy is required in proving the existence of a NE; an assumption that is not true in our game. Instead, we analyze the proposed (nonconvex) smooth game based on the relaxed equilibrium concept of QNE \cite{pangscut}. In the following, a formal definition of QNE is given \cite{pangscut}.

Consider a noncooperative game with $Q$ player each of whose strategies are restricted by some private constraints denoted as
\begin{equation}
\label{nonconvexset}
\mathcal{X}_q = \{x_q \in X_q | h_q(x_q) \le 0\}.
\end{equation}
The set $X_q$ is a convex set, and $h_q: \xi_q \rightarrow \mathbb R^{l_q}$ is a continuously differentiable mapping on the open convex set $\xi_q$ containing $X_q$. No convexity assumption is made on $h_q$. Hence, although $X_q$ is a convex set, $\mathcal{X}_q$ is not necessarily so. Player $q$ has an objective function $g_q: \xi \rightarrow \mathbb R$, assumed to be continuously differentiable where $\xi = \prod_{q = 1}^Q \xi_q$. The action of each player is formulated as follows:
\begin{align}
\underset{x_q \in \mathcal{X}_q}{\text{minimize~}}&g_q(x_q,x_{-q}) \nonumber
\\
\text{s.t.~}&x_q \in \mathcal{X}_q.
\end{align}
Obviously, the equivalent formulation can be written for when the action of each player is maximizing an objective (e.g., utility). Given the actions of other players, i.e., $x_{-q}^*$, and provided that a constraint qualification (CQ) condition holds at a point $x^*_q$, a necessary condition for $x_q^*$ to be an optimal point of player $q$'s optimization problem (i.e., action) is the existence of a nonnegative constant vector $\mu_q^* \in \mathbb R^{l_q}_+$ such that
\begin{subequations}
\label{cq}
\begin{align}
&\nabla_{x_q} L_q(x_q^*,x^*_{-q}, \mu_q^*) = \nabla_{x_{q}} g_{q}(x_q^*,x_{-q}^*)+{\mu_q^*}^T \nabla_{x_{q}}h_q(x^*_q) = 0,
\\
&{\mu_q^*}^Th_q(x^*_q) = 0,
\\
& h_q(x_q^*) \le 0,~x_q \in X_q.
\end{align}
\end{subequations}
If any CQ is satisfied at $x_q^*$, the optimality conditions in \eqref{cq} can be written as a VI over the set $X_q$; that is, the necessary condition for $x^*_q$ to be an optimal solution to player $q$'s optimization problem is if $x^*_q$ solves $\text{VI}(\nabla_{x_q}L_q(\bullet,x^*_{-q}, \mu_q^*),X_q)$ \cite[Proposition 1.3.4]{pang}. Furthermore, the existence of a nonnegative vector $\mu_q^*$ together with the complementarity of $\mu_q^*$ and $h_q(x_q^*)$ can be interpreted as $\mu_q^*$ being such that
\begin{equation}
\label{slackness}
-(\mu_q-\mu_q^*)^Th_q(x_q^*) \ge 0,~\forall \mu_q \in \mathbb R^{l_q}_+.
\end{equation}
Clearly, if $h_q(x_q^*)$ is not binding, i.e., $h_q(x^*_q) < 0$, then 
$\mu_q^* = 0$ satisfies \eqref{slackness}. Furthermore, when $h_q(x^*_q)$ is binding, i.e., 
$h_q(x^*_q) = 0$, inequality \eqref{slackness} is trivially satisfied for all $\mu_q\in \mathbb R^{l_q}_+$. Hence, using \eqref{slackness} and the fact that
 $x^*_q$ solves $\text{VI}(\nabla_{x_q}L_q(\bullet,x^*_{-q}, \mu_q^*),X_q)$, the pair $(x_q^*, \mu_q^*)$ solves the following VI:
\begin{equation}
\label{kktmap}
\left(\begin{array}{c}
x_q-x_q^*
\\
\mu_q-\mu_q^*
\end{array}\right)^T\Gamma_q(x,\mu_q) \ge 0,~\forall (x_q,\mu_q) \in \mathcal{R}_q = X_q \times \mathbb R^{l_q}_+
\end{equation}
where
\begin{equation}
\Gamma_q(x,\mu_q) = \left(\begin{array}{c}
\nabla_{x_q}L_q(\bullet,x^*_{-q}, \mu_q^*)
\\
-h_q(x_q^*)
\end{array}\right).
\end{equation}
Notice that although it might seem that $\text{VI}(\nabla_{x_q}L_q(\bullet,x^*_{-q}, \mu_q^*),X_q)$ and \eqref{slackness} cannot be combined to build \eqref{kktmap}, using the fact that VI is a generalized definition of a set-valued mapping\footnote{A point-to-set map, also called a multifunction or a set-valued map, is a map $N$ from $\mathbb R^n$ into the power set of $\mathbb R^n$, i.e., for every $x \in \mathbb{R}^n$, $N_{\mathbb R^n}(x)$ is a (possibly empty) subset of $\mathbb{R}^n$ \cite[Chapter 2.1.3]{pang}. To avoid confusion, note that the definition of a set-valued map is fundamentally different from that of a vector function such as $h_q(x_q)$ defined in \eqref{nonconvexset}.}, we are able to justify \eqref{kktmap}. it can be proved that for the set-valued mappings $N_{X_q}(x_q)$ and $N_{\mathbb R^{l_q}_+}(\mu_q)$, we have $N_{X_q\times\mathbb R^{l_q}_+}(x_q,\mu_q) = N_{X_q}(x_q) \times N_{\mathbb R^{l_q}_+}(\mu_q)$ \cite{uw}. The same conclusion holds for VI problems. Hence, inequality \eqref{kktmap} can be deduced.

Concatenating the inequality in \eqref{kktmap} over the set of players, the QNE can be defined as follows:
\begin{definition}
The QNE is the pair $\left(x^*_q,\mu^*_q\right)$, $q = 1\dots,Q$, that satisfies the following inequality:
\begin{align}
\label{qnedef}
&\left(\left(\begin{array}{c}
x_q-x_q^*
\\
\mu_q-\mu_q^*
\end{array}\right)_{q=1}^Q\right)^T\left(\Gamma_q(x,\mu_q)\right)_{q=1}^Q \ge 0, \nonumber
\\
&\forall (x_q,\mu_q)_{q=1}^Q \in \prod_{q=1}^Q\mathcal{R}_q = \prod_{q=1}^Q (X_q \times \mathbb R^{l_q}_+)
\end{align}
where $(\bullet)_{q=1}^Q$ denotes a column vector.
\end{definition}
Notice that the set $\prod_{q=1}^Q\mathcal{R}_q$ is a convex set, and if the actions of each player is a convex program, the QNE reduces to NE. In our scenario, since the private constraints for each player is a convex set, we embedded the private constraints into the set $\mathcal R_q$ defined in \eqref{qnedef}. We need to emphasize the fact that the constant vectors $\mu_q^*$ for all $q$ can only be defined if the optimization problem of each player satisfies some CQ conditions. For players with convex problems, these constant vectors are trivially satisfied since the K.K.T conditions are necessary and sufficient conditions of optimality in convex programs.

One intuition that can be given on the concept of QNE is as follows. QNE is point where no player has an incentive to unilaterally change his strategy because any change makes a player not satisfy the K.K.T conditions of his problem. This is in contrast with the definition of NE in which the lack of incentives at NE is because of losing optimality. Again, optimality and satisfying the K.K.T conditions are equivalent when players solve convex programs.
\subsection{Analysis of QNE}
According to the aforementioned definition, the QNEs are tuples that satisfy the K.K.T conditions of all players' optimization problems. Under a constraint qualification, stationary points of each player's optimization problem satisfy its K.K.T conditions. To begin the analysis of the QNE, we first show that the stationary point found using AO mentioned previously (i.e., Line 4-10 of Algorithm 1) satisfies the K.K.T conditions of \eqref{secrecy2}.
\begin{proposition}
\label{kktprop}
For the $q$th link, $q\in \mathbb Q$, the stationary point found using AO (i.e., Line 4-10 of Algorithm 1) satisfies the K.K.T conditions of \eqref{secrecy2}. 
\begin{IEEEproof}
See Appendix \ref{appkkt}.
\end{IEEEproof}
\end{proposition}
Now that the K.K.T optimality of the stationary point found by AO iterations is proved, we rewrite the K.K.T conditions of all players to a proper VI problem \cite{pangscut}. The solution(s) to the obtained VI is the QNE(s) of the proposed smooth game. For the proposed smooth game defined using \eqref{reform2}, we can establish the following VI to characterize the QNE points. Let the QNE point be as follows
\begin{equation}
\label{vectorofNE}
\bt Y = \{\bt Y_q\}_{q = 1}^Q \triangleq [{\bf \Sigma}^T, \bt W^T]^T =  \{[{\bf \Sigma}_q^T, \bt W_q^T]^T\}_{q = 1}^Q
\end{equation}
where {$\{[{\bf \Sigma}_q^T, \bt W_q^T]^T\}_{q = 1}^Q = [{\bf \Sigma}_1^T, \bt W_1^T,{\bf \Sigma}_2^T, \bt W_2^T,\dots,{\bf \Sigma}_Q^T, \bt W_Q^T,]^T$}. The function $F^\mathbb{C}(Z)$ is written as
\begin{align}
\label{FofNE}
&F^\mathbb{C} = F^\mathbb{C}({\bf \Sigma},\bt W, \bt S) = \left\{F^\mathbb{C}_q({\bf \Sigma}_q,\bt W_q,\{\bt S_{q,k}\}_{k=0}^K)\right\}_{q = 1}^Q \triangleq \nonumber
\\
&\left\{\left[-(\nabla_{{\bf \Sigma}_q}\bar{f}_q)^T,-(\nabla_{\bt W_q}\bar{f}_q)^T\right]^T\right\}_{q = 1}^Q
\end{align}
where the terms $\nabla_{{\bf \Sigma}_q}\bar{f}_q$ and $\nabla_{\bt W_q}\bar{f}_q$ are given in \eqref{every}.
Therefore, the system of inequalities indicated as $VI(F^\mathbb{C},\mathcal K)$ can be established according to \eqref{compvi}, where $\mathcal K = \prod_{q=1}^Q\mathcal{F}_q$. Furthermore, for a given response ${\bf \Sigma}_q$ and $\bt W_q$, the solutions of $\{\bt S_{q,k}\}_{k = 0}^K$ are uniquely determined by \eqref{b} and \eqref{c} for all $q$. Hence, from now on, we assume that the values of $\{\bt S_{q,k}\}_{k = 0}^K$ are already plugged into $F^\mathbb{C}_q\left({\bf \Sigma}_q,\bt W_q,\{\bt S_{q,k}\}_{k=0}^K\right)$, so we drop the term $\{\bt S_{q,k}\}_{k=0}^K$ in the subsequent equations for notational convenience.

In order to show that the K.K.T conditions are valid necessary conditions for a stationary solution of \eqref{secrecy2}, an appropriate CQ must hold \cite{abadie}. In this paper, we use the Slater's constraint qualification \cite{abadie} as the strategy set of each player is a convex set. Moreover, at NE (if it exists) all of the players use their best responses, i.e., each player has found the optimal solution to his optimization problem and will not deviate from that. Since the optimal solution for each player also satisfies the K.K.T conditions, then NE must be a QNE \cite{pangscut}. In fact, the set of QNEs includes the NE.
\subsection{Existence and Uniqueness of the QNE}
To begin our analysis in this part, we consider the VI described by \eqref{compvi}, \eqref{vectorofNE}, and \eqref{FofNE} again. In the case of the domain of $\bt Z$ being square complex matrices, the definition of VI in complex domain can be further simplified to achieve the same form of VI in the real case (i.e., \eqref{videf}). More specifically, let $F^\mathbb{C}$ be a $2N\times N$ matrix and let $\ve(F^{\mathbb C}) \triangleq [(F_1)^T,\dots,(F_N)^T]^T$ denote a $2N^2 \times 1$ vector where $F_i \triangleq [F^\mathbb{C}(\bt Z)]_{:,i},~i = 1,\dots,N$, denotes the vector corresponding to the $i$th column of $F^{\mathbb{C}}(\bt Z)$. Furthermore, let $\ve(\bt Z) = [[\bt Z]_{:,1}^T,\dots,[\bt Z]_{:,N}^T ]^T$ be the vector version of the complex matrix $\bt Z$. Hence, the vector version of the VI in complex domain can be expressed as
\begin{equation}
\left(\ve(\bt Z) -\ve(\bt Y)\right)^H\ve(F^{\mathbb C}(\bt Y)) \ge 0,~\forall \bt Z \in \mathcal{K}.
\end{equation}
In order to further simplify the VI in complex domain to be completely identical to the real case, we define $F^\mathbb{R} \triangleq [\re\left\{\ve(F^\mathbb{C})\right\}^T,\im\left\{\ve(F^\mathbb{C})\right\}^T]^T$ and $\bt Z^\mathbb{R} \triangleq [\re\left\{\ve(\bt Z)\right\}^T,\im\left\{\ve(\bt Z)\right\}^T]^T$ where $\re\{...\}$ and $I\{...\}$ are the real and imaginary parts, respectively. Therefore, the real-vectorized representation of \eqref{compvi} can be written as
\begin{equation}
\left(\bt Z^\mathbb{R} -\bt Y^\mathbb{R}\right)^T(F^{\mathbb R}(\bt Y^\mathbb R)) \ge 0,~\forall \bt Z^\mathbb{R} \in \mathcal{K}^\mathbb R,\text{ where } \mathcal{K}^\mathbb R \subseteq \mathbb R^{2N^2}.
\end{equation}
The vector form of \eqref{vectorofNE} and \eqref{FofNE} are as follows:
\begin{align}
&\ve(\bt Z) = [\ve(\bar{{\bf \Sigma}})^T,\ve(\bar{\bt W})^T]^T = \left\{[\ve(\bar{{\bf \Sigma}}_q)^T,\ve(\bar{\bt W}_q)^T]^T\right\}_{q = 1}^Q
\end{align}
\begin{align}
\label{semifinal1}
&\ve(F^\mathbb{C}(\bt Z))\! \!=\!\! 
\left\{\left[\ve(-\nabla_{{\bf \Sigma}_q}\bar{f}_q)^T,\ve(-\nabla_{\bt W_q}\bar{f}_q)^T\right]^T\right\}_{q = 1}^Q.
\end{align}
Hence, the vector form of the complex VI problem $VI(F^\mathbb{C},\mathcal{K})$ can be written as
\begin{align}
\label{semifinal2}
&\left([\ve({\bf \Sigma})^T,\ve(\bt W)^T]^T-[\ve(\bar{{\bf \Sigma}})^T,\ve(\bar{\bt W})^T]^T\right)^H\ve(F^\mathbb{C}(\bar{\bf \Sigma},\bar{\bt W}))  \ge 0.
\\
\label{final}
&\left(\left[{{\bf \Sigma}^\mathbb{R}}^T,{{\bt W}^\mathbb{R}}^T\right]-\left[{\bar{{\bf \Sigma}^\mathbb{R}}}^T,{\bar{{\bt W}^\mathbb{R}}}^T\right]\right)F^{\mathbb R} \ge 0,~\forall ({{\bf \Sigma}^\mathbb{R}},{{\bt W}^\mathbb{R}}) \in \mathcal{K}^\mathbb R,~\mathcal{K}^\mathbb R \subseteq \mathbb R^{m}.
\end{align}
Hence, the equivalent real-vectorized representation of the VI in \eqref{compvi} that complies with the definition in \eqref{videf} can be determined as \eqref{final} where $m \triangleq \sum_{q = 1}^Q 2N_{T_q}^2$. Note that the set of matrices $\left({{\bf \Sigma}}_1,\dots,{{\bf \Sigma}}_Q,{{\bt W}}_1,\dots,{{\bt W}}_Q\right)$ that are in $\mathcal K = \prod_{q=1}^Q\mathcal{F}_q$ are the ones whose real-vectorized versions will be inside $\mathcal{K}^\mathbb{R}$. Now that the proposed smooth game is modeled as a real-vectorized VI, we can use the following theorem to prove the existence of the QNE. 
\begin{theorem}
\label{existancthm}
The proposed smooth game, where the actions of each player is given by \eqref{reform2} admits at least one QNE.
\begin{IEEEproof}
See Appendix \ref{existanceproof}.
\end{IEEEproof}
\end{theorem}
The uniqueness of the QNE is discussed in the following theorem:
\begin{theorem}\label{tm3}
The proposed smooth game characterized by \eqref{reform2} has a unique QNE if
\begin{equation}
\label{domreal}
\lambda_{q,\min} > \sum_{\underset{q \neq l}{q = 1}}^Q {|||D_{Z_l}F_q^\mathbb{C}(Z_q)|||}_2,~q \in \mathbb Q
\end{equation}
where $\lambda_{q,\min}$ is the smallest eigenvalue of $D_{Z_q}F^\mathbb{C}_q(Z_q)$, and $D_{Z_l}F_q^\mathbb{C}(Z_q)\triangleq \frac{\partial~\ve\left(F_q^\mathbb{C}(Z_q)\right)}{\partial~\ve(Z_l)^T}$, for all $q,l \in {\mathbb Q}^2$, is defined as
\begin{equation}
\label{jacobzentry}
D_{Z_l}F_q^\mathbb{C}(Z_q) \triangleq \left[\begin{array}{cc}D_{{\bf \Sigma}_l}(-\nabla_{{\bf \Sigma}_q}\bar{f}_q)&D_{\bt W_l}(-\nabla_{{\bf \Sigma}_q}\bar{f}_q)
\\
D_{{\bf \Sigma}_l}(-\nabla_{\bt W_q}\bar{f}_q)&D_{\bt W_l}(-\nabla_{\bt W_q}\bar{f}_q)
\end{array}\right].
\end{equation}
\begin{IEEEproof}
See Appendix \ref{sec:apla}.
\end{IEEEproof}
\end{theorem}
\section{Analysis of the Proposed Game in the Presence of Multiple QNEs}
\label{gra}
\subsection{On the Convergence of Algorithm 1}
The conditions for the uniqueness of the QNE do not guarantee the convergence of Algorithm 1 to a (unique) QNE. Since the optimization of each player is nonconvex, only stationary points of players' utilities could be achieved. Hence, solving each player's optimization problem using AO does not necessarily lead to the best response of each player. This hinders us from proving the convergence of Algorithm 1. However, we verified the convergence via simulations. In this section, we present a slightly modified algorithm, namely, the gradient-response algorithm with proof of convergence. Furthermore, the gradient-response algorithm paves the way for further performance improvements introduced later in this paper.
\subsection{Gradient-Response Algorithm}
\label{graa}
A solution to the VI in \eqref{final} can be characterized by the following iteration \cite[Chapter 12]{pang}:
\begin{equation}
\label{iteration}
x^{(i+1)} = \Pi_{\mathcal{K}^{\mathbb{R}}}\left(x^{(i)}-\gamma F^{\mathbb{R}}(x^{(i)},\{S^{(i)}_{q,k}\}_{k = 0}^K)\right)
\end{equation}
where $\Pi_{\mathcal{K}^\mathbb{R}}$ is the projection to set $\mathcal{K}^{\mathbb{R}}$, $x = \left[{{{\bf \Sigma}^\mathbb{R}}}^T,{{{\bt W}^\mathbb{R}}}^T\right]^T$, the superscript $(i)$ is the number of iterations, and $\gamma = \text{diag}([\gamma_1,\dots,\gamma_m]^T)$ is a diagonal matrix which indicates the step size that each player takes in the improving direction of his utility function. The solutions to $\{\bt S^{(i)}_{q,k}\}_{k = 0}^K$ are as follows:
\begin{subequations}
\label{aoit}
\begin{align}
\label{bb2}&\bt S^{(i)}_{q,0} \triangleq ({\bt M_q^{(i)}})^{-1}= \nonumber
\\
&\Big(\bt I + \bt H_{qq}\bt W^{(i)}_q \bt H^H_{qq}+\sum_{\underset{r\neq q}{r = 1}}^Q\bt H_{rq}\left({\bf \Sigma}^{(i-1)}_r+\bt W^{(i-1)}_r\right) \bt H^H_{rq}\Big)^{-1},
\\
\label{cc2}&\bt S^{(i)}_{q,k\neq 0} \!\!\triangleq\!\! \left(\bt M^{(i)}_{e,q,k}+\bt G_{qk}{\bf \Sigma_{q}}^{(i)}\bt G_{qk}^H\right)^{-1}\!\!= \nonumber
\\
&\!\!\Big(\bt I + \bt G_{qk}\bt ({\bf \Sigma}^{(i)}_q+W^{(i)}_q) \bt G^H_{qk}\!+\!\sum_{\underset{r\neq q}{r = 1}}^Q\bt G_{rk}\left({\bf \Sigma}^{(i-1)}_r\!+\!\bt W^{(i-1)}_r\right) \bt G^H_{rk}\Big)^{-1}
\end{align}
\end{subequations}
where \eqref{cc2} holds for $k\neq 0$. It is easy to confirm that the iteration in \eqref{iteration} is a simplified version of the projection done by each user in \eqref{PG} and \eqref{PG2}. Notice that the only difference of the gradient-response algorithm, characterized by iteration in \eqref{iteration}, from Algorithm 1 is that at each round of the gradient-response algorithm, a player only does one iteration of the PG method (i.e., \eqref{PG}) and one iteration according to \eqref{aoit}. The real-vectorized version of the gradient-response algorithm is shown in \eqref{iteration}. Since the values of $\{\bt S^{(i)}_{q,k}\}_{k=0}^K$ are uniquely determined for a given $x^{(i)}$, we drop the term $\{\bt S^{(i)}_{q,k}\}_{k=0}^K$ from the argument of $F^\mathbb{R}$ for notational convenience.

Assuming that $F^\mathbb{R}$ is \textit{strongly monotone} (with modulus $c_{s}/2$)\footnote{The notion of strong monotonicity is a basic definition in the topic of VI (see Appendix \ref{sec:apla}).
} and \textit{Lipschitz continuous} (with constant $L$)\footnote{It can be seen from \eqref{PG} and \eqref{PG2} that the power constraint of each user makes the variations of $\nabla_{{\bf \Sigma}_q}\bar{f}_q$ and $\nabla_{\bt W_q}\bar{f}_q$ bounded for all $q \in \mathbb Q$. Hence, $F^\mathbb{R}$ is Lipschitz continuous on $\mathcal{K}^\mathbb{R}$.} w.r.t $({\bf \Sigma}_q,\bt W_q)$, the convergence to a unique solution follows if $\gamma_{i^\prime} = d < \frac{c_{s}}{L^2},~\forall i^\prime = 1,\dots,m$, where $d$ is constant. Hence, the mapping $x \rightarrow \Pi_{\mathcal{K}^{\mathbb{R}}}\left(x-\gamma F^{\mathbb{R}}(x)\right)$ becomes a contraction mapping and the fixed points of this map are solutions of the VI in \eqref{final}  \cite[Chapter 12]{pang}. It turns out that sufficient conditions for the strong monotonicity of $VI(F^\mathbb R, \mathcal K^\mathbb{R})$ are in fact the same as the conditions derived in \eqref{domreal} for the uniqueness of the QNE\footnote{More explanation can be found in Appendix \ref{sec:apla}.}.
Therefore, based on \eqref{iteration}, a pseudo-code of the gradient-response algorithm is given in Algorithm 2. Note that the operation in Line 6 of Algorithm 2 is the same as the iteration in \eqref{iteration}. In fact, since the set $\mathcal{K}^\mathbb{R}$ is a Cartesian product of players' strategies, the iteration in \eqref{iteration} can be easily converted back to its matrix form to have the following iteration:
\begin{align}
\label{PG22}
\left(\begin{array}{c} 
{\bf \Sigma}_q^{(i+1)}
\\
\bt W_q^{(i+1)}
\end{array}\right) = \texttt{Proj}_{\mathcal{F}_q}\left(\begin{array}{c}{\bf \Sigma}_q^{(i)}+\gamma^\prime_q\nabla_{{\bf \Sigma}_q}\bar{f}_q({\bf \Sigma}^{(i)}_q,\bt W^{(i)}_q,\left\{\bt S^{(i)}_{q,k}\right\}_{k=0}^K)
\\
\bt W_q^{(i)}+\gamma^\prime_q\nabla_{\bt W_q}\bar{f}_q({\bf \Sigma}^{(i)}_q,\bt W^{(i)}_q,\left\{\bt S^{(i)}_{q,k}\right\}_{k=0}^K)
\end{array}\right), \forall q \in \mathbb{Q}.
\end{align}
Notice that $\gamma^\prime_q$ is a diagonal matrix that can obtained by dividing the matrix $\gamma$ into $Q$ block-diagonal matrices. That is, with a slight abuse of notations, $\gamma = \text{diag}([\gamma_1,\dots, \gamma_m]^T) = \gamma^\prime = \text{diag}(\gamma^\prime_1,\dots,\gamma^\prime_Q)$, $Q<m$. Therefore, the gradient response in \eqref{iteration} can be shown as an iteration that is done in each link, independent of other links. This is essentially a distributed implementation. The gradient-response algorithm is given in Algorithm 2.
\begin{algorithm}[H]
\caption{Gradient-Response Algorithm}
\label{alg2}
\footnotesize
\setstretch{1.5}
{\bf Initialize:} ${\bf \Sigma}^{(1)}_q$, $\bt W_q^{(1)}$, $\tr({\bf \Sigma}^{(1)}_q+\bt W_q^{(1)}) < P_q$, $\forall q$
\begin{algorithmic}[1]
\Repeat\hspace{5mm}\% superscript $(i)$ indicates the iterations starting from here
\\
Compute $\bt M_q$, $\bt M_{e,q,k}$, $\forall (q,k)\in \mathbb Q\times \mathbb K$
\\
Compute $\bt S^{(i)}_{q,k}$, $\forall (q,k)\in \mathbb Q\times \mathbb K$
\\
Compute $\varphi_{e,q,k}({\bf \Sigma}^{(i)}_q,\bt W_q^{(i)},\bt S^{(i)}_{q,k})$, $\forall (q,k)\in \mathbb Q\times \mathbb K$
\For {q =1,\dots,Q}
\\
\hspace{5mm}Compute $({\bf \Sigma}^{(i+1)}_q, \bt W_q^{(i+1)})$ using \eqref{PG22}
\EndFor
\Until{Convergence to QNE}
\end{algorithmic}
\end{algorithm}

The convergence point of Algorithm 2 is a QNE of the game where players' actions are defined by \eqref{reform2}. Specifically, assume that for $i\rightarrow \infty$, the convergence point is denoted as $(\bar{\bf \Sigma}, \bar{\bt W})$. Hence, we have for all $q\in \mathbb Q$
\begin{subequations}
\begin{align}
\label{b2}
&\bar{\bt S}_{q,0} = \arg \max_{\bt S_{q,0}\succeq 0} \varphi_{q}(\bar{\bf \Sigma}_q,\bar{\bt W}_q,\bt S_{q,0})
\\
\label{c2}
&\bar{\bt S}_{q,k} = \arg \max_{\bt S_{q,k}\succeq 0} \varphi_{e,q,k}(\bar{\bf \Sigma}_q,\bar{\bt W}_q,\bt S_{q,k}),~k\neq 0.
\end{align}
\end{subequations}
The solution of \eqref{b2} and \eqref{c2} is the same as \eqref{bb2} and \eqref{cc2} for $i \rightarrow \infty$. By plugging the solutions of \eqref{b2} and \eqref{c2} in $\nabla_{{\bf \Sigma}_q}\bar{f}_q(\bar{\bf \Sigma}_q,\bar{\bt W}_q,\left\{\bt S_{q,k}\right\}_{k=0}^K)$ and $\nabla_{\bt W_q}\bar{f}_q(\bar{\bf \Sigma}_q,\bar{\bt W}_q,\left\{\bt S_{q,k}\right\}_{k=0}^K)$, the convergence point of Algorithm 2 is a QNE of the proposed game. Overall, by using the gradient-response algorithm, the uniqueness of the QNE and $\gamma_{i^\prime} = d < \frac{c_{s}}{L^2},~\forall i^\prime = 1,\dots,m$ directly suggest the convergence of the iteration in \eqref{iteration}. Hence, a separate proof for the convergence of Algorithm 2 is not needed.

The iteration proposed in \eqref{iteration} has two major issues. First, the Lipschitz constant of $F^{\mathbb{R}}(x)$ has to be known. Apart from being difficult to derive, the knowledge of Lipschitz constant requires a centralized computation. Second, the strong monotonicity of $F^{\mathbb{R}}$ cannot be always guaranteed. In fact, the conditions derived in \eqref{domreal} are very dependent on the channel gains and network topology. Hence, in most typical network scenarios, the inequality in \eqref{domreal} cannot be satisfied. This means that in some situations, the game might have more than one QNE. Consequently, the convergence of Algorithm 2 is in jeopardy. However, on the condition that $F^{\mathbb{R}}$ is \textit{monotone}\footnote{See Appendix \ref{sec:apla}
to recall the difference between monotonicity and strong monotonicity.}, which is a weaker condition than strong monotonicity, the ability to choose between multiple QNEs is possible. This means that the users are able to select the QNE that satisfies a certain design criterion, thus guaranteeing convergence in the case of multiple QNEs. Moreover, depending on the design criterion, the performance of the resulting QNE in terms of the achieved secrecy sum-rate can be improved. To do this, we first review the regularization methods proposed for VIs.

\subsection{Tikhonov Regularization}
The general idea of regularization techniques is to modify the players' utility functions such that the VI becomes strongly monotone (and hence easily solvable by using Algorithm 2), and the limit point of a sequence of solutions for the modified VI converges to some solution of the original VI. In Tikhonov regularization, the process of regularizing $\text{VI}(F^{\mathbb{R}},\mathcal{K}^{\mathbb{R}})$ involves solving a sequence of VIs, where the following iteration is characterized for a given $\epsilon$ \cite[chapter 12]{pang}:
\begin{equation}
\label{itertikhonov}
x^{(i+1)} = \Pi_{\mathcal{K}^{\mathbb{R}}}\left(x^{(i)}-\gamma^T \left(F^{\mathbb{R}}(x^{(i)})+\epsilon x^{(i)}\right)\right).
\end{equation}
The solution to \eqref{itertikhonov} when $i \rightarrow \infty$ is denoted as $x(\epsilon)$. Given that $F^{\mathbb{R}}$ is monotone, solving a sequence of (strongly monotone) $\text{VI}(F^{\mathbb{R}}(x)+\epsilon x,\mathcal{K}^\mathbb{R})$'s while $\epsilon \rightarrow 0$ has a limit point, (i.e., $\lim_{\epsilon \rightarrow 0}x(\epsilon)$ exists) and that limit point is equal to least-norm solution of the $\text{VI}(F^\mathbb{R},\mathcal{K}^\mathbb R)$ \cite[Theorem 12.2.3]{pang}.
\subsection{QNE Selection using Tikhonov Regularization}
Generalizing the applicability of Tikhonov regularization, we are more interested in converging to the QNE that is more beneficial to the users. In our approach to QNE selection, we define benefit as when the selected QNE satisfies a particular design criterion. Let the set of solutions of $\text{VI}(F^{\mathbb{R}},\mathcal{K}^\mathbb {R})$ be denoted as $\text{SOL}(F^{\mathbb{R}},\mathcal{K}^\mathbb {R})$. We want to select the NE that minimizes a strongly convex\footnote{A strongly convex function is a function whose derivative is strongly monotone. We use the definitions of \cite{paloscut} to distinguish between different types of convexity.} function $\Phi(x): \mathcal{K^{\mathbb{R}}}\rightarrow \mathbb{R}$. In fact, the QNE selection satisfies the following design criterion\footnote{The discussion on how we determine the function $\Phi(x)$ will be tackled in Section \ref{choiceofdesign}.}
\begin{align}
\label{criterion}
\underset{}{\text{minimize}}~~~&\Phi (x) \nonumber
\\
\text{s.t.}~~~& x \in \text{SOL}(F^{\mathbb{R}},\mathcal{K}^\mathbb {R}).
\end{align}
The optimization in \eqref{criterion} is convex because the monotonicity of $F^\mathbb{R}$ suggests that $SOL(F^\mathbb{R},\mathcal K^\mathbb{R})$ is a convex set \cite[Chapter 2]{pang}. The unique point that solves problem \eqref{criterion}, is the solution to $\text{VI}(\nabla \Phi(x),\text{SOL}(F^{\mathbb{R}},\mathcal{K}^\mathbb {R}))$. However, as there is no prior knowledge on $\text{SOL}(F^{\mathbb{R}},\mathcal{K}^\mathbb {R})$ (i.e., QNEs are not known), this optimization cannot be solved easily. To overcome this issue, we modify the function $F^\mathbb{R}$ in $\text{VI}(F^{\mathbb{R}},\mathcal{K}^\mathbb {R})$ to
\begin{equation}
\label{tikh}
F_\epsilon^{\mathbb{R}} \triangleq F^{\mathbb{R}}+\epsilon \nabla \Phi(x).
\end{equation}
As the function $\Phi(x)$ is a strongly convex function, its derivative w.r.t $x$ is strongly monotone. Assuming that $F^{\mathbb{R}}$ is monotone, then the function $F_\epsilon^{\mathbb{R}}$ is strongly monotone and the solution to $\text{VI}(F_\epsilon^{\mathbb{R}},\mathcal{K}^\mathbb {R})$, namely, $x(\epsilon)$, is unique for all values of $\epsilon > 0$ (i.e., convergence to a QNE can be guaranteed). The iteration used for QNE selection is written as
\begin{equation}
\label{itertikhonov2}
x^{(i+1)} = \Pi_{\mathcal{K}^{\mathbb{R}}}\left(x^{(i)}-{\gamma^{(i)}} \left(F^{\mathbb{R}}(x^{(i)})+\epsilon^{(j)} \nabla \Phi(x)+\theta^{(i)} (x^{(i)}-x^{(i-1)}\right)\right).
\end{equation}
The iteration in \eqref{itertikhonov2} is the same as \eqref{itertikhonov} with the difference that the multiplier of $\epsilon$ in \eqref{itertikhonov} is replaced by $\nabla \Phi(x)$. The following theorem shows the potential of using \eqref{itertikhonov2}
for QNE selection:
\begin{theorem} \cite[ pp. 1128 and Theorem 12.2.5]{pang}
\label{tikhonthm}
Consider $\text{VI}(F_\epsilon^{\mathbb{R}},\mathcal{K}^\mathbb {R})$ with $x(\epsilon)$ as its solution. Assume that $\mathcal{K}^\mathbb{R}$ is closed and convex, and $\text{SOL}(F^{\mathbb{R}},\mathcal{K}^\mathbb {R})$ is nonempty. The following claims hold:
\begin{itemize}
\item
The assumption that $\mathcal{K}^\mathbb{R}$ is closed and convex together with the nonemptiness of $\text{SOL}(F^\mathbb{R},\mathcal{K}^\mathbb{R})$ (i.e., the existence of the QNE, proved in Theorem \ref{existancthm}) are necessary and sufficient for $x_\infty = \underset{{\epsilon \rightarrow 0}}{\lim}~x(\epsilon)$ to exist.
\item
Assuming that $F^\mathbb{R}$ is monotone\footnote{Later on, we elaborate on the monotonicity assumption for $F^\mathbb{R}$ (cf. Section \ref{overhead}).}, $x_\infty$ is the solution of $\text{VI}(\nabla \Phi(x),\text{SOL}(F^{\mathbb{R}},\mathcal{K}^\mathbb {R}))$. This means that a QNE among several QNEs can be selected\footnote{We emphasize that by QNE selection, the players are maximizing their (modified) utility functions. Hence, the noncooperative nature of the game is preserved.}.\qed
\end{itemize}
\end{theorem}
{\subsection{Guaranteeing Monotoncity of $F^\mathbb{R}$ in Tikhonov Regularization}
Theorem \ref{tikhonthm} requires $F^\mathbb{R}$ to be monotone to be applicable. However, the monotonicity of $F^\mathbb{R}$, as highlighted by Theorem \ref{tm3}, depends on many factors such as the channels between different nodes in the network, meaning that it is not possible to always guarantee the monotonicity of $F^\mathbb{R}$. In order to guarantee the monotonicity, we add a strongly concave term to the utility of each player. Let this term be $-\frac{\tau_q}{2}\left(||{\bf \Sigma}_q-Y_{{\bf \Sigma}_q}||_F^2+||\bt W_q-Y_{\bt W_q}||_F^2\right)$ where ${||.||}_F$ indicates the Frobenius norm. Hence, the utility of each player defined in \eqref{reform2} will change to 
\begin{align}
\label{reform2}
\underset{{\bf\Sigma}_q,\bt W_q, \bt S_q}{\text{maximize}}&~\bar{f}_q({\bf \Sigma}_q,\bt W_q,\left\{\bt S_{q,k}\right\}_{k=0}^K)-\frac{\tau_q}{2}\left(||{\bf \Sigma}_q-Y_{{\bf \Sigma}_q}||_F^2+||\bt W_q-Y_{\bt W_q}||_F^2\right), \nonumber
\\
\text{s.t.~~~~~~}& ({\bf \Sigma}_q,\bt W_q)\in \mathcal{F}_q,~\bt S_k \succeq 0,~q \in \mathbb{Q},~k \in \mathbb{K}
\end{align}
where $Y_{{\bf \Sigma}_q}$ and $Y_{\bt W_q}$ are complex constants which will be explained later. With this modification on the utility of each player, a new VI problem, $VI(F_\tau^\mathbb{R},\mathcal{K}^\mathbb{R})$ is established where:
\begin{equation}
\label{pert}
F_\tau^\mathbb{R}(x) = F^\mathbb{R}(x) + \tau (x-y)
\end{equation}
where $y$ is the vector that contains the vectorized versions of $Y_{{\bf \Sigma}_q}$ and $Y_{\bt W_q}$, and $\tau = diag(\tau_1,\tau_2,\dots, \tau_m)$ is an $m\times m$ diagonal matrix, and $F^\mathbb{R}$ is not a function of $y$. This perturbation is also known as Proximal Point regularization method \cite[Chapter 12.3.2]{pang}. Recalling Definition 1, the augmented Jacobian matrix of $F_\tau^\mathbb{R}(x)$, namely as $\mathcal{J}_\tau$, is as follows
\begin{equation}
\mathcal{J}_\tau \triangleq \mathcal{J} + \tau I
\end{equation}
where $\mathcal{J}$ is the augmented Jacobian matrix of $F^\mathbb{R}$ and $I$ is the identity matrix. Considering the matrix $\tau$ as a free parameter, we can choose a suitable value for each diagonal element of $\tau$, such that the matrix $\mathcal{J}_\tau$ becomes a diagonally dominant matrix. In the following we exploit the diagonal dominance of $\mathcal{J}_\tau$ to establish the monotonicity property of $F^\mathbb{R}_\tau$\footnote{Later as we proceed, we present the equivalent regularization for the complex version of $F^\mathbb{R}$, i.e., $F^\mathbb{C}$ as well.}.

Let $D(d_i, [\mathcal{J}_\tau]_{ii}),~i = 1,\dots,m$ be the closed disc centered at $ [\mathcal{J}_\tau]_{ii}$ with radius $d_i = \sum_{j\neq i}\left|[\mathcal{J}_\tau]_{ij}\right|$, where $[.]_{ii}$ denotes the diagonal element and $[\mathcal{J}_\tau]_{ii} = [\mathcal{J}]_{ii}+\tau_i$. Using the Gerschgorin circle theorem \cite{horn}, 
for all $i = 1,\dots,m$, every eigenvalue of $\mathcal{J}_\tau$ is within at least one of the discs. We also know that for the function $F^\mathbb{R}_\tau$, in order to be monotone, the matrix $\mathcal{J}_\tau$ has to be APSD (cf. Appendix \ref{sec:apla}). Hence, provided that a suitable value for $\tau_i$ is chosen for all $i = 1,\dots,m$, all the radii of the Gershgorin circles must be less than their respective diagonal elements, ensuring that $\mathcal{J}_\tau$ remains APSD. Using this fact, the value for $\tau_i$ that guarantees $\mathcal{J}_\tau$ to be APSD is
\begin{equation}
\label{gersh}
\tau_i \ge d_i-\mathcal{J}_{ii},~\forall i.
\end{equation}
Therefore, using the condition \eqref{gersh} with equality, the matrix $\mathcal{J}_\tau$ becomes an APSD matrix, and consequently, $F^\mathbb{R}_\tau$ becomes monotone. Therefore, the Tikhonov regularization changes to solving the problem $VI(F^\mathbb{R}_{\tau,\epsilon},\mathcal{K}^\mathbb{R})$ where
\begin{equation}
F^\mathbb{R}_{\tau,\epsilon} \triangleq F^\mathbb{R}(x) +\tau(x-y)+\epsilon^{(j)} \nabla \Phi(x)
\end{equation}

Building upon the perturbation in \eqref{pert}, we can now use $F^\mathbb{R}_\tau$ instead of $F^\mathbb{R}$ in the original VI in \eqref{final} which makes us able to use Tikhonov regularization and perform equilibrium selection. One might argue that using $F^\mathbb{R}_\tau$ instead of $F^\mathbb{R}$ is actually creating a new game with different solutions. In the following we give a property that makes the use of $F^\mathbb{R}_\tau$ reasonable. It can be easily seen that the perturbation $F^\mathbb{R}_\tau$ does not change the fact that the NE in $VI(F^\mathbb{R}_\tau,\mathcal{K}^\mathbb{R})$ still exists, i.e., the set $\text{SOL}(F^\mathbb{R}_\tau,\mathcal{K}^\mathbb{R})$ is nonempty (cf. Theorem \ref{thmvi}). Furthermore, the addition of a monotone term (i.e., $\tau(x-y)$ does not change the convexity of utilities to their actions. We set the vector $y$ to be $y = x(\epsilon^{(j-1)})$, which means that 
while computing the $j$-th member of solutions of $VI(F_{\tau,\epsilon}^\mathbb{R},\mathcal{K}^\mathbb{R})$, namely as $x(\epsilon^{(j)})$, the vector $y$ is the same as the solution found for $VI(F_{\tau,\epsilon}^\mathbb{R},\mathcal{K}^\mathbb{R})$ when $\epsilon = \epsilon^{(j-1)}$. Therefore, in the limit point where $x_\infty \in \text{SOL}(F^\mathbb{R}_\tau,\mathcal{K}^\mathbb{R})$, we have
\begin{align}
&x_\infty \in \text{SOL}(F^\mathbb{R}_\tau,\mathcal{K}^\mathbb{R}) \Rightarrow (x-x_\infty) F^\mathbb{R}_\tau(x_\infty) > 0 \nonumber
\\
&\Rightarrow (x-x_\infty) \left(F^\mathbb{R}(x_\infty)+\tau(x_\infty-x_\infty)\right) > 0 \nonumber
\\
&\Rightarrow x_\infty \in \text{SOL}(F^\mathbb{R},\mathcal{K}^\mathbb{R}).
\end{align}
Hence, the term $\tau(x_\infty-x_\infty)$ vanishes since the limit point is guaranteed to be reached.}
\subsection{Distributed Tikhonov Regularization}
Tikhonov regularization (QNE selection) is done in two nested loops. In the inner loop, for a given $\epsilon^{(j)}$, the solution to $\text{VI}(F^\mathbb R_{\epsilon},\mathcal{K}^\mathbb{R})$ will be found from the iteration in \eqref{itertikhonov} (where the multiplier of $\epsilon$ is replaced with $\nabla \Phi(x)$). In the outer loop, the next value of $\epsilon^{(j)}$ will be chosen (according to a predefined sequence such that $\lim_{j\rightarrow \infty}\epsilon^{(j)} = 0$) until the solution to $\text{VI}(\nabla \Phi(x),\text{SOL}(F^{\mathbb{R}},\mathcal{K}^\mathbb {R}))$ is reached (cf. Theorem \ref{tikhonthm}).

Despite having the ability to select a specific QNE among multiple QNEs, 
QNE selection requires heavy signaling and centralized computation because still the Lipschitz Continuity constant $L$ and strong monotonicity modulus of $F^\mathbb{R}_{\epsilon} (x)$ must be known (cf. Section \ref{graa}). In order to address these issues, we introduce another regularization method, namely, proximal point regularization. In this regularization, a term $\theta^{(i)} (x^{(i)}-x^{(i-1)})$ is added to the function $F_{\epsilon}^\mathbb R (x)$ to build a function $F_{\tau, \epsilon,\theta}^\mathbb R (x) \triangleq F_{\tau,\epsilon}^\mathbb R (x)+\theta^{(i)} (x^{(i)}-x^{(i-1)})$ where $\theta^{(i)}$ is a diagonal matrix.
 Considering this modification, the following property can be used:
\begin{proposition}
\label{propprox}
Let $F_{\tau,\epsilon}^{\mathbb R}(x)$ be a strictly monotone and Lipschitz continuous mapping\footnote{Note that Lipschitz continuity of $F_{\tau,\epsilon}^{\mathbb R}(x)$ requires both $F^{\mathbb R}(x)$ and $\nabla \Phi (x)$ to be Lipschitz continuous. Hence, the proposed choices for $\Phi(x)$ in the next section are all Lipschitz continuous.}; $\max_{z\in \mathcal{K}^\mathbb R}||x|| \le C$, and $\max_{z\in \mathcal{K}^\mathbb R}||F^\mathbb{R}_{\tau,\epsilon}|| \le B$ where $C$ and $B$ are positive constants. Furthermore, suppose that for a given $\epsilon^{(j)}$, the solution to $\text{VI}(F^\mathbb R_{\tau,\epsilon},\mathcal{K}^\mathbb R)$ is denoted as $x(\epsilon^{(j)})$. Let $x^{(i)}$ denote the set of iterates defined by
\begin{equation}
\label{itertikhonov3}
x^{(i+1)} = \Pi_{\mathcal{K}^{\mathbb{R}}}\left(x^{(i)}-{\gamma^{(i)}} \left(F^{\mathbb{R}}(x^{(i)})+\tau(x^{(i)}-x(\epsilon^{(j-1)}))+\epsilon^{(j)} \nabla \Phi(x^{(i)})+\theta^{(i)} (x^{(i)}-x^{(i-1)}\right)\right)
\end{equation}
where the step size matrix $\gamma^{(i)}$ is changing with the iterations. Lastly, set $\gamma^{(i)}\theta^{(i)} = c = \text{diag}([c_1,\dots,c_m])$ where $c_{i^\prime} \in (0,1), \forall i^\prime = 1,\dots,m$ is a constant,  and let the following hold:
\begin{equation}
\label{cond}
\sum_{i = 1} \gamma^{(i)} = \infty,~~\sum_{i=1}^\infty \left(\gamma^{(i)}\right)^2 < \infty,~~\text{and~}\sum_{i = 1}^\infty (\gamma_{max}^{(i)}-\gamma_{min}^{(i)})<\infty.
\end{equation}
where $\gamma_{max}^{(i)}$ and $\gamma_{min}^{(i)}$ are respectively the maximum and minimum diagonal elements of the matrix $\gamma^{(i)}$. Therefore, we have $\lim_{i\rightarrow \infty}x^{(i)} = x(\epsilon^{(j)})$.
\qed
\end{proposition}
The proof of Proposition \ref{propprox} can be found in \cite[Proposition 3.4]{uday}. However, note that the assumption of strict monotonicity of $F_{\tau, \epsilon}^\mathbb{R}(x)$ is immediately satisfied as $F_{\tau, \epsilon}^\mathbb{R}(x)$ is already strongly monotone (cf. \eqref{tikh}). The conditions $\max_{z\in \mathcal{K}^\mathbb R}||x|| \le C$ and $\max_{z\in \mathcal{K}^\mathbb R}||F^\mathbb{R}_{\tau, \epsilon}|| \le B$ can also be satisfied due to having power constraints on each link. According to \cite[Proposition 3.4]{uday}, the step size $\gamma^{(i)}$ can be chosen as $\gamma_{i^\prime}^{(i)} = (i+\alpha_{i^\prime})^{-\omega}$ where $\alpha_{i^\prime}$ is a positive integer for $i^\prime = 1,\dots,N$ and $0<\omega <1$. Hence, we can write
\begin{equation}
\label{stepsize}
\gamma_{max}^{(i)} = (i+\alpha_{max})^{-\omega},~~\gamma_{min}^{(i)} = (i+\alpha_{min})^{-\omega}.
\end{equation}
Note that in Proposition \ref{propprox}, $\theta^{(i)}$ is already set to $\theta^{(i)} = \frac{c}{\gamma^{(i)}}$. Using Proposition \ref{propprox}, we can design a distributed transmit optimization algorithm without the knowledge of Lipschitz constant and strong monotonicity modulus of $F^\mathbb R_{\tau,\epsilon}$. The next section discusses the implementation of QNE selection using \eqref{itertikhonov3}\footnote{Note that in all of the proposed algorithms throughout this paper, it was assumed that at each round of the game, all of the players are maximizing the utilities. This update fashion is also known as Jacobi implementation. The feasibility of implementing the algorithms using other update fashions (e.g., Gauss-Seidel or Asynchronous) can be a subject of future research.}.
{\section{The QNE Selection Algorithm: Design and Discussion}
\label{sec6}}
In this section of the paper, we propose the QNE selection algorithm together with three possible choices for the design criterion (i.e., $\Phi(x)$). Each of these choices imposes a certain amount of signaling overhead as well as a certain amount of improvement on the performance of Algorithm 1 and Algorithm 2.

\subsection{QNE Selection Algorithm}
The pseudo-code for the QNE selection algorithm is shown in Algorithm 3. As mentioned previously, it can be seen in Algorithm 3 that the modified game (i.e., QNE selection algorithm) is comprised of two nested loops: outer loop (i.e., line 1), and inner loop (i.e., line 3). In the outer loop the $j$th member of $\epsilon^{(j)}$'s is selected. In the inner loop, the game s played among the players, and the players update their strategies according to \eqref{itertikhonov3}. The sequence $\epsilon^{(j)}$ must be a decreasing sequence such that $\lim_{j\rightarrow \infty}\epsilon^{(j)} = 0$. {The operation in line 10 of Algorithm 3 can be written as
{\small \begin{align}
\label{PG33}
\left(\begin{array}{c} 
{\bf \Sigma}_q^{(i+1)}
\\
\bt W_q^{(i+1)}
\end{array}\right) = \texttt{Proj}_{\mathcal{F}_q}\left(\begin{array}{c}{\bf \Sigma}_q^{(i)}+\gamma^\prime_q\left(\nabla_{{\bf \Sigma}_q}\bar{f}_q+\tau_q\left({\bf \Sigma}_q^{(i)}-{\bf \Sigma}_q(\epsilon^{(j-1)})\right)+\epsilon^{(j)}\nabla_{{\bf \Sigma}_q}\Phi(x^{(i)})-\theta_q^{(i)}({\bf \Sigma}_q^{(i)}-{\bf \Sigma}_q^{(i-1)})\right)
\\
\bt W_q^{(i)}+\gamma^\prime_q\left(\nabla_{\bt W_q}\bar{f}_q+\tau_q\left(\bt W_q^{(i)}-\bt W_q(\epsilon^{(j-1)})\right)+\epsilon^{(j)}\nabla_{\bt W_q}\Phi(x^{(i)})-\theta_q^{(i)}(\bt W_q^{(i)}-\bt W_q^{(i-1)})\right)
\end{array}\right).
\end{align}}}
Notice that {$\theta_q^{(i)}$} is a diagonal matrix that can obtained via dividing the matrix $\theta^{(i)}$ into $Q$ block-diagonal matrices. That is, (with a slight abuse of notations) $\theta^{(i)} =  \text{diag}(\theta^{(i)}_1,\dots,\theta^{(i)}_Q)$. In the next subsection, we specifically explain the terms $\nabla_{{\bf \Sigma}_q}\Phi(x)$ and $\nabla_{\bt W_q}\Phi(x)$ in line 10, so that Algorithm 3 will be completely defined. Lastly, notice that all of our analysis on VI problems were under the assumption that every player is solving a minimization problem as his strategy. Hence, if maximization is the strategy of each player, the proximal terms in \eqref{PG33} appear as a negative values. Furthermore, the addition of $\nabla_{{\bf \Sigma}_q}\Phi(x)$ and $\nabla_{\bt W_q}\Phi(x)$ means that $\Phi(x)$ must be a strongly concave function of $x$.
\begin{algorithm}[H]
\caption{QNE Selection Algorithm}
\label{alg3}
\footnotesize
\setstretch{1.5}
{\bf Initialize:} ${\bf \Sigma}^{(1)}_q$, $\bt W_q^{(1)}$, $\tr({\bf \Sigma}^{(1)}_q+\bt W_q^{(1)}) < P_q$, $\forall q$, and j = 1
\begin{algorithmic}[1]
\Repeat\hspace{5mm}\% Outer loop: superscript $(j)$ indicates the iterations starting from here
\\
Choose the $j$th member of the sequence $\epsilon^{(j)}$
\Repeat\hspace{5mm}\% Inner loop: superscript $(i)$ indicates the iterations starting from here
\\
Compute $\bt M_q$, $\bt M_{e,q,k}$, $\forall (q,k)\in \mathbb Q\times \mathbb K$
\\
Compute $\bt S^{(i)}_{q,k}$, $\forall (q,k)\in \mathbb Q\times \mathbb K$
\\
Compute $\varphi_{e,q,k}({\bf \Sigma}^{(i)}_q,\bt W_q^{(i)},\bt S^{(i)}_{q,k})$, $\forall (q,k)\in \mathbb Q\times \mathbb K$
\For {q = 1,\dots,Q}
\\
Update the values of $\tau_q$ for all $q = 1,\dots,Q$ such that the inequality in \eqref{domreal} is satisfied
\\
\hspace{10mm}Replace $\nabla_{{\bf \Sigma}_q}\bar{f}_q$ with $\nabla_{{\bf \Sigma}_q}\bar{f}_q-\tau_q\left({\bf \Sigma}_q^{(i)}-{\bf \Sigma}_q(\epsilon^{(j-1)})\right)+\epsilon^{(j)}\nabla_{{\bf \Sigma}_q}\Phi(x^{(i)})-\theta_q^{(i)}\left({\bf \Sigma}_q^{(i)}-{\bf \Sigma}_q^{(i-1)}\right)$
\\
\hspace{10mm}Replace $\nabla_{\bt W_q}\bar{f}_q$ with $\nabla_{\bt W_q}\bar{f}_q-\tau_q\left(\bt W_q^{(i)}-\bt W_q(\epsilon^{(j-1)})\right)+\epsilon^{(j)}\nabla_{\bt W_q}\Phi(x^{(i)})-\theta_q^{(i)}\left(\bt W_q^{(i)}-\bt W_q^{(i-1)}\right)$
%
\\
\hspace{10mm}Compute $({\bf \Sigma}^{(i+1)}_q, \bt W_q^{(i+1)})$ using \eqref{PG33}
%
\EndFor
\Until{Convergence to QNE}\hspace{5mm} \% $x(\epsilon^{j})$ is found
\\
j = j+1
\Until{Convergence to limit point of $x(\epsilon^{j})$'s}
\end{algorithmic}
\end{algorithm}
\subsection{On the Choice of Design Criterion for QNE Selection}
\label{choiceofdesign}
Assume that the derivatives of $\Phi(x)$ are described as:
\begin{subequations}
\label{realvectorized}
\begin{align}
&\nabla \Phi(x) \triangleq [\nabla_{{\bf \Sigma}_1, \bt W_1}^\mathbb{R} \Phi(x)^T,\dots,\nabla_{{\bf \Sigma}_Q, \bt W_Q}^\mathbb{R} \Phi(x)^T]^T,
\\
&\nabla_{{\bf \Sigma}_q, \bt W_q}^\mathbb{R} \Phi(x) \triangleq [\nabla_{{\bf \Sigma}_q}^\mathbb{R} \Phi(x)^T,\nabla_{\bt W_q}^\mathbb{R} \Phi(x)^T]^T,~q \in \mathbb Q,
\\
&\nabla_{{\bf \Sigma}_q}^\mathbb{R} \Phi(x) \triangleq [\re\{\ve(\nabla_{{\bf \Sigma}_q} \Phi(x))\}^T, \im\{\ve(\nabla_{{\bf \Sigma}_q} \Phi(x))\}^T]^T,
\\
&\nabla_{\bt W_q}^\mathbb{R} \Phi(x) \triangleq [\re\{\ve(\nabla_{\bt W_q} \Phi(x))\}^T, \im\{\ve(\nabla_{\bt W_q} \Phi(x))\}^T]^T.
\end{align}
\end{subequations}
We are now ready to present the possible choices of $\Phi(x)$:
\subsubsection{Maximizing the sum of information rates}
We aim to select the QNE that maximizes the sum-rate of all links. Recalling the reformulated information rate (i.e., $\varphi_q({\bf \Sigma}_q,\bt W_q,\bt S_{q,k})$) in \eqref{bb}, $\Phi(x)$  can be described as (with $q \in \mathbb Q$):
\begin{subequations}
\label{crisumrate}
\begin{align}
&\nabla_{{\bf \Sigma}_q} \Phi(x) = \sum_{\underset{r\neq q}{r = 1}}^Q \bt H_{qr}^H \left((\bt M_r+\bt H_{rr}{\bf \Sigma}_{r}\bt H_{rr}^H)^{-1}-\bt S_{r,0}\right)\bt H_{qr},
\\
&\nabla_{\bt W_q} \Phi(x) = \sum_{\underset{r\neq q}{r = 1}}^Q \bt H_{qr}^H \left((\bt M_r+\bt H_{rr}{\bf \Sigma}_{r}\bt H_{rr}^H)^{-1}-\bt S_{r,0}\right)\bt H_{qr}.
\end{align}
\end{subequations}
Notice that although we wrote $\Phi$ as a function of $x$, one can easily relate the vector $x$ to the covariance matrices $\{({\bf \Sigma}_q,\bt W_q)\}_{q=1}^Q$ using \eqref{realvectorized} and \eqref{iteration}. Hence, the derivatives of $\Phi(x)$ at the end of Algorithm 3 would be:
\begin{subequations}
\begin{align}
\label{oneone}
&\nabla_{{\bf \Sigma}_q} \Phi(x) = \sum_{\underset{r\neq q}{r = 1}}^Q \bt H_{qr}^H \left((\bt M_r^\star+\bt H_{rr}{\bf \Sigma}_{r}^\star\bt H_{rr})^{-1}-\bt S_{r,0}^\star\right)\bt H_{qr}
\\
\label{twotwo}
&\nabla_{\bt W_q} \Phi(x) = \sum_{\underset{r\neq q}{r = 1}}^Q \bt H_{qr}^H \left((\bt M_r^\star+\bt H_{rr}{\bf \Sigma}_{r}^\star\bt H_{rr})^{-1}-\bt S_{r,0}^\star\right)\bt H_{qr}
\end{align}
\end{subequations}
where $\bt M_r^\star = \bt I +\bt H_{rr}(\bt W_r^\star) \bt H^H_{rr} +\bt H_{qr}(\bt W_q^\star +{\bf \Sigma}_q^\star) \bt H^H_{qr}+\sum_{\underset{l\neq q,r}{l = 1}}^Q\bt H_{lr}\left({\bf \Sigma}_l^\star+\bt W_l^\star\right) \bt H^H_{lr}$, with ${\bf \Sigma}_q^\star$ and $\bt W_q^\star$ being the limit points of ${\bf \Sigma}_q$ and $\bt W_q$. Integrating \eqref{oneone} w.r.t. ${\bf \Sigma}_q^\star$ and integrating \eqref{twotwo} w.r.t. $\bt W_q^\star$, we end up with $\Phi(x) =  \sum_{q=1}^Q\sum_{\underset{r\neq q}{r = 1}}^Q\varphi_r({\bf \Sigma}_r, \bt W_r, \bt S_{r,0})$. Hence, at the end of Algorithm 3, the QNE that is a stationary point of sum-rate of all links is selected, i.e., the point that is the unique solution of $\text{VI}{(\nabla \Phi(x),\text{SOL}(F^{\mathbb{R}},\mathcal{K}^\mathbb {R}))}$.
\subsubsection{Minimizing the received rates at Eves}
We can describe $\Phi(x)$ by (with $q \in \mathbb Q$)
\begin{subequations}
\label{crieves}
\begin{align}
&\nabla_{{\bf \Sigma}_q} \Phi(x) = \sum_{\underset{r\neq q}{r = 1}}^Q\sum_{k = 1}^K \rho_{r,k}\bt G_{rk}^H\left(({\bt M_{e,r,k}}^{-1}-\bt S_{r,k}\right)\bt G_{rk}
\\
&\nabla_{\bt W_q} \Phi(x) = \sum_{\underset{r\neq q}{r = 1}}^Q\sum_{k = 1}^K \rho_{r,k}\bt G_{rk}^H\left(({\bt M_{e,r,k}}^{-1}-\bt S_{r,k}\right)\bt G_{rk}
\end{align}
\begin{align}
\bt M_{e,r,k} \triangleq \bt I + \bt G_{rk}\bt W_r \bt G^H_{rk}+\bt G_{qk}&\left({\bf \Sigma}_q+\bt W_q\right) \bt G^H_{qk} +
\\
\sum_{\underset{l\neq q,r}{l = 1}}^Q&\bt G_{lk}\left({\bf \Sigma}_l+\bt W_l\right) \bt G^H_{lk}
\end{align}
\end{subequations}
where the term $\rho_{r,k}$ is defined in \eqref{rhoorig}. Following the same reasoning used in the previous QNE selection, at the limit point of $x(\epsilon^{(j)})$, we end up with $\Phi(x) = \sum_{q=1}^Q \sum_{\underset{r\neq q}{r = 1}}^Q-\frac{1}{\beta}\ln(\sum_{k=1}^K \exp\{\beta \varphi_{e,r,k}({\bf \Sigma}_r,\bt W_r,\bt S_{r,k})\})$, where $\varphi_{e,r,k}({\bf \Sigma}_r,\bt W_r,\bt S_{r,k})$ is defined in \eqref{cc}. Hence, the selected QNE guides the game to the stationary point of minimizing Eves' received rates, i.e., the point that is the unique solution of $\text{VI}{(\nabla \Phi(x),\text{SOL}(F^{\mathbb{R}},\mathcal{K}^\mathbb {R}))}$.
\subsubsection{Maximizing the sum of secrecy rates}
In this criterion, a simple addition of previous design criteria gives us another QNE selection method, in which the QNE that is a stationary point of secrecy sum-rate is selected.
\section{Centralized Algorithm}
\label{seccent}
An appropriate measure of efficiency (i.e., social welfare) in our game would be the sum of utilities of all players or the secrecy sum-rate. The price of anarchy (PoA) can be defined as the ratio between the performance of the optimal centralized solution for the secrecy sum-rate maximization problem and the \emph{worst} NE. However, such definition of PoA requires us to solve the secrecy sum-rate maximization problem, which is a nonconvex problem. Moreover, all of the proposed algorithms converge to the QNEs of the proposed game, which are not necessarily NEs. Hence, direct PoA analysis is not feasible. Instead, to measure the efficacy of QNEs, we design a centralized algorithm that provides locally optimal solutions for the (network-wide) secrecy sum-rate maximization problem. We refer to this algorithm as Centralized Secrecy Sum-rate Maximization method (CSSM). The objective value of solving \eqref{CSSMsrmreform} via the CSSM is considered as the social welfare in our game.

In CSSM, the objective is to find a stationary solution for the following optimization problem:
\begin{align}
\label{CSSMsrm}
\underset{({\bf \Sigma}_q,\bt W_q) \in \mathcal F_q,~\forall q}{\text{maximize}~~}\sum_{q = 1}^Q\bar{R}_{s,q}({\bf \Sigma}_q,\bt W_q).
\end{align}
Using the reformulation techniques given in Section \ref{sec3}, the problem in \eqref{CSSMsrm} can be rewritten as
\begin{align}
\label{CSSMsrmreform}
\underset{\underset{\forall q,k}{{(\bf \Sigma}_q,\bt W_q, \bt S_{q,k})}}{\text{maximize}~~}&\sum_{q = 1}^Q\bar{f}_{q}({\bf \Sigma}_q,\bt W_q, \{\bt S_{q,k}\}_{k = 0}^K) \nonumber
\\
\text{s.t.~~}&({\bf \Sigma}_q,\bt W_q) \in \mathcal F_q, \forall q \in \mathbb Q, \nonumber
\\
& \bt S_{q,k} \succeq 0,~\forall (q,k) \in \mathbb Q \times \{0\} \cup \mathbb K.
\end{align}
Problem \eqref{CSSMsrmreform} can be shown to be convex w.r.t either {$[{\bf \Sigma}, \bt W] = \{{\bf \Sigma}_q, \bt W_q\}_{q= 1}^Q = \left[[{\bf \Sigma}_1, \bt W_1]^T,\dots,[{\bf \Sigma}_Q, \bt W_Q]^T\right]$} or {$\bt S = \{\bt S_{q,k}\}_{\forall q,k} = \left[\bt S_{1,0},\dots,\bt S_{1,K},\bt S_{2,0},\dots,\bt S_{2,K},\dots, \bt S_{Q,K}\right]^T$}. Hence, a stationary point of \eqref{CSSMsrm} can be found by solving \eqref{CSSMsrmreform} sequentially w.r.t. {$[{\bf \Sigma}, \bt W]$} and {$\bt S$} until reaching a convergence point. That is, in one iteration, problem \eqref{CSSMsrmreform} is solved w.r.t. {$\bt S$} to find an optimal solution {$\bt S^*$}. Next, with {$\bt S^*$} plugged in the objective of \eqref{CSSMsrmreform}, problem \eqref{CSSMsrmreform} can be optimized w.r.t. {$[{\bf \Sigma}, \bt W]$} to find an optimal solution {$[{\bf \Sigma}^*,\bt W^*] = \{{\bf \Sigma}_q^*,\bt W_q^*\}_{q = 1}^Q$}. Problem \eqref{CSSMsrmreform} is separable w.r.t. every element of {$\bt S$}. Hence,
\begin{subequations}
\begin{align*}
&\bt S^*_{q,0} \triangleq \arg \max_{\bt S_{q,0}\succeq 0} \sum_{q = 1}^Q\bar{f}_{q}({\bf \Sigma}_q,\bt W_q,\bt S_{q,k}) = ({\bt M_q})^{-1}\nonumber
\\
&\bt S^*_{q,k} \!\triangleq \!\arg \max_{\bt S_{q,k}\succeq 0} \sum_{q = 1}^Q\bar{f}_{q}({\bf \Sigma}_q,\bt W_q,\bt S_{q,k}) \!\!=\!\!
\\
& \left(\bt M_{e,q,k}\!+\!\bt G_{qk}{\bf \Sigma_{q}}^n\bt G_{qk}^H\right)^{-1}\!\!\!=\!\! \nonumber
\end{align*}
\end{subequations}
Now, we can solve \eqref{CSSMsrmreform} w.r.t $[{\bf \Sigma}, \bt W]$. We use the augmented Lagrangian multiplier method \cite{bert1} to derive a centralized solution for $[{\bf \Sigma}^*, \bt W^*]$. Let $\text{\bf c}_q = \tr({\bf \Sigma}_q + \bt W_q)-P_q < 0$. The augmented Lagrangian of \eqref{CSSMsrmreform} is \cite{bert1}\footnote{We converted the problem in \eqref{CSSMsrmreform} to a minimization problem by considering the negative of the objective function.}
\begin{align}
L({\bf \Sigma}, \bt W, \text{\bf a}, \text{\bf p}, \bt S^*&) = -\sum_{q = 1}^Q\bar{f}_{q}({\bf \Sigma}_q,\bt W_q, \{\bt S^*_{q,k}\}_{k = 0}^K)+\nonumber
\\
&\frac{1}{2\text{\bf p}} \sum_{q = 1}^Q\left\{(\max\{\text{\bf a}_q+\text{\bf p}\text{\bf c}_q,0\})^2+\text{\bf a}_q^2\right\}
\end{align}
where {$\text{\bf p}$} is a positive penalty (to prevent constraint violations) and {$\text{\bf a}_q$, $q  = 1,\dots, Q$}, are the nonnegative Lagrange multipliers. At a stationary point, the following equalities 
hold for all {$q \in \mathbb Q$} 
\begin{subequations}
\label{lagCSSMsrmreform}
\begin{align}
\label{lagCSSMsrmreformsig}
\frac{\partial }{\partial {\bf \Sigma}_q}L({\bf \Sigma}, \bt W, \text{\bf a}, \text{\bf p}, \bt S^*) = -\sum_{r = 1}^Q\frac{\partial }{\partial {\bf \Sigma}_q}\bar{f}_{r}({\bf \Sigma}_r,\bt W_r, \{\bt S^*_{r,k}\}_{k = 0}^K)+\frac{1}{2\text{\bf p}} \sum_{r = 1}^Q\frac{\partial }{\partial {\bf \Sigma}_q}(\max\{\text{\bf a}_r+\text{\bf p}\text{\bf c}_r,0\})^2 = 0
\\
\label{lagCSSMsrmreformw}
\frac{\partial }{\partial \bt W_q}L({\bf \Sigma}, \bt W, \text{\bf a}, \text{\bf p}, \bt S^*) = -\sum_{r = 1}^Q\frac{\partial }{\partial \bt W_q}\bar{f}_{r}({\bf \Sigma}_r,\bt W_r, \{\bt S^*_{r,k}\}_{k = 0}^K)+\frac{1}{2\text{\bf p}} \sum_{r = 1}^Q\frac{\partial }{\partial \bt W_q}(\max\{\text{\bf a}_r+\text{\bf p}\text{\bf c}_r,0\})^2 = 0
\end{align}
\end{subequations}
where 
{\small \begin{align}
\label{oneaux}
& \frac{\partial }{\partial {\bf \Sigma}_q}\bar{f}_{r}({\bf \Sigma}_r,\bt W_r, \{\bt S^*_{r,k}\}_{k = 0}^K) = 
\left\{\begin{aligned}
&\bt H^H_{qq}(\bt M_q+\bt H_{qq}{\bf \Sigma}_q\bt H_{qq}^H)^{-1}\bt H_{qq}-\sum_{k = 1}^K\rho_{q,k}\bt G_{q,k}^H\bt S^*_{q,k}\bt G_{q,k},~r = q,
\\
&\bt H_{qr}^H \left((\bt M_r+\bt H_{rr}{\bf \Sigma}_{r}\bt H_{rr}^H)^{-1}-\bt S^*_{r,0}\right)\bt H_{qr}+\sum_{k = 1}^K \rho_{r,k}\bt G_{rk}^H\left(({\bt M_{e,r,k}}^{-1}-\bt S^*_{r,k}\right)\bt G_{rk},~r \neq q
\end{aligned}\right.
\end{align}
\begin{align}
\label{twoaux}
& \frac{\partial }{\partial \bt W_q}\bar{f}_{r}({\bf \Sigma}_r,\bt W_r, \{\bt S^*_{r,k}\}_{k = 0}^K) = 
\left\{\begin{aligned}
\bt H_{qq}^H \left((\bt M_q+\bt H_{qq}{\bf \Sigma}_{q}\bt H_{qq})^{-1}-\bt S^*_{q,0}\right)\bt H_{qq}+
\sum_{k = 1}^K \rho_{q,k}\bt G_{qk}^H&\left(({\bt M_{e,q,k}})^{-1}-\bt S^*_{q,k}\right)\bt G_{qk},~r = q,
\\
\bt H_{qr}^H \left((\bt M_r+\bt H_{rr}{\bf \Sigma}_{r}\bt H_{rr}^H)^{-1}-\bt S^*_{r,0}\right)\bt H_{qr}+
\sum_{k = 1}^K \rho_{r,k}\bt G_{rk}^H&\left(({\bt M_{e,r,k}})^{-1}-\bt S^*_{r,k}\right)\bt G_{rk},~r \neq q,
\end{aligned}\right.
\end{align}}
and 
\begin{align}
\label{rhoorig}
& \rho_{q,k} = \frac{e^{\beta \varphi_{e,q,k}({\bf \Sigma}_q,\bt W_q,\bt S^*_{q,k})}}{\sum_{j = 1}^Ke^{\beta \varphi_{e,q,j}({\bf \Sigma}_q,\bt W_q,\bt S^*_{q,j})}}.
\end{align} The second term in the RHS of \eqref{lagCSSMsrmreformsig} is continuously differentiable w.r.t {${\bf \Sigma}_q$} when {$r = q$} \cite[pp. 397]{bert1}. Thus,
{\begin{align}
&\frac{\partial }{\partial {\bf \Sigma}_q}(\max\{\text{\bf a}_r+\text{\bf p}\text{\bf c}_r,0\})^2\hspace{-1mm}=\hspace{-1mm}\left\{\begin{aligned}
& 2 \text{\bf p}(\text{\bf a}_q+\text{\bf p}\text{\bf c}_q){\bf \Sigma}_q,&&\hspace{-3mm}r = q,\text{\bf a}_q+\text{\bf p}\text{\bf c}_q\hspace{-1mm}> 0
\\
&0,&&\mbox{ow}.
\end{aligned}\right.
\end{align}}
and
{\begin{align}
&\frac{\partial }{\partial \bt W_q}(\max\{\text{\bf a}_r+\text{\bf p}\text{\bf c}_r,0\})^2\hspace{-1mm}=\hspace{-1mm}\left\{\begin{aligned}
& 2 \text{\bf p}(\text{\bf a}_q+\text{\bf p}\text{\bf c}_q)\bt W_q,&&\hspace{-3mm}r = q,\text{\bf a}_q+\text{\bf p}\text{\bf c}_q\hspace{-1mm}> 0
\\
&0,&&\mbox{ow}.
\end{aligned}\right.
\end{align}}
To satisfy the conditions in \eqref{lagCSSMsrmreform}, we used gradient descent with a line search satisfying Armijo rule. The details of the centralized algorithm is presented in Algorithm CSSM. The centralized nature of Algorithm CSSM can be seen in Line 12, where the equalities in \eqref{lagCSSMsrmreform} are checked for all $q \in \mathbb Q$ and  Line 11, where the Armijo rule is applied. The convergence of this algorithm can be proved by extending the proof of Theorem 1  in the paper and \cite[Corollary 2]{park}, which is skipped here for the sake of brevity. Note that Algorithm CSSM is sensitive to the initial values of $[{\bf \Sigma}, \bt W]$. Thus, we simulated this algorithm with random initializations and averaged its performance over the total number of initializations.
\begin{algorithm}[H]
	\floatname{algorithm}{Algorithm CSSM \hspace{-70mm}}
	\caption{\hspace{66mm}The Centralized Secrecy Sum-rate Maximization Algorithm (CSSM)}
	\footnotesize
\setstretch{1.5}
	{\bf Initialize:} ${\bf \Sigma}^{(1)}_q$, $\bt W_q^{(1)}$, $\tr({\bf \Sigma}^{(1)}_q+\bt W_q^{(1)}) < P_q$, $\forall q$, $i = 0$
	\begin{algorithmic}[1]
		\Repeat\hspace{5mm}i = i+1~~~~~~\% superscript $(i)$ indicates the iterations starting from here.
		\\
		Compute $\bt S^{(i)}_{q,k}$, $\forall (q,k)\in \mathbb Q\times \mathbb K$, $\text{\bf p} = 1$, $\text{\bf a}_q = 0,~\forall q$, and $s_t$ (Armijo step size)
		\Repeat \hspace{5mm}Set $m = 1$
		\Repeat\hspace{5mm}Set $n = 1$~~~~~~\% superscript $(m)$ indicates the iterations starting from here.
		\\
		\hspace{10mm}Set \hspace{-1mm}$[d_{{\bf \Sigma}_q}, d_{\bt W_q}]^T$ $\hspace{-2mm}=\hspace{-1mm}$
		$-[{\frac{\partial }{\partial {\bf \Sigma}_q}L^{(m)}}^T, {\frac{\partial }{\partial \bt W_q}L^{(m)}}^T],~\forall q \Rightarrow d = \{d_{{\bf \Sigma}_q}, d_{\bt W_q}\}_{q=1}^Q$
		\\
		\hspace{10mm}Set $[\hat{\bf \Sigma}, \hat{\bt W}] = [\bf \Sigma^{(m)}, \bt W^{(m)}]+d$
		\\
		\hspace{10mm}Set $ [\bf \Sigma^{(m+1)}, \bt W^{(m+1)}] =  [\bf \Sigma^{(m)}, \bt W^{(m)}]+s_t^n ([\hat{\bf \Sigma}, \hat{\bt W}]-[\bf \Sigma^{(m)}, \bt W^{(m)}])$
		\Repeat\hspace{10mm}\% superscript $(n)$ indicates the iterations starting from here.
		\\
		\hspace{15mm}$s_t^{n+1} = s_t (s_t^n)$
		\\
		\hspace{15mm}Set $ [\bf \Sigma^{(m+1)}, \bt W^{(m+1)}] =  [\bf \Sigma^{(m)}, \bt W^{(m)}]+s_t^{n+1} ([\hat{\bf \Sigma}, \hat{\bt W}]-[\bf \Sigma^{(m)}, \bt W^{(m)}])$
		\Until{${L({\bf \Sigma}^{(m+1)}, \bt W^{(m+1)}, \text{\bf a}^{(m+1)}, \text{\bf p}, \bt S^{(i)})<}$$~~~$
		${L({\bf \Sigma}^{(m)}, \bt W^{(m)}, \text{\bf a}^{(m)}, \text{\bf p}, \bt S^{(i)})+ s_t^n d^T\{\frac{\partial }{\partial {\bf \Sigma}_q}L^{(m)}, \frac{\partial }{\partial \bt W_q}L^{(m)}\}_{q}}}$
		\Until{$\frac{\partial }{\partial {\bf \Sigma}_q}L = \frac{\partial }{\partial \bt W_q}L = 0,~\forall q$}
		\\
		\hspace{10mm}$\text{\bf a}_q = \max\{\text{\bf a}_q+p\text{\bf c}_q,0\}$
		\\
		\hspace{10mm}$\text{\bf p} = \text{\bf p}\times u$~~~~~~ \% $u \ge 1$ increase the penalty.
		\Until{$\max\{\text{\bf c}_1,\dots,\text{\bf c}_q\} \le 0$}
		\Until{Convergence of $L({\bf \Sigma}, \bt W, \text{\bf a}, \text{\bf p}, \bt S)$}
	\end{algorithmic}
\end{algorithm}
\section{Simulation Results and Discussion}
\label{sec7}
In this section, we simulate and compare all the algorithms presented so far. In these simulations, we set the noise power to $0$ dBm. $Q$ links as well as $K$ eavesdroppers are randomly placed in a circle, namely, the simulation region, with radius $r_{circ}$. The distance between the transmitter and the receiver of each link is set to be a constant $d_{link} = 10 \text{ m}$. The path-loss exponent is set to 2.5. For all simulated algorithms, $\beta = 5$ (cf. \eqref{logsumexp}) and the termination criterion is set to when the normalized relative difference in each link's secrecy rate for two consecutive iterations is less than $10^{-3}$. For the QNE selection algorithms, we set their parameters as follows: The step size matrix (i.e., $\gamma^\prime$) is set such that ${\gamma^\prime}_j^{(i)} = \gamma_0 i^{(-0.6)}$, $j = 1,\dots,m$, where $\gamma_0$ is a positive constant\footnote{We found out that setting the maximum value of $\gamma_0 = 20000$ brings the best performance for our algorithms.}, $c = 0.08I_{m\times m}$, and $\epsilon^{(j)} = \frac{1}{j}$.

\subsection{Signaling Overhead and Running Time}
\label{overhead}
\label{overhead}
While the distributed implementation of our proposed algorithms is now complete (cf. \eqref{PG22} and \eqref{PG33}), we still need to make sure that the amount of coordination that each link has to do (to make each QNE selection method possible) is reasonably low. That is, we need to check how much (if any) information a link needs to know about other links' corresponding channels and transmission attributes (i.e., covariance matrices of information signal and AN) in order to execute one iteration of each algorithm.

Algorithm 1 presented in the previous manuscript only requires each link to measure the interference at its receiver to perform the optimization in \eqref{reform2}. By examining the iteration in \eqref{PG22} for each link, where $\nabla_{{\bf \Sigma}_q}\bar{f}_q$ and $\nabla_{\bt W_q}\bar{f}_q$ are given in \eqref{every}, we can deduce that Algorithm 2 requires the same amount of coordination as Algorithm 1. The amount of coordination for Algorithm 3, however, depends on the choice of the function $\Phi(x)$. Here, we compare all of the flavors of Algorithm 3 in terms of how much signaling overhead they impose on the network. If maximizing sum-rate is the criterion, from \eqref{crisumrate} it can be seen that during the computation of $x(\epsilon^{(j)})$, at each iteration, the $q$th link, $q \in \mathbb Q$, needs the values of received signal, noise-plus-interference, and $\bt S_{r,0}$ ($r\in \mathbb Q, r \neq q$) of other links. Furthermore, the cross-channel gains of the $q$th link with other (unintended) legitimate receivers (i.e., $\bt H_{qr},\forall r\in \mathbb Q, r\neq q$) should also be available. Note that the cross-channel gains need not to be acquired multiple times at each iteration, as they are fixed throughout the coherence time of the channels\footnote{Note that all aforementioned algorithms must run during the coherence time of the channels.}. If the $r$th receiver sends training signals to its corresponding transmitter, for (implicit) channel estimation, $r\in \mathbb Q, r \neq q$, the channel gains $\bt H_{qr}$ can be estimated by the $q$th transmitter using channel reciprocity. Moreover, it should be noted that while the $q$th link, $q = 1,\dots,Q$, is using this criterion, it does not need to know any information about the channel gains between other links and eavesdroppers. This feature makes this design criterion more favorable than other criteria, which require obtaining the eavesdropping channel gains (i.e., $\bt G_{rk}$ and $\bt S_{r,k},\forall r\neq q,\forall k$) of all other links.

\begin{figure}
\begin{center}
\includegraphics[scale = 0.4, trim = -70mm 10mm 0mm 15mm]{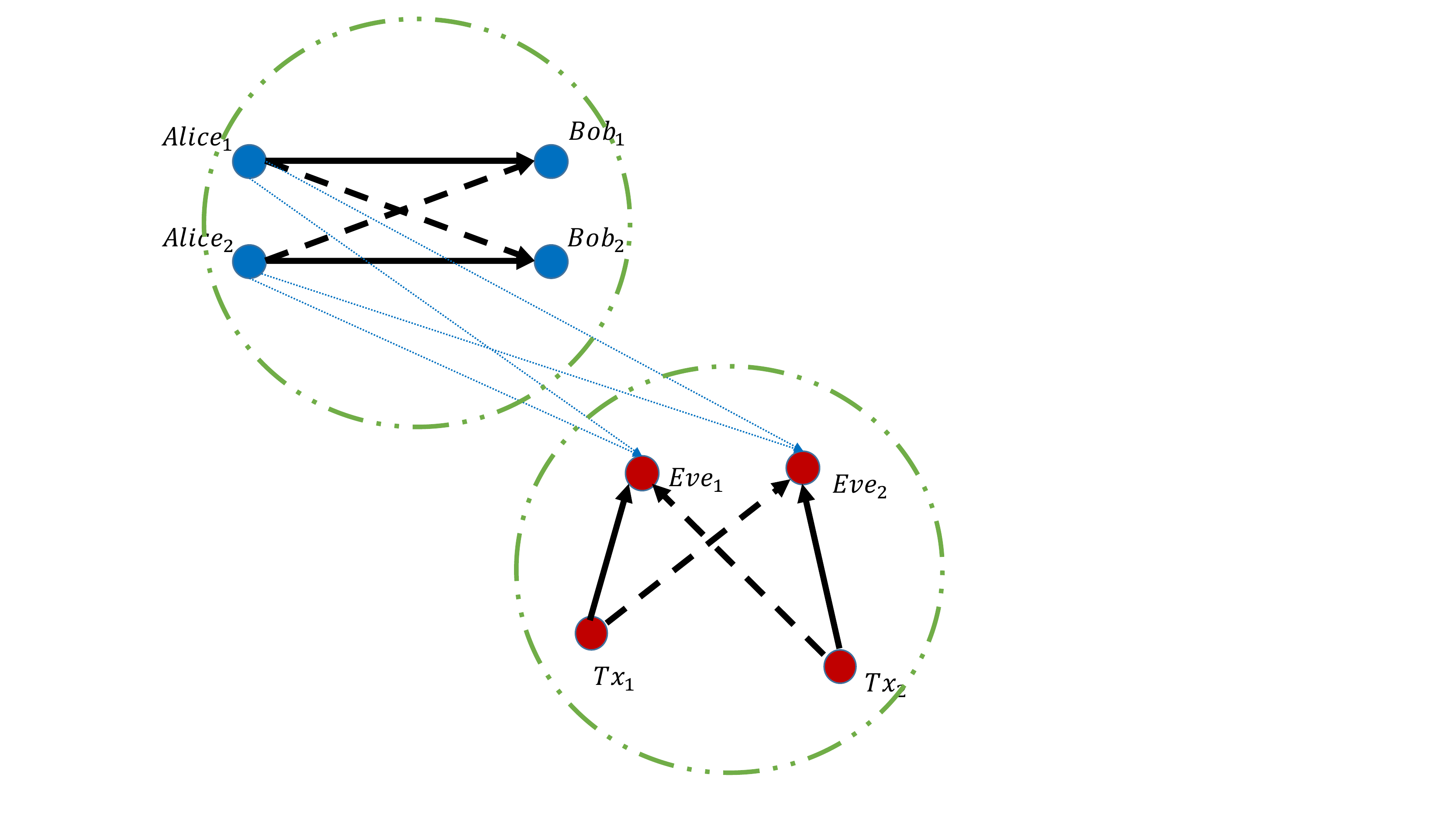}
\end{center}
\centering
\caption{A (clustered) MANET where two clusters (indiacted by green circle) of ad-hoc nodes are near each other. Hence, one cluster might be interested in the ongoing communications of the other cluster.}
\label{examplary}
\end{figure}

For the case of passive eavesdroppers, it does not seem difficult to derive the responses (or gradients) while assuming the knowledge of only statistics of the eavesdropping channels. This can be done if in \eqref{reform3} we replace the term $\varphi_{e,q,k}$ with $E\left[\varphi_{e,q,k}\right]$ where the expectation is w.r.t $G_{qk},\forall q,k \in \mathbb Q\times \mathbb K$. Note that including the expectation operator in the utilities, does not compromise the generality of any of the analyses done in previous sections.
Despite general difficulties in acquiring eavesdroppers's CSI (ECSI), some applications can be considered as practical examples where the knowledge of ECSI can be easily captured. One such example is mobile ad-hoc networks (MANETs) where the ad-hoc links of one cluster are interfering with one another, and can be considered as the legitimate links of our setup (See Fig. \ref{examplary}). On the other hand, the receivers of another cluster may try to overhear the communications of the legitimate links in the nearby clusters. These receivers can be considered as the external eavesdroppers of our setup. The clustering may have been done to ease the routing process in the network. It is possible that the clustering algorithm requires the links to exchange their location, power, and (possibly) channel state information (CSI). Hence, provided that the coherence time of the channels are long enough, each link can maintain the CSI between itself and the links from another cluster. Hence, the ECSI can be known to the links.

Another instance of our setup involves the downlink scenario of current cellular networks. Specifically, assume that the communication of the BS of a cell is interfering with other nearby cells.
Each BS-user pair can be assumed as a legitimate link in our scenario. We assume that no MU-MIMO technique is done in this scenario, so a BS is only communicating with one receiver (i.e., UE) at a given time. There might be other idle users in such network that are interested in overhearing the current communications. We can consider these idle users as the external eavesdroppers. It is possible that during the cell association phase, the idle users --which are now the external eavesdroppers-- exchange their location information (using known packets) with all the nearby BSs to eventually select a cell for their respective communications. Hence, the BSs can extract the CSI between themselves and the external eavesdroppers and maintain it (till the end of one coherence time) for use in PHY-layer security optimizations.

The issue of knowledge of ECSI has also been investigated in the recent literature. One example is when Eve is acting as a reactive jammer. That is, after some eavesdropping on the current transmissions, Eve injects her jamming signal to disrupt the ongoing communications. In such a case when jamming happens, assuming that the jamming signal of Eves are previously known, the ECSI can be extracted by the legitimate links using channel reciprocity. Moreover, in \cite{commmag}, it was shown that in a massive MIMO scenario, a passive Eve cannot be very dangerous and must therefore be active and attack the training phase. This active attack can make Eve exposed, and hence the legitimate links can acquire some knowledge about ECSI. Recently, the authors in \cite{swindle1} proposed a method with which the legitimate nodes can detect the passive eavesdropper from the local oscillator power leaked from its RF front end. Hence, an approximation on the location of Eve can be acquired. Lastly, in some scenarios where the legitimate nodes can detect the transmissions from Eves (e.g., active eavesdropping attacks), blind channel estimation techniques can be exploited to capture ECSI \cite{blind1, blind2}.

Lastly, regarding the computation of the proximal term $\tau_q$ as described by \eqref{gersh}, through numerous simulations we found that regardless of the topology of the network and the channel gains, the value found for $\tau_q$ is always a vary small value (i.e., $\tau_q < 10^{-4}$). This does not compromise the validity of inequality \eqref{gersh}. However, in practice it seems that the transmit optimization game is always a monotone VI problem. The derivation of inequality \eqref{gersh} was done because of the fact that it is not that obvious to see the monotonicity of $VI (F^\mathbb{C},\mathcal{K})$.

It is also interesting to understand how the choice of design criterion changes the running time of our proposed algorithm. To do this, we start from analyzing the computational complexity of Algorithm 1 and extend it to the analysis of our proposed algorithms.
\subsubsection{Algorithm 1}
In Line 2 of Algorithm 1, there is no need to compute every term of $\bt M_q$ and $\bt M_{e,q,k}$; that is, in measuring the interference, only the aggregate value is needed. Hence, the complexity of Line 2 is equivalent to the complexity of calculating the covariance matrix of the received interference. More specifically, at the receivers of legitimate links, the covariance matrix calculation of the $N_{R_q}\times 1$ received interference vector (i.e., $\bt M_q$) yields a complexity of $\mathcal O(N_{R_q}^2)$. Similar computation is needed to obtain $\bt M_{e,q,k}$, which has the complexity of $\mathcal O(\sum_{k=1}^K N_{ek}^2)$. Line 5 of Algorithm 1 involves a matrix inversion for $\bt S_{q,0}$ and a matrix multiplication together with a matrix inversion for $\{\bt S_{q,k}\}_{k=1}^K$. The total complexity of this line is $\mathcal O(\sum_{k = 1}^K(N_{T_q}N_{e,k}^2+N_{e,k}N_{T_q}^2+N_{e,k}^3)+N_{R_q}^3)$. Computation of the gradients in Line 8 requires the computation of $\varphi_{e,q,k}$ for all $k \in \mathbb K$ and $(\bt M_{q}^{n,l} +\bt H_{qq} {\bf \Sigma}_q\bt H_{qq}^H)^{-1}$. Computation of $\varphi_{e,q,k}$ for all $k \in \mathbb K$ has the complexity of $\mathcal O(\sum_{k=1}^K N_{T_q}N_{e,k}^2+N_{e,k}N_{T_q}^2+N_{e,k}^3)$ due to matrix multiplications and determinant calculations (cf. \eqref{cc}). The inverse of $(\bt M_{q}^{n,l}+ \bt H_{qq} {\bf \Sigma}_q\bt H_{qq}^H)$ yields an additional complexity of $\mathcal O(N_{R_q}^3+N_{R_q}^2N_{T_q}+N_{R_q}N_{T_q}^2)$. 
Notice that in calculating $\bt M_q^{n,l}$ and $ \bt M_{e,q,k}^{n,l}$ for all $k \in \mathbb K$, an additional computation for calculating $\bt H_{qq}\bt W^{n,l}_q \bt H^H_{qq}$ and $\bt G_{qk}\bt W^{n,l}_q \bt G^H_{qk}$ must be carried at each iteration of the PG method (i.e., Line 6 of Algorithm 1), which respectively have complexities of $\mathcal O(N_{R_q}^2N_{T_q}+N_{R_q}N_{T_q}^2)$ and $\mathcal O(\sum_{k=1}^K N_{T_q}N_{e,k}^2+N_{e,k}N_{T_q}^2))$. The other computations that were not mentioned in gradient derivation are redundant and do not affect the general complexity. Apart from the gradient derivations, the Euclidean projection also has its own complexity.
The projection in \eqref{projector} requires eigenvalue decomposition, and thus has $\mathcal O(N_{T_q}^3)$ complexity. Adding all of the aforementioned computations, the complexity of Algorithm 1 for each user $q$ is $\mathcal O\left(N_{R_q}^3+N_{T_q}^3+N_{R_q}^2N_{T_q}+N_{R_q}N_{T_q}^2+K (N_{T_q}N_{e,k}^2+N_{e,k}N_{T_q}^2+N_{e,k}^3)\right)$ or simply $\mathcal O\left(N_{R_q}^3+N_{T_q}^3+K N_{e,k}^3\right)$.
Note that one might also multiply this complexity by the amount of iterations in the PG method and the AO process. Let the constants $N_{PG}$ and $N_{AO}$ denote the iterations taken in the PG method and AO process, respectively. Hence, the total complexity for each player $q$ is\footnote{Notice that this result only makes sense when the QNE is unique. Otherwise if QNE is not unique, Algorithm 2 might not even converge, taking the running time to infinity.} $\mathcal O\left(N_{PG}N_{AO}\left(N_{R_q}^3+N_{T_q}^3+K N_{e,k}^3\right)\right)$.

\subsubsection*{Algorithm 2}
This algorithm can also be handled with the same complexity as Algorithm 1 with the difference that the number of iterations in Algorithm 2 (i.e., repeating the loop at Line 1 of Algorithm 2) was shown in Fig. \ref{fig123} (a) to be more than Algorithm 1, and hence a slower algorithm compared to Algorithm 1. Let the convergence time of the loop in Line 1 of Algorithm 2 be $N_{GR}$. Thus, the total complexity of Algorithm 2 for each player $q$ is $\mathcal O\left(N_{GR}\left(N_{R_q}^3+N_{T_q}^3+K N_{e,k}^3\right)\right)$.

\subsubsection*{Algorithm 3}
In this algorithm, some additional calculations are generally required. For the criterion of sum-rate maximization, the derivation of the gradients of $\Phi(x)$ are shown in \eqref{crisumrate}, which has the additional complexity of $\mathcal O(\sum_{\underset{r\neq q}{r=1}}^Q N_{R_r}^3+N_{R_r}^2N_{T_r}+N_{R_r}N_{T_r}^2)$. In the case of minimizing Eves' rates as the QNE selection method, according to \eqref{crieves}, computing $\Phi(x)$ would have the complexity of $\mathcal O(\sum_{\underset{r\neq q}{r=1}}^Q\sum_{k=1}^K N_{T_r}N_{e,k}^2+N_{e,k}N_{T_r}^2+N_{e,k}^3)$. The convergence time of Algorithm 3 is generally different from that of Algorithm 2 due to the presence of criterion function in Algorithm 3. Setting $N_{QNE}$~as the convergence time of the loop in Line 1 of Algorithm 3, the total complexity of Algorithm 3 is obtained as follows:\begin{itemize}
\item
Under sum-rate maximization as the QNE selection method, for every player $q$, the computational complexity is
$$
\mathcal O\left(N_{QNE}N_{GR}\left(N_{T_q}^3+Q(N_{R_q}^3+N_{R_q}^2N_{T_q}+N_{R_q}N_{T_q}^2)+K (N_{T_q}N_{e,k}^2+N_{e,k}N_{T_q}^2+N_{e,k}^3)\right)\right),$$
or simply
\begin{align}
\label{complsumrate}
\mathcal O\left(N_{QNE}N_{GR}\left(N_{T_q}^3+Q N_{R_q}^3+K N_{e,k}^3\right)\right).
\end{align}
\item
Under the minimization of Eves' rates as the QNE selection method, for every player $q$, the complexity is
$$
\mathcal O\left(N_{QNE}N_{GR}\left(N_{T_q}^3+N_{R_q}^3+N_{R_q}^2N_{T_q}+N_{R_q}N_{T_q}^2+QK (N_{T_q}N_{e,k}^2+N_{e,k}N_{T_q}^2+N_{e,k}^3)\right)\right),$$
or simply
\begin{align}
\label{complsumeve}
\mathcal O\left(N_{QNE}N_{GR}\left(N_{T_q}^3+N_{R_q}^3+Q K N_{e,k}^3\right)\right)
\end{align}
\item
Under the maximization of the secrecy sum-rate as the QNE selection method, for every player $q$, the complexity is
$$\mathcal O\left(N_{QNE}N_{GR}Q\left(N_{R_q}^3+N_{T_q}^3+N_{R_q}^2N_{T_q}+N_{R_q}N_{T_q}^2+K (N_{T_q}N_{e,k}^2+N_{e,k}N_{T_q}^2+N_{e,k}^3)\right)\right),$$
or simply
\begin{align}
\label{complsumsec}
\mathcal O\left(N_{QNE}N_{GR}Q\left(N_{R_q}^3+N_{T_q}^3+K N_{e,k}^3\right)\right)
\end{align}
\end{itemize}
\begin{figure*}
\begin{center}
\begin{tabular}{cc}
\hspace*{-11mm}\includegraphics[scale = 0.375]{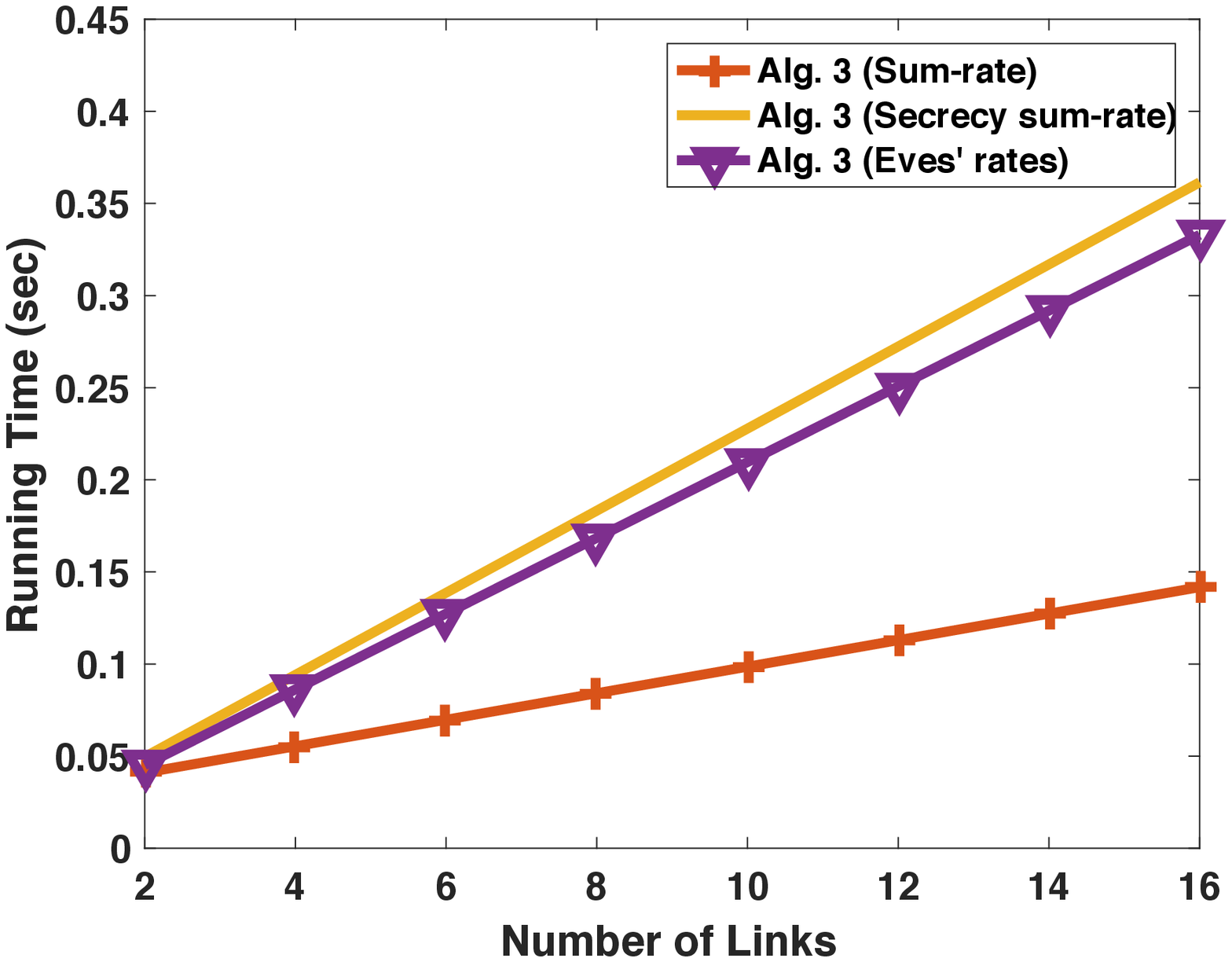} & \includegraphics[scale = 0.375]{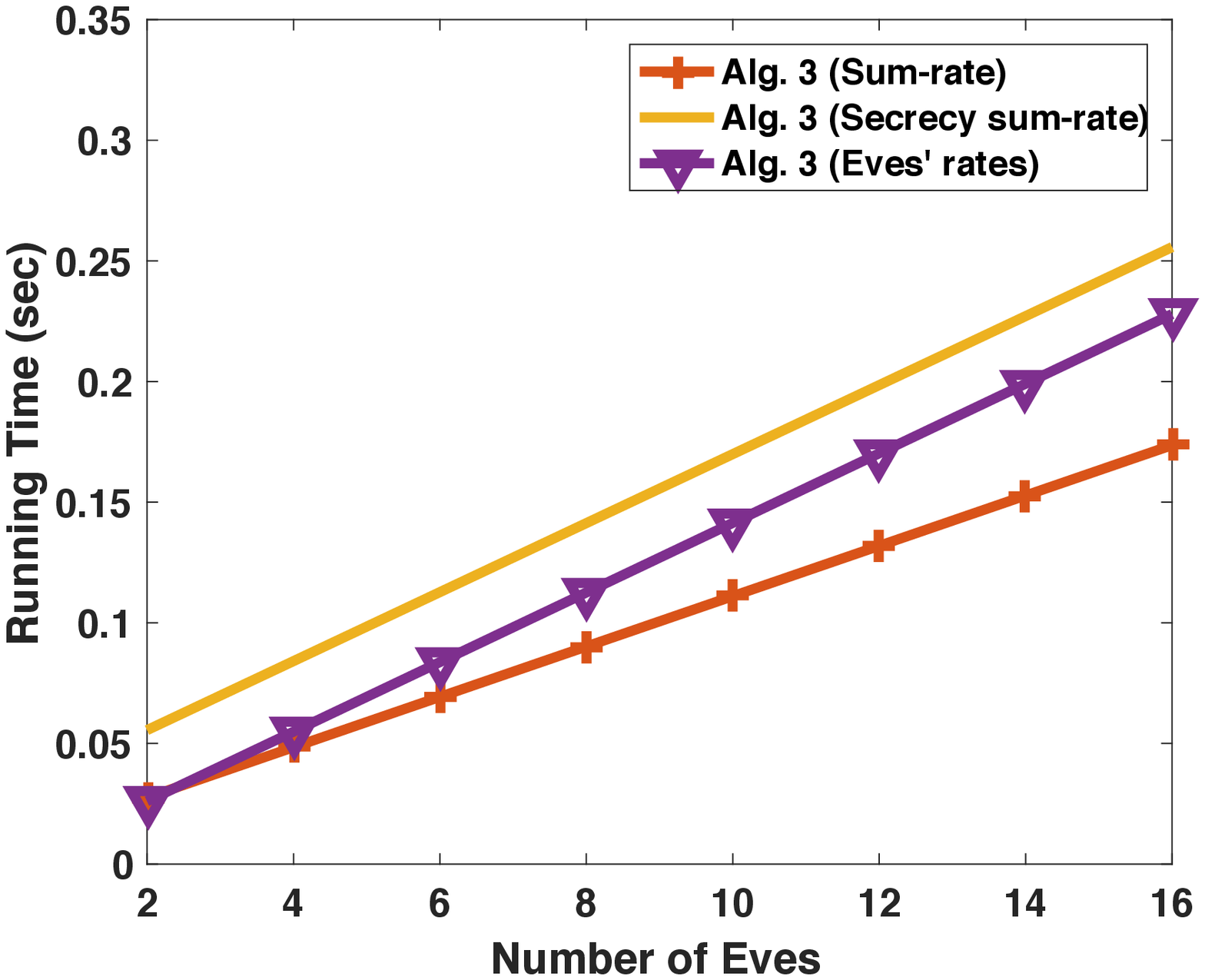}\\
\hspace*{-11mm}(a) & (b)
\end{tabular}
\end{center}
\caption{(a) Comparison of actual running time of proposed algorithms vs. (a) number of links, and (b) number of Eves: $r_{circ} = 30\text{ m}, K = 5, N_{T_q} = 5,$ $N_{r_q} = 5,N_{e,k} = 5, P_{q} = 40 \text{ dBm},~\forall q,k$.}
\label{fig00}
\end{figure*}
\begin{figure}
\begin{center}
\includegraphics[scale = 0.5, trim = 25mm 15mm 0mm 10mm]{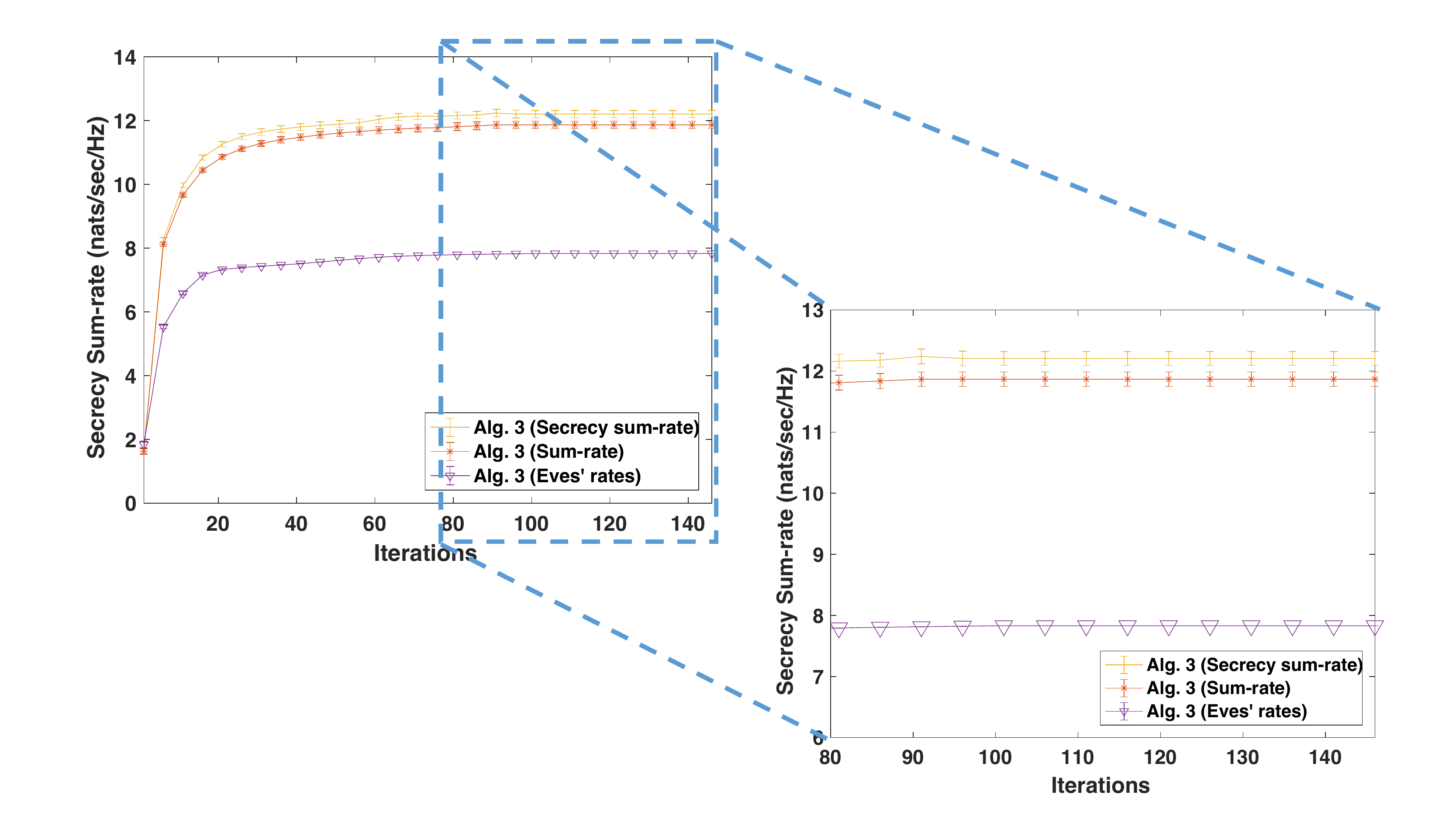}
\\
\hspace{-20mm}
\end{center}
\centering
\caption{Comparison of convergence trend of the proposed QNE selection methods: $\left(\text{8 links }(Q = 8)\right.$
\\
$\left.\text{ and 7 Eves }(K = 7), r_{circ} = 30\text{ m}, N_{T_q} = 5, N_{r_q} = 2~\forall q, N_{e,k} = 2~\forall k,d_{link} = 10\text{ m}, P_{q} = 40 \text{ dBm}\right)$.}
\label{initpoint}
\end{figure}
\begin{figure*}
\begin{center}
\begin{tabular}{ccc}
\hspace*{-17mm}\includegraphics[scale = 0.375]{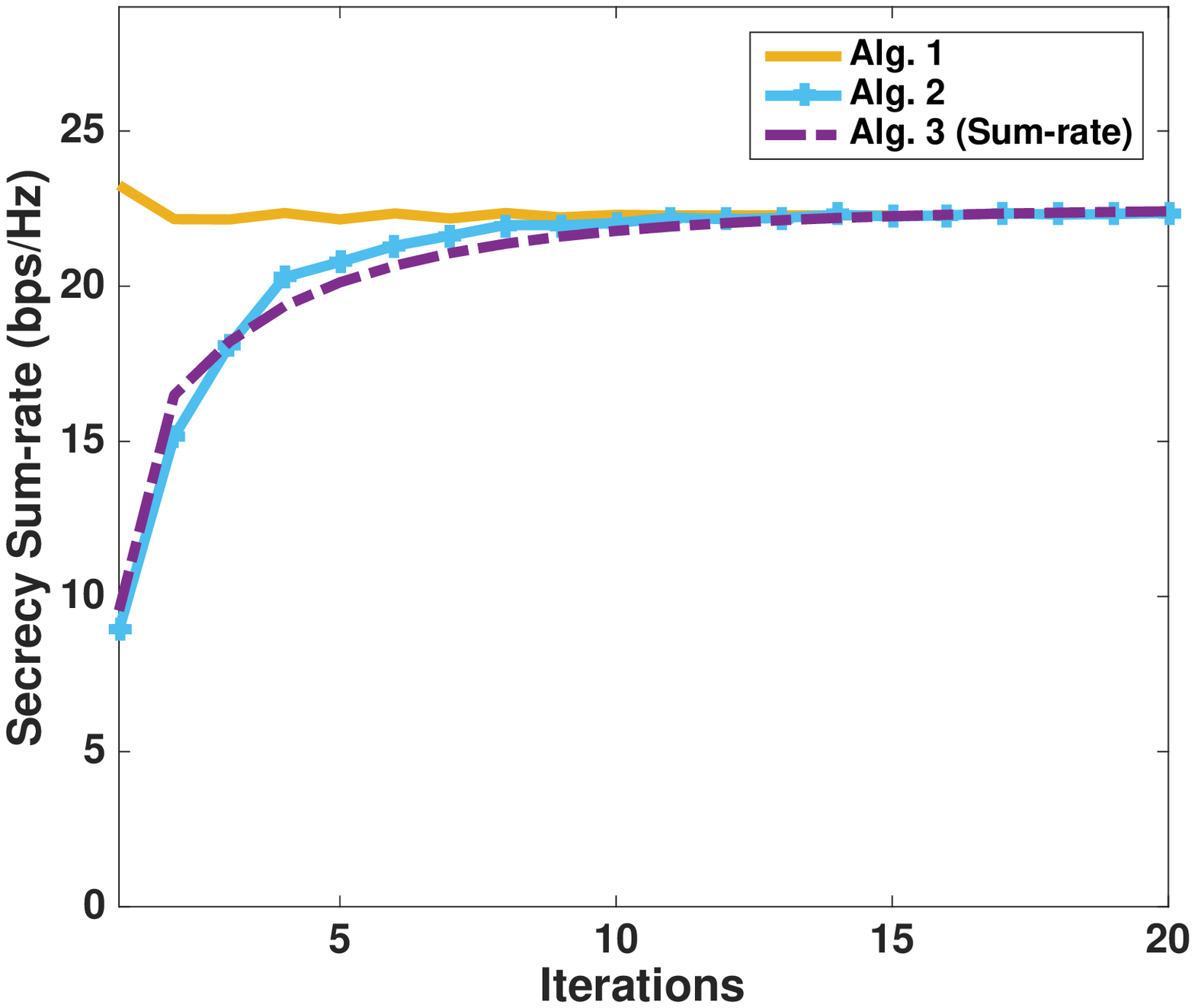}\hspace*{-8mm}  & \includegraphics[scale = 0.375,trim = 0mm 0mm 0mm -4mm]{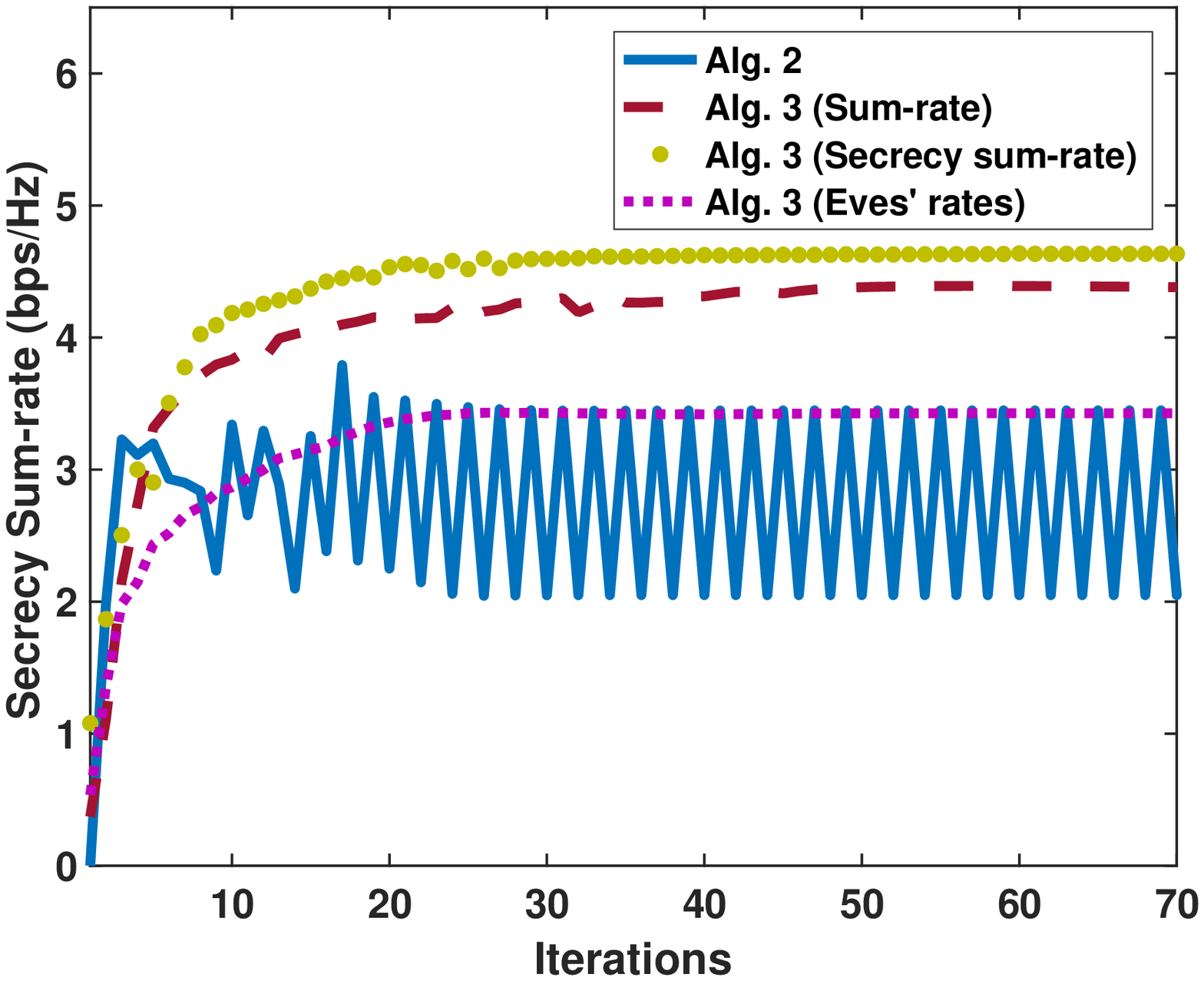}\hspace*{-5mm}  &  \includegraphics[scale = 0.375,trim = 0mm 0mm 0mm 6mm]{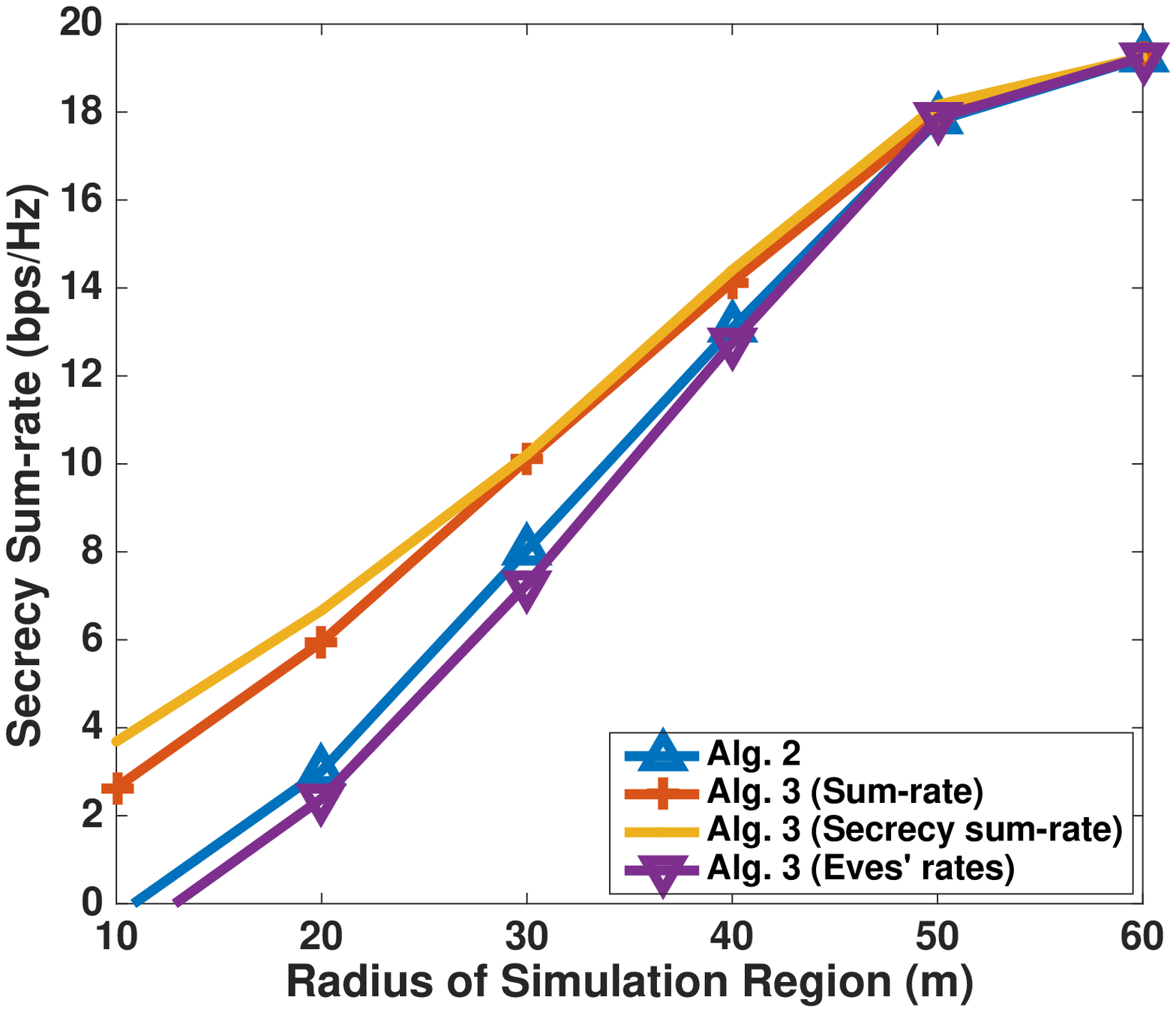}\\
\hspace{4mm}(a) & \hspace{10mm}(b) & \hspace*{2mm}(c)
\end{tabular}
\end{center}
\caption{(a) Convergence of secrecy sum-rate when QNE is unique; (b) convergence of secrecy sum-rate when multiple QNEs exist; (c) secrecy sum-rate vs. $r_{circ}:$ $Q = 8, K = 5, N_{T_q} = 5, N_{r_q} = 2~\forall q, N_{e,k} = 2~\forall k, r_{circ} = \text{(a) } 100\text{ m},~\text{(b) } 20\text{ m},$ $P_{q} = \text{(a) } 20 \text{ dBm},~\text{(b) }30 \text{ dBm},~ \text{(c) }40 \text{ dBm}$.}
\label{fig123}
\end{figure*}
We also computed the actual running time of our algorithm using MATLAB on a commercial PC with the following specifications: 1) 2.4 GHz Intel Core i5 CPU, 
2) 8 GB 1333 MHz DDR3 RAM,
3) Mac OS X El Capitan v. 10.11.6.
We show the results in Fig. \ref{fig00} for one iteration of Algorithm 3 while using different criteria. Hence, in comparing these results with the theoretical derivations, one should skip the term $N_{GR}$ and $N_{QNE}$. Each point in the presented curves is averaged over the number of iterations and also over 100 channel realizations of a given (random) network topology\footnote{While we tried to generate the results that are as close as possible to the theoretical derivations, we ended up with non-smooth curves at some points during this simulation. This is mainly due to the fact that in different channel realizations and different initial points, the convergence behavior, and thus the total number of iterations, is not consistent. In order to tackle these problems and generate clean figures, at some points we used linear regression of the actual complexity curve.}. The results in Fig. \ref{fig00} (a) and Fig. \ref{fig00} (b) show that the running time of the QNE selection when secrecy sum-rate is the criterion (i.e., Alg. 3 (Secrecy sum-rate)) is relatively higher than the other two QNE selection methods. It can be seen in Fig. \ref{fig00} (a) that the difference in the computational complexity of Alg. 3 (Eves' rates) (i.e., QNE selection when minimizing Eves' rates is the criterion) and Alg. 3 (Secrecy sum-rate) appears to be in the slope of the curves, which complies with theoretical derivations in \eqref{complsumeve} and \eqref{complsumsec}. However, this difference becomes clear when the number of links/antennas are high enough\footnote{Note that the theoretical derivations are derived for the worst case.}. It can be seen from Fig. \ref{fig00} (b) that both Alg. 3 (Secrecy sum-rate) and Alg. 3 (Eves' rates) have the same slope. This can be seen in the theoretical derivation for the complexity of both QNE selection methods in \eqref{complsumeve} and \eqref{complsumsec}, where for both criteria, the complexity w.r.t $K$ is a multiple $Q N_{e,k}^3$. For the case of Alg. 3 (Sum-rate) (i.e., QNE selection when maximizing sum-rate is the criterion) the complexity w.r.t $K$ is only a multiple of $N_{e,k}^3$. The gap between the Alg. 3 (Secrecy sum-rate) and Alg. 3 (Eves' rates) in Fig. \ref{fig00} (b) is because of the additional complexity of Alg. 3 (Secrecy sum-rate), which is independent of the number eavesdroppers (i.e., $K$).
\subsection{Effect of Initial Points}
In general, the initial values for the covariance matrices of information and AN signals can affect the results. Given the non-convexity of links' optimization problems, and the fact that at a QNE links operate at their stationary points, which are not necessarily unilaterally optimal, it is theoretically expected that different initial values can make the algorithm converge to different stationary points, thus affecting the final results. However, in our simulations, we did not see any significant variations in the secrecy sum-rate when the initial values of information and AN covariance matrices are changed. For example, by changing the initial values, for networks with 10 to 16 links, a maximum difference of $3~\text{nats/sec/Hz}$ and maximum of 150 iterations until convergence were observed. The results can be seen in Fig. \ref{initpoint}, where the simulated convergence behavior of all three QNE selection methods is depicted for one channel realization. A point at the $n$th iteration of a curve represents the resulting secrecy sum-rate of that particular QNE selection method at the $n$th iteration, averaged over 100 random initial points. The corresponding 95\% confidence intervals are also shown. The tightness of the confidence intervals indicate that while the performance varies when the initial points change, this variation is negligible. Note that in all of our simulations, we considered random initializations for each channel realization of a given (random) network topology.
\begin{figure*}
\begin{center}
\begin{tabular}{ccc}
\hspace*{-17mm}\includegraphics[scale = 0.36]{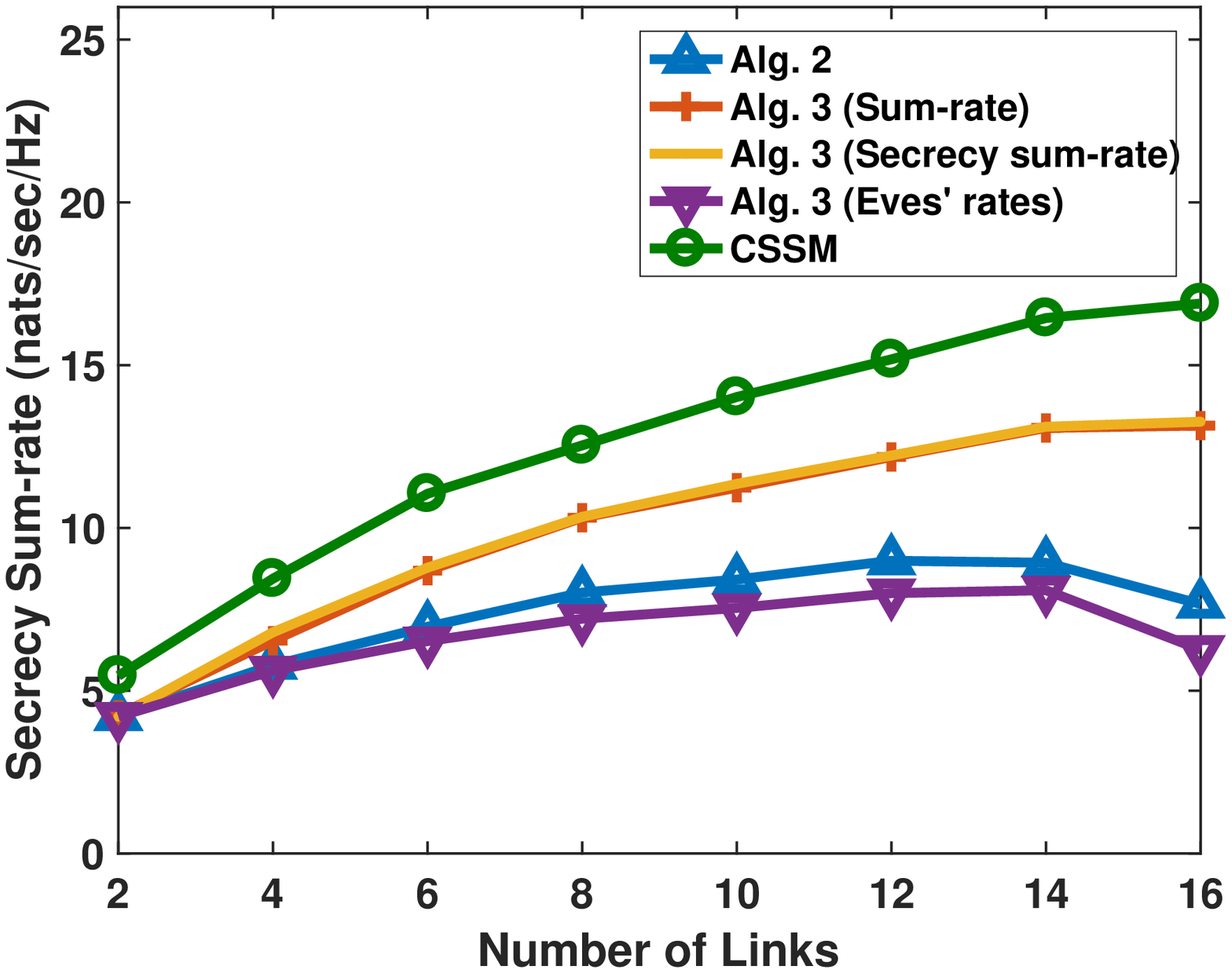}\hspace*{-8mm}  & \includegraphics[scale = 0.36,trim = 0mm 0mm 0mm -4mm]{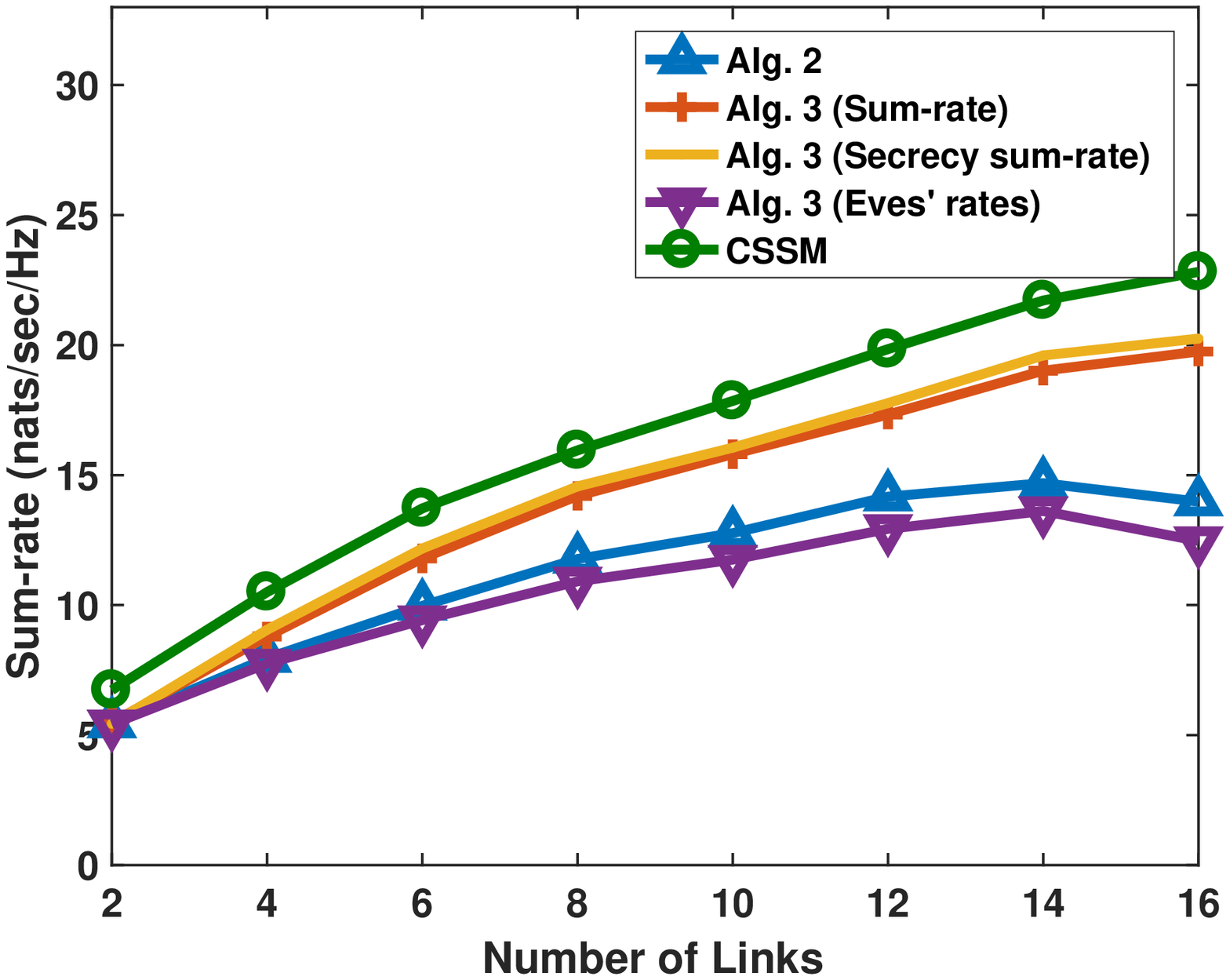}\hspace*{-5mm}  &  \includegraphics[scale = 0.36,trim = 0mm 0mm 0mm 6mm]{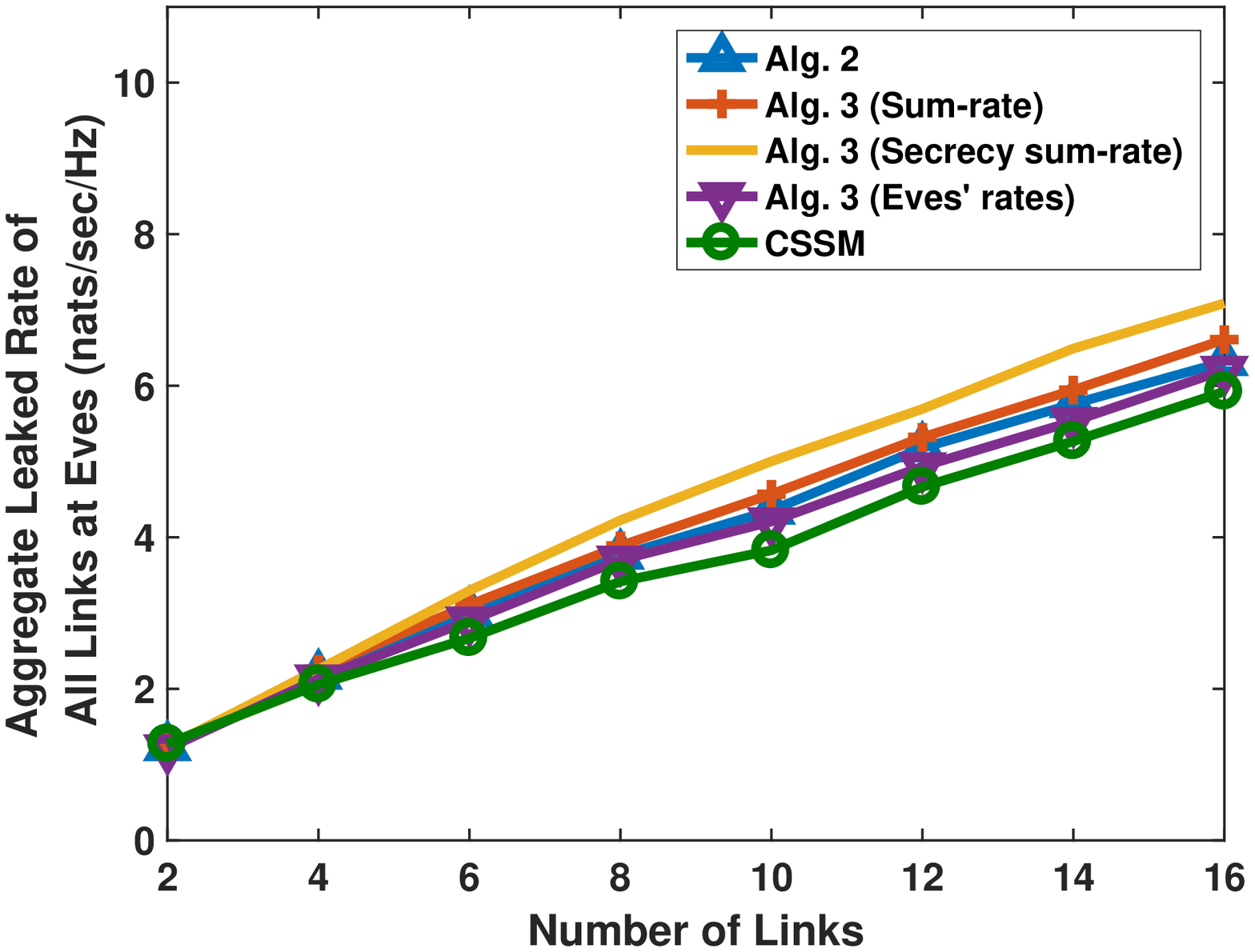}\\
\hspace{-5mm}(a) & \hspace{10mm}(b) & \hspace*{2mm}(c)
\end{tabular}
\end{center}
\caption{Comparison of (a) secrecy sum-rate, (b) sum-rate, (c) sum of eavesdroppers' received rates vs. number of links: 
$r_{circ} = 30\text{ m}, K = 5, N_{T_q} = 5, N_{r_q} = 2~\forall q, N_{e,k} = 2~\forall k,d_{link} = 10\text{ m}, P_{q} = 40 \text{ dBm}$.}
\label{fig456}
\end{figure*}
\subsection{Overall Performance and Energy Efficiency}
Fig. \ref{fig123} (a) compares the three proposed algorithms in a channel realization for the case when the QNE is unique. According to the uniqueness condition in Theorem \ref{tm3}, it is generally expected that if links are far enough from each other, then the resulting QNE is likely to be unique. We simulate this scenario by increasing $r_{circ}$ significantly. We consider the secrecy sum-rate as the measure of comparison between the algorithms. It can be seen that all of the algorithms converge to almost the same point. This result indicates the equivalence between the QNEs found by both Algorithms 1 and 2. Furthermore, it can be concluded that the QNE selection algorithm with sum-rate as its design criterion (indicated by Alg. 3 (Sum-rate)) does not outperform Algorithm 2 when the QNE is unique (i.e., the condition in Theorem \ref{tm3} is satisfied). That is, if the QNE is unique the QNE selection algorithms only have one QNE to choose from. It should be noted that Algorithm 1 converges faster than other algorithms. This might be because Algorithms 2 and 3 use smaller steps towards the QNE at each iteration.

Fig. \ref{fig123} (b) compares the achieved secrecy sum-rate in a channel realization between Algorithm 2 and different versions of Algorithm 3, indicated by \lq\lq Alg. 3 (Secrecy sum-rate)\rq\rq~when secrecy sum-rate is the design criterion, \lq\lq Alg. 3 (Eves' rates)\rq\rq~when reducing Eves' rates is the design criterion, and \lq\lq Alg. 3 (Sum-rate)\rq\rq~when sum-rate is the design criterion. Furthermore, due to the existence of multiple QNEs, Algorithm 2 is oscillating between QNEs and never converges even after 70 iterations\footnote{Recall that convergence of Algorithm 2 is tied to the uniqueness of the QNE. Furthermore, due to the similarity in the behavior of Algorithms 1 and 2, we only showed Algorithm 2 in subsequent simulations.}. We increased the number of iterations to 1000, but did not see the convergence of Algorithm 2. However, all of the versions of Algorithm 3 converge to a QNE\footnote{The result in Fig. \ref{fig123} (b) should not be confused with the previous simulation in Fig. \ref{fig123} (a). In fact, equal secrecy sum-rate for all of the algorithms happen only when QNE is unique (i.e., the condition in Theorem \ref{tm3} is satisfied). However, Fig. \ref{fig123} (b) is showing results when the condition in Theorem \ref{tm3} is not likely to be satisfied.}.

\begin{figure*}
\begin{center}
\begin{tabular}{ccc}
\hspace*{-17mm}\includegraphics[scale = 0.365]{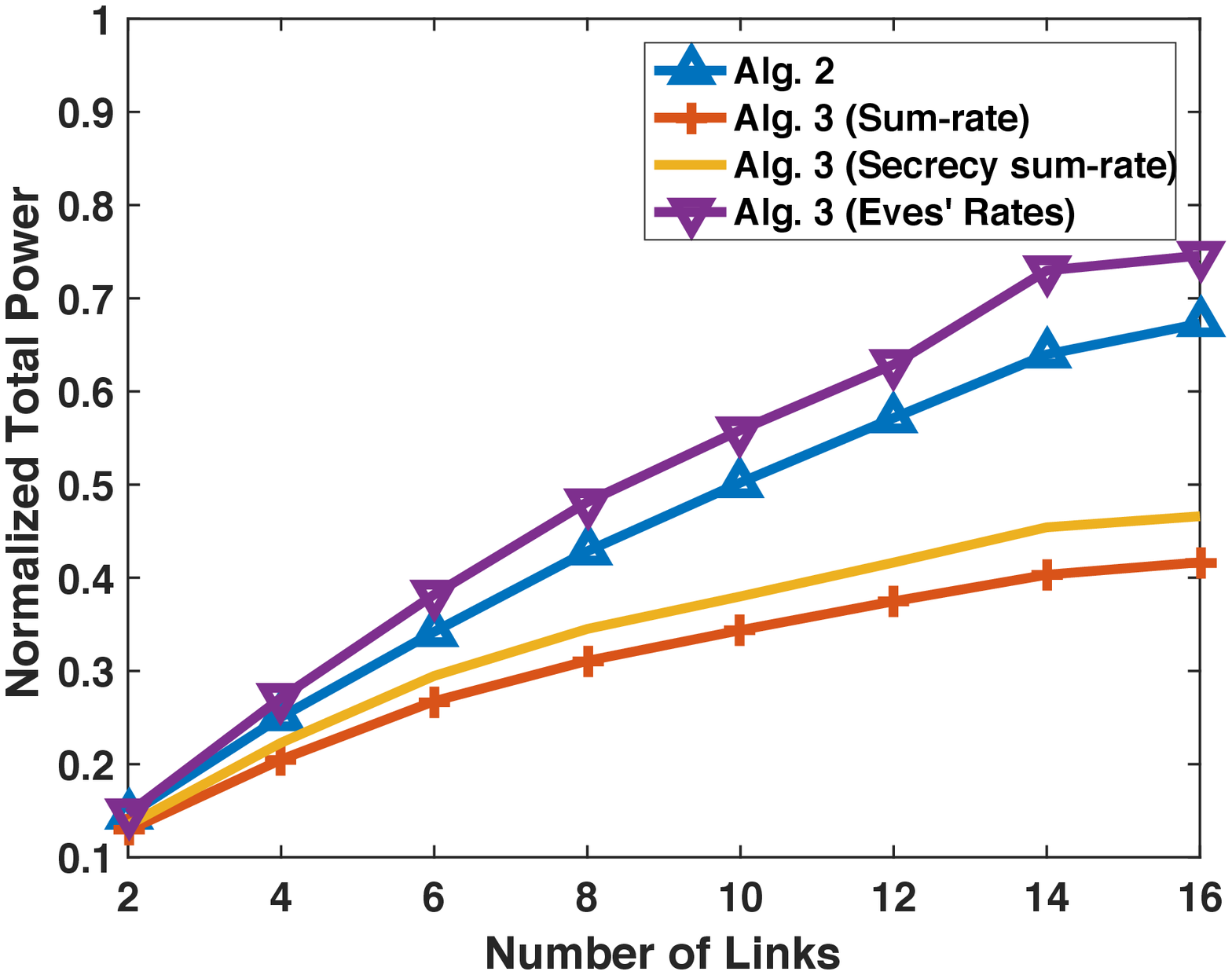}\hspace*{-8mm}  & \includegraphics[scale = 0.365,trim = 0mm 1mmmm 0mm -15mm]{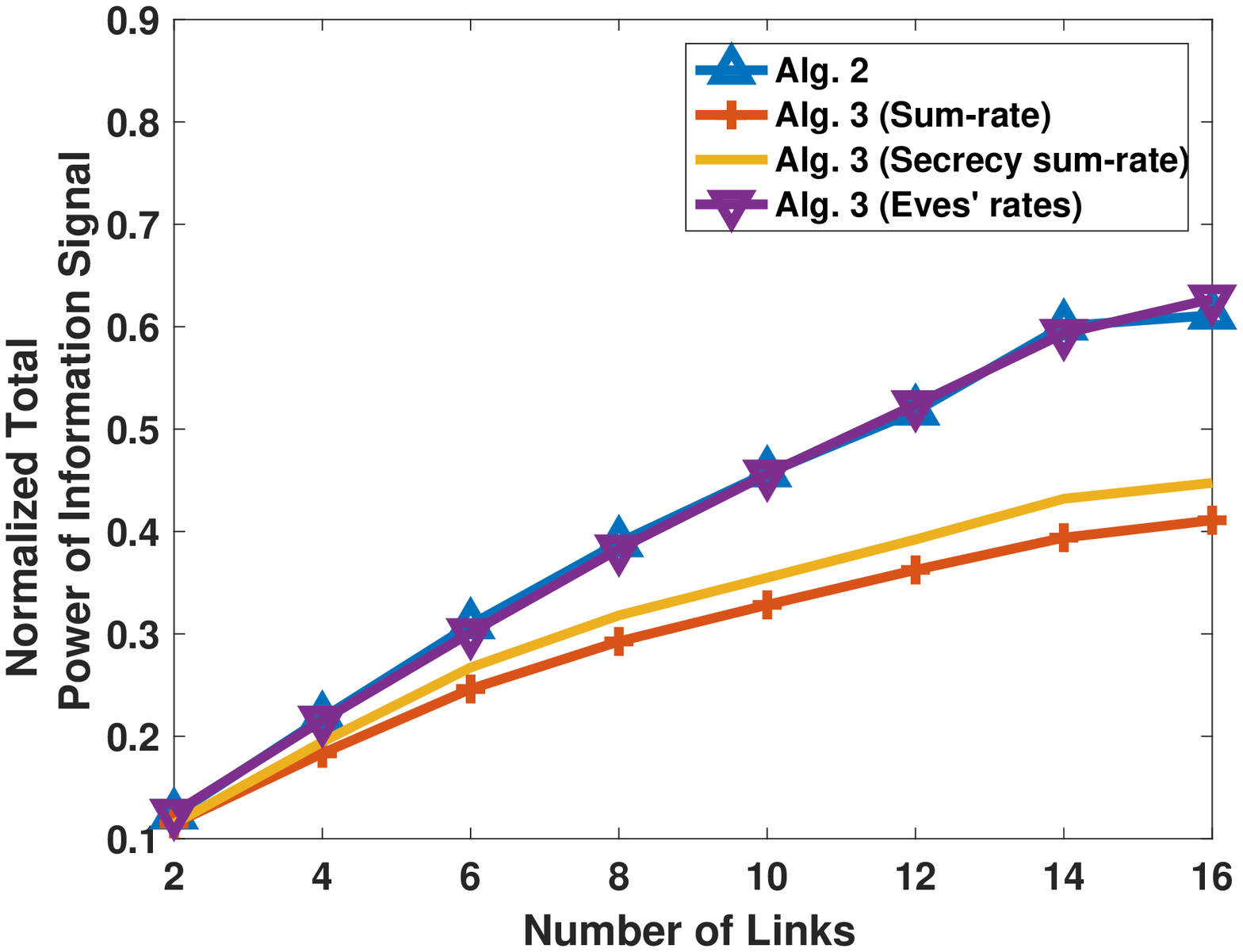}\hspace*{-9mm}  &  \includegraphics[scale = 0.36,trim = 0mm 0mm 0mm 6mm]{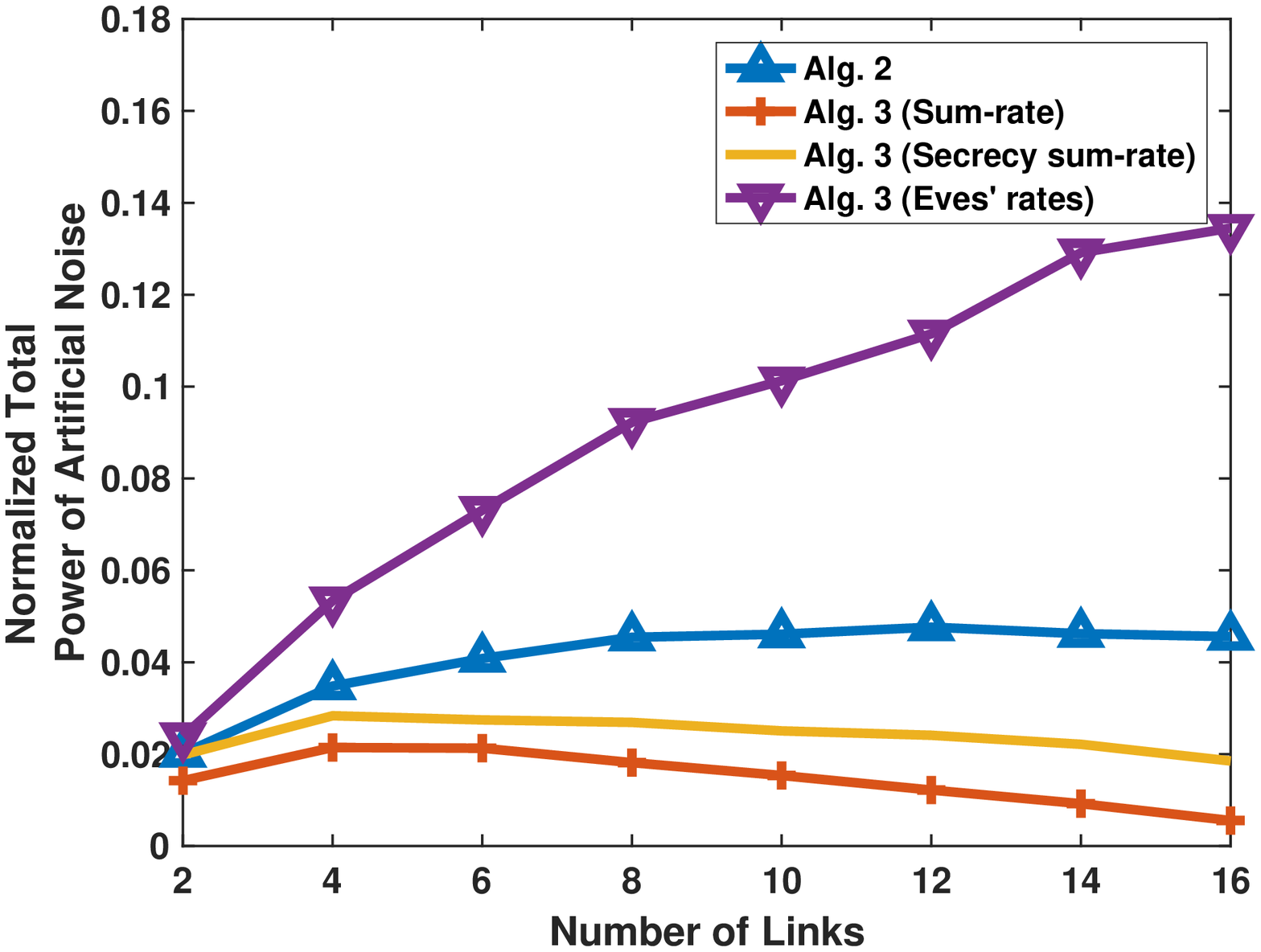}\\
\hspace{-5mm}(a) & \hspace{10mm}(b) & \hspace*{2mm}(c)
\end{tabular}
\end{center}
\caption{Comparison of (a) total power (b) total power of information signal (c) total power of AN vs. number of links: $r_{circ} = 30\text{ m}, K = 5, N_{T_q} = 5, N_{r_q} = 2~\forall q, N_{e,k} = 2~\forall k,d_{link} = 10\text{ m}, P_{q} = 40 \text{ dBm}$.}
\label{fig789}
\end{figure*}
Fig. \ref{fig123} (c) shows the secrecy sum-rate resulting from different algorithms vs. $r_{circ}$. For Algorithm 2, we limit the iterations to 100. For Algorithm 3, we limit the iterations of the inner loop (i.e., line 3 in Algorithm 3) and the outer loop (i.e., line 1 in Algorithm 3) to 50  and 3, respectively. Each point in the figure is the result of averaging over 50 random network topologies, where in each topology, 200 channel realizations are simulated and averaged. It can be seen that when $r_{circ}$ is small (i.e., high interference), Alg. 3 (Sum-rate) and Alg. 3 (Secrecy sum-rate) have higher secrecy sum-rate than Algorithm 2. This is due to the fact that the myopic maximization of secrecy rates in Algorithm 2 is not guaranteed to converge to a QNE. Moreover, it can be seen that in Alg. 3 (Eves' rates), we cannot increase the secrecy rate as much as other versions of Algorithm 3. This is due to the fact that in minimizing the received rate at eavesdroppers, too much AN power creates unwanted interference on legitimate receivers, preventing any improvement on the secrecy sum-rate. 

Fig. \ref{fig456} (a) compares the secrecy sum-rate of Algorithms 2 and 3 for different number of links. Alg. 3 (Secrecy sum-rate) and Alg. 3 (Sum-rate) consistently outperform Algorithm 2 in terms of secrecy sum-rate (Fig. \ref{fig456} (a)) and sum-rate (Fig. \ref{fig456} (b)), and Alg. 3 (Eves' rates) does not result in a secrecy sum-rate as high as the other two flavors of Algorithm 3. As shown in Fig. \ref{fig456} (c), using Alg. 3 (Eves' rates) slightly reduces sum of Eves' received rates by increasing interference at Eves, but this directly affects legitimate transmissions as well. Furthermore, Alg. 3 (Secrecy sum-rate) does not have a significant advantage over Alg. 3 (Sum-rate). Another interesting point is that Alg. 3 (Secrecy sum-rate) has slightly higher sum-rate and higher leaked rate compared to Alg. 3 (Sum-rate). Hence, the performance of Alg. 3 (Secrecy sum-rate) is not necessarily a combination of Alg. 3 (Sum-rate) and Alg. 3 (Eves' rates), but rather a good tradeoff point. Lastly, it can be seen that the proposed algorithms have lower secrecy sum-rates compared to CSSM. We conjecture that this might be due to the fact that CSSM has a larger solution space compared to our methods. Note that the solution space of CSSM may contain some points that are not necessarily the QNEs of the game, whereas both Algorithms 2 and 3 can only converge to QNEs of the game. The difference between Algorithms 2 and 3 is that Algorithm 3 selects the best QNE (according to a criterion), but Algorithm 2 does not. As can be seen in Fig. \ref{fig456} (a), for the case of 16 links, the loss of Algorithm 3 compared to CSSM is less than 25\% when either secrecy sum-rate or sum-rate is the criterion for the QNE selection phase of Algorithm 3. Despite this loss, using Algorithm 3 facilitates not only a distributed implementation, but also the flexibility in the amount of coordination. The latter gives us freedom to keep the coordination as low as possible. Neither of these features are available in CSSM.

In Fig. \ref{fig789} (a)--(c) the power consumption of different algorithms are compared. The total power in Fig. \ref{fig789} (a)--(c) is normalized w.r.t the total power budget $\sum_qP_q$. Generally, Alg. 3 (Sum-rate) is the most energy efficient algorithm. Both Alg. 2  and Alg. 3 (Eves' rates) perform poorly in energy efficiency as the increase in the power of AN creates interference at other legitimate receivers. This makes the links to spend even more power on the information signal which eventually leads to neither a high sum-rate nor a high secrecy sum-rate. Moreover, the increase in the power of AN seems to be more significant in Alg. 3 (Eves' rates), as the design criterion forces the users to carelessly increase the interference at Eves. Lastly, Alg. 3 (Secrecy sum-rate) and Alg. 3 (Sum-rate) decrease the power of AN as the number of links increases because as the links abound, they automatically create additional interference at Eves. Hence, the links do not spend more power on AN.

\begin{figure}
\begin{center}
\hspace{-3mm}\includegraphics[scale = 0.33]{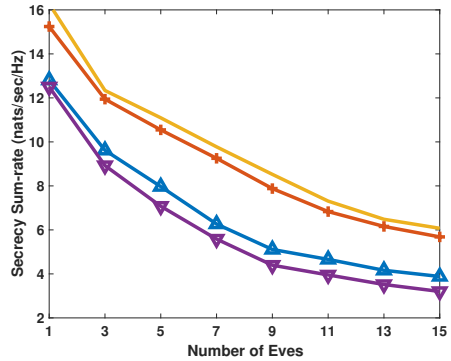}
\end{center}
\centering
\caption{Comparison of secrecy sum-rate vs. number of Eves:
\text{$r_{circ} = 30 \text{ m}, Q = 8, N_{T_q} = 5, N_{r_q} = N_{e,k} = 2, P_{q} = 40 \text{ dBm}$}}
\label{everate}
\end{figure}

Fig. \ref{everate} shows that as the number of eavesdroppers in the network increases, Alg. 3 (Sum-rate) and Alg. 3 (Secrecy sum-rate) outperform Algorithm 2 in terms of secrecy sum-rate, and Alg. 3 (Eves' rates) still achieves a low secrecy sum-rate. Overall, in these simulations, maximizing sum-rate as a design criterion seems to be the best to increase the secrecy sum-rate because other proposed criteria cannot add significant improvements despite requiring more extensive signaling between the links (e.g., the knowledge of all eavesdropping channel gains). Lastly, minimizing Eves' rates as the design criterion although brings poor performance to the QNE selection, it gives us valuable insights on the importance of interference management such that if it is overlooked, the secrecy sum-rate in the network can be severely decreased.
\section{Conclusions}
\label{sec8}
We designed a game theoretic secure transmit optimization for a MIMO interference network with several MIMO-enabled eavesdroppers. We proposed three algorithms to increase secrecy sum-rate. In the first algorithm, the links myopically optimize their transmission until a quasi-Nash equilibrium (QNE) is reached. Because of the inferior performance of first algorithm in case of multiple QNEs, we designed the second algorithm based on the concept of variational inequality. The second algorithm enables us to analytically derive convergence conditions, but achieves the same secrecy sum-rate as the first algorithm. To increase the secrecy sum-rate, we proposed the third algorithm in which the links can select the best QNE according to a certain design criterion. Simulations showed that not every criterion is good for the performance improvement. Specifically, reducing co-channel interference is a better criterion compared to increasing interference at the eavesdroppers to improve secrecy sum-rate.
\section*{Acknowledgment}
This research was supported in part by NSF (grants 1409172 and CNS-1513649), the Army research Office (grant W911NF-13-1-0302), the Qatar National Research Fund (grant NPRP 8-052-2-029), and the Australian Research Council (Discovery Early Career Researcher Award DE150101092). Any opinions, findings, conclusions, or recommendations expressed in this paper are those of the author(s) and do not necessarily reflect the views of NSF, ARO, QNRF, or ARC.
\appendices
\section{Proof of Proposition \ref{kktprop}}
\label{appkkt}
Let $({\bf \Sigma}_q^*,\bt W_q^*,\{\bt S_{q,k}^*\}_{k=0}^K)$ denote the limit point  of AO iterations found in Line 10 of Algorithm 1 for the $q$th link, $q \in \bt Q$. As mentioned earlier, problem \eqref{reform2} is convex w.r.t either $({\bf \Sigma}_q,\bt W_q)$ or $\{\bt S_q\}_{k=0}^K$. Then, recalling the minimum principle in \eqref{minprinc}, we have the following\footnote{AO iterations converge to a stationary point of \eqref{reform2} \cite[Section IV-B]{winkin}, \cite[Corollary 2]{park}.}:
{\begin{subequations}
\label{similar}
\begin{align}
&X_q = [{{\bf \Sigma}^*_q}^T,{\bt W_q^*}^T]^T, Z_q = [{{\bf \Sigma}_q}^T,{\bt W_q}^T]^T, \nabla_{Z_q}\bar{f}_q({\bf \Sigma}_q^*,\bt W_q^*,\{\bt S_{q,k}^*\}_{k=0}^K) = \left[-(\nabla_{{\bf \Sigma}_q}\bar{f}_q)^T,-(\nabla_{\bt W_q}\bar{f}_q)^T\right]^T,
\\
&\label{minprinc2} \left<Z_q-X_q,\nabla_{Z_q}\bar{f}_q({\bf \Sigma}_q^*,\bt W_q^*,\{\bt S_{q,k}^*\}_{k=0}^K)\right> \ge 0,~ \forall ({\bf \Sigma}_q,\bt W_q) \in \mathcal{F}_q,
\\
\label{minprincs}&\left<\bt S_{q,k}-\bt S_{q,k}^*,\nabla_{\bt S_{q,k}}\bar{f}_q({\bf \Sigma}_q^*,\bt W_q^*,\{\bt S_{q,k}^*\}_{k=0}^K)\right> \ge 0,~\forall \bt S_{q,k} \succeq 0,~\forall k\in \bt K.
\end{align}
\end{subequations}}
It should be noted that for a given $({\bf \Sigma}^*_q,\bt W^*_q)$, the value of $\{\bt S^*_{q,k}\}_{k=0}^K$ are uniquely determined (cf. \eqref{b} and \eqref{c}). Hence, using Danskin's theorem \cite{bert1}, the function $\bar{f}_q({\bf \Sigma}_q,{\bt W}_q,\left\{{\bt S}^*_{q,k}\right\}_{k=0}^K)$ is differentiable w.r.t $({\bf \Sigma}_q,\bt W_q)$, and inequality \eqref{minprinc2} holds\footnote{Similar reasoning for $\bar{f}_q({\bf \Sigma}^*_q,\bt W^*_q, \bt S_{q,k})$ can be used to justify the inequality in \eqref{minprincs}.}. Moreover, it can be verified that
{\begin{align}
\label{equivalent}
&\nabla_{{\bf \Sigma}_q}\bar{f}_q({\bf \Sigma}_q^*,{\bt W}_q^*,\left\{{\bt S}^*_{q,k}\right\}_{k=0}^K) = \nabla_{{\bf \Sigma}_q}\bar{R}_{s,q}({\bf\Sigma}_q^*, {\bt W_q}^*),
\\
&\nabla_{\bt W_q}\bar{f}_q({\bf \Sigma}_q^*,{\bt W}_q^*,\left\{{\bt S}^*_{q,k}\right\}_{k=0}^K) = \nabla_{\bt W_q}\bar{R}_{s,q}({\bf\Sigma}_q^*, {\bt W_q}^*)
\end{align}}
where $\bar{R}_{s,q}$ is the smooth approximation of secrecy rate mentioned in \eqref{secrecy2}. Then, according to \eqref{equivalent},
{\begin{equation}
\left<Z_q-X_q,\nabla_{Z}\bar{R}_{s,q}({\bf \Sigma}_q^*,\bt W_q^*)\right> \le 0,~ \forall ({\bf \Sigma}_q,\bt W_q) \in \mathcal{F}_q
\end{equation}}
where $\nabla_{Z}\bar{R}_{s,q}({\bf \Sigma}_q^*,\bt W_q^*) = \left[(\nabla_{{\bf \Sigma}_q}\bar{R}_{s,q})^T,(\nabla_{\bt W_q}\bar{R}_{s,q})^T\right]^T$. Hence, $({\bf \Sigma}_q^*,\bt W_q^*)$ is the optimal solution to 
{\begin{align}
\underset{Z_q}{\text{maximize}}~~~&\left<Z_q-X_q,\nabla_{Z}\bar{R}_{s,q}({\bf \Sigma}_q^*,\bt W_q^*)\right>\nonumber
\\
\label{optkkt}\text{s.t.~~~}&Z_q \in \mathcal{F}_q.
\end{align}}
Hence, $({\bf \Sigma}_q^*,\bt W_q^*)$ must satisfy the K.K.T conditions of \eqref{optkkt}, which can be written as
{\begin{subequations}
\begin{align}
&\nabla_{{\bf \Sigma}_q}\bar{R}_{s,q}({\bf \Sigma}_q^*,\bt W_q^*)-\zeta_q I+\Xi_{q,1} = 0
\\
&\nabla_{\bt W_q}\bar{R}_{s,q}({\bf \Sigma}_q^*,\bt W_q^*)-\zeta_q I+\Xi_{q,2} = 0
\\
&\zeta_q(\tr({\bf \Sigma}_q^*+\bt W_q^*)-P_q) = 0, {\bf \Sigma}_q^*\Xi_{q,1} = 0, \bt W_q^* \Xi_{q,2} = 0
\\
& \zeta_q \ge 0, \Xi_{q,1} \succeq 0, \Xi_{q,2} \succeq 0.
\end{align}
\end{subequations}}
where $\zeta_q$, $\Xi_{q,1}$, and $\Xi_{q,2}$ are Lagrange multipliers. Therefore, the stationary point of AO iterations satisfies the K.K.T conditions of \eqref{secrecy2}.
\section{Proof of Theorem \ref{existancthm}}
\label{existanceproof}
To prove the existence of the QNE, we use the following theorem:
\begin{theorem}
\label{thmvi}
\cite[Corollary 2.2.5]{pang}
For a mapping $F: \mathcal{Q} \rightarrow \mathcal{R}^N$ that is continuous on the compact and convex set $\mathcal{Q} \subseteq \mathcal{R}^N$, 
the solution set for $VI(F,\mathcal{Q})$ is nonempty and compact.
\qed
\end{theorem}
The objective in \eqref{reform2} is continuously differentiable on its domain, making $F^\mathbb{R}$ continuous. Furthermore, the set $\mathcal{K}$ is a compact convex set because it is the Cartesian product of compact convex sets (i.e., players' strategy sets). Hence, $\mathcal{K}^\mathbb{R}$, the real-vector version of $\mathcal{K}$, is a convex set. Due to the presence of power constraints, the strategy set of each player is compact,
then the set $\mathcal{K}^\mathbb{R}$ is also compact. Thus, according to Theorem \ref{thmvi}, the solution set to the VI in \eqref{final} is nonempty, meaning that the QNE in the proposed smooth game exists.
\section{Proof of Theorem \ref{tm3}}
\label{sec:apla}

We first introduce following definition:
\begin{definition} \cite[Definition 26]{monotone}
Considering the complex VI in \eqref{compvi}, with $F^\mathbb{C}(Z): \mathcal K \rightarrow \mathbb C^{N^\prime \times N},~\mathcal K \subseteq \mathbb C^{N^\prime \times N}$ being a continuously $\mathbb{R}-$differentiable function and $\mathcal{K}$ being a convex set that has a non-empty interior. The augmented Jacobian matrix for $F^\mathbb{C}(Z)$, namely, $JF^{\mathbb{C}}(Z)$, is defined as follows\footnote{For the case of $\mathcal{K}$ having a possibly empty interior, the equivalent conidtion in \cite[Proposition 28]{monotone} can be used.}:
{\begin{align}
\label{augm}
&JF^{\mathbb{C}}(Z) \triangleq \frac{1}{2}\left[\begin{aligned}&D_ZF^\mathbb{C}(Z)&~~~&D_{Z^*}F^\mathbb{C}(Z)
\\
&D_Z(F^\mathbb{C}(Z)^*)&~~~&D_{Z^*}(F^\mathbb{C}(Z)^*)
\end{aligned}\right]
\end{align}}
where $D_Z(F^\mathbb{C}(Z)) \triangleq \frac{\partial~\ve\left(F^\mathbb{C}(Z)\right)}{\partial~\ve(Z)^T}$ is a $N^\prime N\times N^\prime N$ derivative matrix, $D_{Z^*}F^\mathbb{C}(Z)^* = D_Z(F^\mathbb{C}(Z)^*$, and $D_{Z}\left(F^\mathbb{C}(Z)^*\right) = D_{Z^*}F^\mathbb{C}(Z)$.
\end{definition}
Using this definition, the following proposition holds for $VI(F^\mathbb{C},\mathcal{K})$.
\begin{proposition} \cite[Proposition 27]{monotone} 
\label{prop2}
For the $VI(F^\mathbb{C},\mathcal{K})$ defined in Definition 1, it holds that:
\begin{itemize}
\item{}
$F^\mathbb{C}$ is monotone on $\mathcal{K}$ if and only if $JF^\mathbb{C}(Z)$ is Augmented Positive Semidefinite (APSD) on $\mathcal{K}$. That is, for all $Y \in \mathbb{C}^{N^\prime \times N}$ and $Z \in \mathcal{K}$,
{\begin{equation}
\label{aug}
[\ve(Y^*)^T ,\ve(Y)^T]JF^\mathbb{C}(Z)[\ve(Y)^T ,\ve(Y^*)^T]^T \ge 0
\end{equation}}
Therefore, $VI(F^\mathbb{C},\mathcal{K})$ is called a monotone VI and has a (possibly empty) convex solution set.
\item
If $JF^\mathbb{C}(Z)$ is Augmented Positive Definite (APD) on $\mathcal{K}$, then $F^\mathbb{C}$ is strictly monotone on $\mathcal{K}$. $JF^\mathbb{C}(Z)$ is APD if the inequality in \eqref{aug} is strict. Hence, $VI(F,\mathcal{Q})$ is a strictly monotone VI and has at most one solution (if there exists any).
\item{}
$F^\mathbb{C}$ is strongly monotone on $\mathcal{K}$ with constant $c_{s} > 0$ if and only if $JF^\mathbb{C}(Z)$ is uniformly APD on $\mathcal{K}$ with constant $c_{s}/2$. That is, for all $Y \in \mathbb{C}^{N^\prime \times N}$ and $Z \in \mathcal{K}$, there exists a constant $c_{s}$ such that
{\begin{equation}
\label{strmon}
[\ve(Y^*)^T ,\ve(Y)^T]JF^\mathbb{C}(Z)[\ve(Y)^T ,\ve(Y^*)^T]^T \ge c_{s}||Y||_F^2
\end{equation}}
where $||.||_F$ is the Frobenius norm. Hence, $VI(F,\mathcal{Q})$ is a strongly monotone VI and has a unique solution.
\end{itemize}
\end{proposition}
We write the augmented Jacobian matrix for $F^\mathbb{C}({\bf \Sigma}, \bt W)$ according to \eqref{augm}. Let $D_{Z}F^\mathbb{C}(Z)$ be defined as
{\begin{equation}
\label{jacobz}
D_{Z}F^\mathbb{C}(Z) \triangleq \left[\begin{array}{ccc}
D_{Z_1}F_1^\mathbb{C}(Z_1)&\dots&D_{Z_Q}F_1^\mathbb{C}(Z_1)
\\
\vdots&\ddots&\vdots
\\
D_{Z_1}F_Q^\mathbb{C}(Z_Q)&\dots&D_{Z_Q}F_Q^\mathbb{C}(Z_Q)
\end{array} \right]
\end{equation}}
where $D_{Z_l}F_q^\mathbb{C}(Z_q)$ for all $q,l \in {1,...,Q}^2$ is defined as
{\begin{equation}
\label{jacobzentry}
D_{Z_l}F_q^\mathbb{C}(Z_q) \triangleq \left[\begin{array}{cc}D_{{\bf \Sigma}_l}(-\nabla_{{\bf \Sigma}_q}\bar{f}_q)&D_{\bt W_l}(-\nabla_{{\bf \Sigma}_q}\bar{f}_q)
\\
D_{{\bf \Sigma}_l}(-\nabla_{\bt W_q}\bar{f}_q)&D_{\bt W_l}(-\nabla_{\bt W_q}\bar{f}_q)
\end{array}\right],
\end{equation}}
and $D_{Z^*}F^\mathbb{C}(Z) =  D_Z(F^\mathbb{C}(Z)^* = 0$ (cf.  \eqref{FofNE}). Thus the matrix $JF^\mathbb{C}$ becomes a block diagonal matrix. {For a QNE to be unique, the matrix $JF^\mathbb{C}$ has to satisfy inequality \eqref{aug} with strict inequality. Since the game is proved to have at least one QNE (using Theorem 2), and since a strictly monotone VI has at most one solution (if there exists any), then the strict monotonicity of the resulting VI from the game is sufficient to prove the uniqueness of QNE.
The strict monotonicity property requires $JF^\mathbb{C}$ to be APD. In order to satisfy this condition, we only need $ D_ZF^\mathbb{C}(Z)$ to be Positive Definite (PD). }Given $F^\mathbb{C}$ in \eqref{FofNE}, the entries of $D_{Z_l}F_q^\mathbb{C}(Z_q)$ are:
{\begin{align}
&D_{{\bf \Sigma}_l}(-\nabla_{{\bf \Sigma}_q}\bar{f}_q) \triangleq \sum_{k=1}^K\left(\Lambda_{q,l,k}\otimes \bt G_{qk}^H \bt S_{q,k}\bt G_{qk}\right)-\Psi_{ql}^* \otimes \Psi_{ql}.
\\[-20mm] \nonumber
\end{align}}

where:
{\begin{align}
& \Psi_{ql} \triangleq -\bt H_{qq}^H\left(\bt M_q+\bt H_{qq}{\bf \Sigma}_q \bt H_{qq}^H\right)^{-1}\bt H_{ql},
\\[-13mm]\nonumber
\end{align}}
{\small \begin{align}
&\Lambda_{q,l,k} \triangleq \left\{\begin{aligned}
&
\frac{\beta e^{\beta \varphi_{e,q,k}}}{\left(\sum_{k^\prime = 1}^Ke^{\beta\varphi_{e,q,k^\prime}}\right)^2}\bt G_{lk}^H \left(\bt S_{q,k}-\bt M_{e,q,k}^{-1}\right)\bt G_{lk}-\frac{\beta e^{\beta\varphi_{e,q,k}}}{\left(\sum_{k^\prime = 1}^Ke^{\beta\varphi_{e,q,k^\prime}}\right)^2}\sum_{k^\prime = 1}^K\left(e^{\beta\varphi_{e,q,k^\prime}}\bt G_{lk}^H \left(\bt S_{q,k}-\bt M_{e,q,k}^{-1}\right)\bt G_{lk}\right),~~l \neq q,
\\
&\frac{\beta e^{\beta \varphi_{e,q,k}}}{\left(\sum_{k^\prime = 1}^Ke^{\beta\varphi_{e,q,k^\prime}}\right)^2}\bt G_{qk}^H \bt S_{q,k}\bt G_{qk}-\frac{\beta e^{\beta\varphi_{e,q,k}}}{\left(\sum_{k^\prime = 1}^Ke^{\beta\varphi_{e,q,k^\prime}}\right)^2}\sum_{k^\prime = 1}^K\left(e^{\beta\varphi_{e,q,k^\prime}}\bt G_{qk^\prime}^H \bt S_{q,k^\prime}\bt G_{qk^\prime}\right),~~l = q,
\end{aligned}\right.
\end{align}}
and the operator $\otimes$ represents the Kronecker product. Furthermore,
{\begin{align}
\label{hesssymm}
&D_{\bt W_l}(-\nabla_{{\bf \Sigma}_q}\bar{f}_q) \triangleq D_{{\bf \Sigma}_l}(-\nabla_{\bt W_q}\bar{f}_q) =
\sum_{k=1}^K\left(\Omega_{q,l,k} \otimes \bt G_{qk}^H \bt S_{q,k}\bt G_{qk}\right) -\Psi_{ql}^* \otimes \Psi_{ql}
\end{align}}
where {$\forall (l,q) \in \{1,\dots,Q\}^2$,}
{\begin{align}
&\Omega_{q,l,k} \triangleq \frac{\beta e^{\beta \varphi_{e,q,k}}}{\left(\sum_{k^\prime = 1}^Ke^{\beta\varphi_{e,q,k^\prime}}\right)^2}\bt G_{lk}^H \left(\bt S_{q,k}-\bt M_{e,q,k}^{-1}\right)\bt G_{lk}-\frac{\beta e^{\beta\varphi_{e,q,k}}}{\left(\sum_{k^\prime = 1}^Ke^{\beta\varphi_{e,q,k^\prime}}\right)^2}\sum_{k^\prime = 1}^K\left(e^{\beta\varphi_{e,q,k^\prime}}\bt G_{lk^\prime}^H \left(\bt S_{q,k^\prime}-\bt M_{e,q,k^\prime}^{-1}\right)\bt G_{lk^\prime}\right),
\end{align}}
and the first inequality in \eqref{hesssymm} holds because both of the derivatives $D_{\bt W_l}(-\nabla_{{\bf \Sigma}_q}\bar{f}_q)$ and $D_{{\bf \Sigma}_l}(-\nabla_{\bt W_q}\bar{f}_q)$ are continuous which implies the symmetry of the Hessian matrix (i.e., equality of mixed derivatives). Lastly,
{\begin{align}
&D_{\bt W_l}(-\nabla_{\bt W_q}\bar{f}_q) \triangleq \sum_{k=1}^K\left(\Omega_{q,l,k}\otimes\bt G_{qk}^H \bt S_{q,k}\bt G_{qk} -\Omega_{q,l,k}\otimes\bt G_{qk}^H \bt M_{e,q,k}^{-1}\bt G_{qk} +\pi_{q,l,k}\otimes \pi_{q,l,k}\right)-\Psi_{ql}^*\otimes\Psi_{ql}
\\
\text{where} \nonumber
\\
&\pi_{q,l,k} \triangleq \bt G_{qk}^H \bt M_{e,q,k}^{-1}\bt G_{lk}.
\end{align}}
{Recalling equations \eqref{jacobz} and \eqref{jacobzentry} again, to prove $D_{Z}F^\mathbb{C}(Z)$ is PD, we rely on the generalized Gerschgorin circle theorem \cite{horn}.}
\label{hornn}
Specifically, for a block matrix $\bf{A}$ in which the blocks $A_{ij},~(i,j) = 1,\dots,M$ are $N\times N$ matrices with complex entries, define the matrix norm $|||\bullet|||$ in $\mathbb{C}^{N\times N}$as follows:
{\begin{equation}
|||A_{ij}||| \triangleq \underset{x \in \mathbb{C}^N}{\sup}\frac{||A_{ij}x ||}{||x||}.
\end{equation}}
where $||\bullet||$ is a vector norm on $\mathbb{C}^N$.
Using the Gerschgorin circle theorem, every eigenvalue $\lambda$ of $\bf{A}$  satisfies
{\begin{equation}
\label{ger}
|||(A_{ii}-\lambda I)^{-1}|||^{-1} \le \sum_{\underset{k \neq i}{k = 1}}^M |||A_{i,k}|||
\end{equation}}
for at least one $1\le i\le M$, where $|||A^{-1}|||^{-1} \triangleq \inf_{x \in \mathbb{C}^N}\frac{||Ax||}{||x||}$, and $I$ is the identity matrix.
\begin{proposition}\cite{horn}
\label{diagequiv}
If the diagonal block $A_{ii},~i = 1,\dots,M$ of the block matrix $\bf{A}$ are nonsingular and if
{\begin{equation}
\label{dominance}
|||A_{i,i}^{-1}|||^{-1} \ge \sum_{\underset{k \neq i}{k = 1}}^M |||A_{i,k}|||,~i = 1,\dots,M
\end{equation}}
for norm $|||\bullet|||$ in $\mathbb{C}^{N\times N}$ (where $|||A_{i,i}^{-1}|||^{-1} = \underset{x \in \mathbb{C}^N}{\inf}\frac{||A_{ij}x ||}{||x||}$), then $\bf{A}$ is a diagonally dominant matrix.
Also if the diagonal blocks are PSD, the condition in \eqref{dominance} is sufficient for the matrix $\bf{A}$ to be PSD.
\end{proposition}
We can use the above Gerschgorin circle theorem, Proposition \ref{prop2}, and Proposition \ref{diagequiv} on $D_ZF^\mathbb{C}(Z)$ defined in \eqref{jacobz} to obtain the set of conditions with which the augmented Jacobian matrix $JF^\mathbb{C}$ is APSD.
We also set the norm $|||\bullet|||$ to be the spectral norm.
(i.e., $ |||A|||_2 = \sqrt{\lambda_{\max}\left(A^HA\right)}$ where $\lambda_{\max}(\bullet)
$ denotes the spectral radius of a matrix). Therefore, for $JF^\mathbb{C}$ to satisfy the condition in \eqref{dominance}, we must have \cite[Chapter 6.1]{horn}
{\begin{equation}
\label{domreal2}
|\lambda_{q,\min}| \ge \sum_{\underset{q \neq l}{q = 1}}^Q |||D_{Z_l}F_q^\mathbb{C}(Z_q)|||_2,~q = 1,\dots,Q
\end{equation}}
where $\lambda_{q,\min}$ is the smallest eigenvalue of $D_{Z_q}F^\mathbb{C}_q(Z_q)$. {Using the strict inequality to \eqref{domreal2} --as required by the strict monotonicity-- and since the diagonal blocks of $D_{Z}F^\mathbb{C}(Z)$ are already PSD (i.e., $\lambda_{q,\min} \ge 0$ due to concavity of $q$th player's utility to $({\bf \Sigma}_q,\bt W_q)$), 
then the condition in \eqref{domreal2} changes to
{\begin{equation}
\lambda_{q,\min} > \sum_{\underset{q \neq l}{q = 1}}^Q |||D_{Z_l}F_q^\mathbb{C}(Z_q)|||_2,~q = 1,\dots,Q
\end{equation}}}
\ifCLASSOPTIONcaptionsoff
  \newpage
\fi
\ifCLASSOPTIONcaptionsoff
  \newpage
\fi
{\bibliographystyle{IEEEtran}
%
\bibliography{reff2}
}
$\vspace{-20mm}$
\begin{IEEEbiography}[{\includegraphics[width=1in,height=1.25in,clip,keepaspectratio]{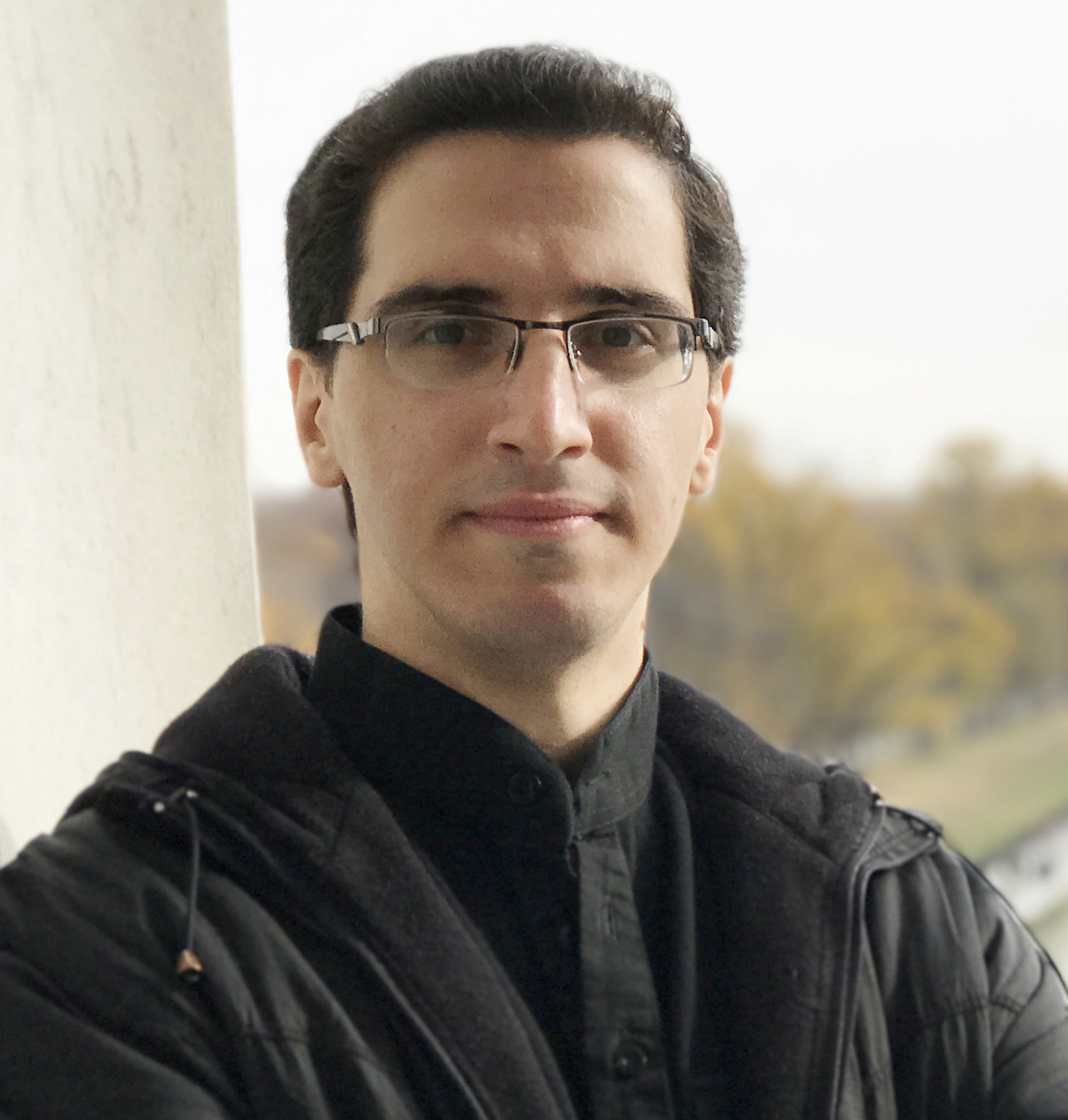}}]{Peyman Siyari} received the B.Sc. degree from Semnan University, Iran, in 2011, and the M.Sc. degree from AmirKabir University of Technology, Iran, in 2013, all in Electrical Engineering. He is working toward his Ph.D. at the University of Arizona. Peyman's research interests include wireless communications and networking, physical layer security, applications of convex optimization in signal processing, and game theory.
\end{IEEEbiography}
$\vspace{-20mm}$
\begin{IEEEbiography}[{\includegraphics[width=1in,height=1.25in,clip,keepaspectratio]{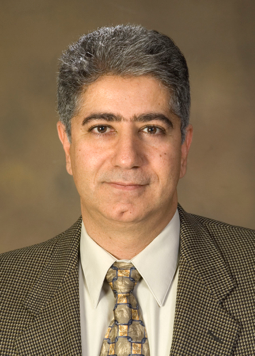}}]{Marwan Krunz}
is the Kenneth VonBehren Endowed Professor in the Department of ECE at the University of Arizona. He  also holds a joint appointment as a professor of computer science. He co-directs the Broadband Wireless Access and Applications Center, a multi-university industry-focused NSF center that includes 16+ industry affiliates. He previously served as the UA site director for Connection One, an NSF IUCRC that focuses on wireless communication circuits and systems. He received his Ph.D. degree in electrical engineering from Michigan State University in 1995 and joined the University of Arizona in January 1997, after a brief postdoctoral stint at the University of Maryland. In 2010, he was a Visiting Chair of Excellence at the University of Carlos III de Madrid. He previously held various visiting research positions at University Technology Sydney, INRIA-Sophia Antipolis, HP Labs, University of Paris VI, University of Paris V, University of Jordan, and US West Advanced Technologies. Dr. Krunz's research interests lie in the areas of wireless communications and networking, with emphasis on resource management, adaptive protocols, and security issues. He has published more than 245 journal articles and peer-reviewed conference papers, and is a co-inventor on several US patents. He is an IEEE Fellow, an Arizona Engineering Faculty Fellow (2011-2014), and an IEEE Communications Society Distinguished Lecturer (2013 and 2014). He was the recipient of the 2012 IEEE TCCC Outstanding Service Award. He received the NSF CAREER award in 1998. He currently serves on the editorial board for the IEEE Transactions on Cognitive Communications and Networks. Previously, he served on the editorial boards for IEEE/ACM Transactions on Networking, IEEE Transactions on Mobile Computing (TMC), IEEE Transactions on Network and Service Management, Computer Communications Journal, and IEEE Communications Interactive Magazine. Effective January 2017, he will be the next EiC for TMC. He was the general vice-chair for WiOpt 2016 and general co-chair for WiSec'12. He was the TPC chair for WCNC 2016 (Networking Track), INFOCOM'04, SECON'05, WoWMoM'06, and Hot Interconnects 9. He has served and continues to serve on the steering and advisory committees of numerous conferences and on the panels of several funding agencies. He was a keynote speaker, an invited panelist, and a tutorial presenter at numerous international conferences. See http://www2.engr.arizona.edu/$\sim$ krunz/ for more details.
\end{IEEEbiography}
$\vspace{-20mm}$
\begin{IEEEbiography}[{\includegraphics[width=1in,height=1.25in,clip,keepaspectratio]{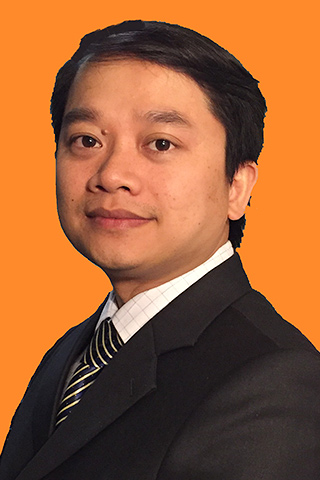}}]{Diep N. Nguyen}
is a faculty member of the School of Computing and Communications, University of Technology Sydney (UTS). He received M.E. and Ph.D. in Electrical and Computer Engineering from University of California, San Diego (UCSD) and The University of Arizona (UA), respectively. Before joining UTS, he was a DECRA Research Fellow at Macquarie University, a member of technical staff at Broadcom (California), ARCON Corporation (Boston), consulting the Federal Administration of Aviation on turning detection of UAVs and aircraft, US Air Force Research Lab on anti-jamming. He has received several awards from LG Electronics, University of California, San Diego, The University of Arizona, US National Science Foundation, Australian Research Council, including nominations for the outstanding graduate student (2012), outstanding RA (2013) awards, the best paper award finalist at the WiOpt conference (2014), discovery early career researcher award (DECRA, 2015). His recent research interests are in the areas of computer networking, wireless communications, and machine learning, with emphasis on systems' performance and security/privacy.
\end{IEEEbiography}
\end{document}